\documentclass[12pt]{article}
\usepackage[utf8]{inputenc}
\usepackage[T1]{fontenc}

\usepackage{multicol}
\usepackage{mathtools} 
\usepackage{amssymb}
\usepackage{pdflscape} 
\usepackage{multirow}
\usepackage{colortbl}
\usepackage[dvipsnames]{xcolor}
\definecolor{burgundy}{rgb}{0.5, 0.0, 0.13}
\definecolor{darkblue}{rgb}{0.0, 0.0, 0.55}
\RequirePackage[colorlinks,citecolor=burgundy,linkcolor=darkblue,urlcolor=darkblue]{hyperref}
\usepackage[autostyle]{csquotes}
\usepackage[english]{babel}
\usepackage{longtable} 
\usepackage{ragged2e}
\usepackage{array}
\usepackage{lmodern}
\usepackage{rotating}
\usepackage{graphicx}
\usepackage{pdfpages}
\usepackage{lscape}
\usepackage{afterpage}
\usepackage{fancyhdr}
\usepackage{geometry}
\usepackage{wrapfig}
\usepackage{subcaption}
\usepackage{booktabs}
\usepackage{tabularx}
\usepackage{tikz}
\usetikzlibrary{decorations.pathreplacing,positioning}
\usepackage{threeparttable}
\usepackage{dcolumn}
\usepackage{caption}
\usepackage{float} 
\usepackage[export]{adjustbox} 
\usepackage{url}
\usepackage[round]{natbib}
\usepackage{setspace}
\usepackage{indentfirst}
\usepackage{enumitem}
\setlist{nosep}
\setlist[itemize]{noitemsep, topsep=0pt}
\setlist[enumerate]{noitemsep, topsep=0pt}
\usepackage{titlesec}
\titleformat{\subsection}[runin]{\normalfont\normalsize\bfseries}{\thesubsection.}{0.5em}{}[:]
\titlespacing*{\subsection}{0pt}{12pt plus 3pt minus 2pt}{0.7em}
\titleformat{\paragraph}[runin]{\normalfont\normalsize\bfseries}{}{0pt}{}
\titlespacing*{\paragraph}{0pt}{10pt plus 2pt minus 2pt}{0.5em}
\renewcommand{\thefigure}{\Roman{figure}}
\makeatletter
\renewcommand{\fnum@figure}{\textup{FIGURE \thefigure}}
\makeatother
\fancypagestyle{plain}{
  \fancyhf{}
  \fancyhead[LE,RO]{\thepage}

}
\newcounter{prop}
\renewcommand{\theprop}{\arabic{prop}}
\newcounter{assump}
\renewcommand{\theassump}{A\arabic{assump}}
\newcounter{hyp}
\renewcommand{\thehyp}{H\arabic{hyp}}
\newcommand{\proposition}[1]{%
  \refstepcounter{prop}%
  \textbf{Proposition \theprop}\label{#1}%
}
\makeatletter
\let\old@footnotetext@footnotetext
\makeatother
\renewcommand{\thefigure}{\Roman{figure}}

\makeatletter
\renewcommand\footnoterule{%
  \kern-3\p@
  \hrule\@width 0.4\textwidth
  \kern2.6\p@}
\makeatother

\makeatletter
\patchcmd{\pprintMaketitle}{\vskip18pt}{\vskip4pt}{}{}
\patchcmd{\pprintMaketitle}{\vskip36pt}{\vskip6pt}{}{}
\patchcmd{\MaketitleBox}{\vskip18pt}{\vskip4pt}{}{}
\patchcmd{\MaketitleBox}{\vskip36pt}{\vskip6pt}{}{}
\makeatother
\makeatletter
\patchcmd{\pprintMaketitle}{\hrule\vskip12pt}{\hrule\vskip30pt}{}{}
\patchcmd{\pprintMaketitle}{\unvbox\absbox\par\vskip10pt}{\unvbox\absbox\par\vskip4pt}{}{}
\patchcmd{\pprintMaketitle}{\ifvoid\keybox\else\unvbox\keybox\par\vskip10pt\fi\hrule\vskip12pt}{\ifvoid\keybox\else\unvbox\keybox\par\vskip10pt\fi\vskip12pt}{}{}
\patchcmd{\MaketitleBox}{\ifvoid\keybox\else\unvbox\keybox\par\vskip10pt\fi\hrule\vskip12pt}{\ifvoid\keybox\else\unvbox\keybox\par\vskip10pt\fi\vskip12pt}{}{}
\makeatother
\usepackage{latexml}
\begin{document}

\newcommand{\PaperTitleA}{Where Has the Foreign Signal Gone?}
\newcommand{\PaperTitleB}{Contributor Composition and the Reliability of Social Media Sentiment}
\newcommand{\PaperAck}{We thank participants at the 2026 North American Summer Meeting of the Econometric Society at Emory University and at the 20th CIREQ-CIRANO Conference in Montr\'eal for helpful comments. Hounyo gratefully acknowledges financial support from the 2025--2026 Faculty Research Award Program, which funded the purchase of the Context Analytics sentiment data from Dewey Data. We thank the Massry School of Business for access to CRSP and Compustat data through Wharton Research Data Services (WRDS). The views expressed herein are those of the authors.}
\newcommand{\PaperCopyright}{\copyright\ 2026 by Piyush Akimitsu and Ulrich Hounyo. All rights reserved. Short sections of text, not to exceed two paragraphs, may be quoted without explicit permission provided that full credit, including \copyright\ notice, is given to the source.}
\newcommand{\PaperJEL}{JEL No. G12, G14, G15, G41}
\newcommand{\PaperKeywords}{Keywords: social media sentiment, information production, investor attention, home bias, noisy rational expectations}
\newcommand{\PaperAbstract}{Social media sentiment is increasingly used to predict stock returns. Its reliability depends on the composition of the crowd generating the feed. For foreign firms, the home-market investors who know them best are underrepresented on English-language Twitter. Hence, their feed should carry a less precise signal. Reduced-form analysis of firm-level Twitter sentiment shows that sentiment noise predicts next-quarter losses less consistently for foreign stocks than for domestic stocks. The loss-on-noise slope is negative in 78 percent of quarters for domestic stocks but in only 48 percent for foreign stocks, with a significant mean difference in slopes. We interpret this difference through a noisy rational expectations model in which the feed carries a public signal produced by three types of contributors. Specialists post repeatedly with information about the firm, attention-driven contributors post once with information about the firm and a transient fad, and automated accounts post repeatedly with no information about the firm. The share of each contributor type and the size of the fad determine the precision of the signal, which, in turn, determines how much of the feed's noise reaches prices. The model gives repeat-author share, fad and precision from the vendor's aggregate metrics, without account-level data, whereas prior work backs out precision from price informativeness. The structural estimates show that feeds for foreign stocks have a higher share of repeat authors, a larger fad component and a less precise signal. Thus, the foreign signal did not disappear but got diluted.}

\iflatexml
\begin{center}
{\large\bfseries \PaperTitleA\\
\PaperTitleB}\\[1ex]
Piyush Akimitsu\footnote{Department of Economics, State University of New York at Albany. \href{mailto:econpiyush@icloud.com}{econpiyush@icloud.com}.}\quad
Ulrich Hounyo\footnote{Department of Economics, State University of New York at Albany. \href{mailto:khounyo@albany.edu}{khounyo@albany.edu}.}\\[0.5ex]
State University of New York at Albany, Department of Economics\\
November 2026\footnote{\PaperAck\ \PaperCopyright\ \PaperJEL. \PaperKeywords.}
\end{center}

\begin{center}\underline{ABSTRACT}\end{center}
\noindent \PaperAbstract
\else
\pagenumbering{gobble}
\addtocontents{toc}{\protect\setcounter{tocdepth}{-10}}

\begin{singlespace}
\begin{center}
\vspace*{1.6in}
\MakeUppercase{\PaperTitleA}\\
\MakeUppercase{\PaperTitleB}\\[1.5\baselineskip]
Piyush Akimitsu \qquad Ulrich Hounyo\\[2.2\baselineskip]
\MakeUppercase{State University of New York at Albany}\\
Department of Economics\\
November 2026
\end{center}
\vfill
\noindent \PaperAck

\vspace{\baselineskip}
\noindent \PaperCopyright
\end{singlespace}
\clearpage

\begin{singlespace}
\noindent \PaperTitleA\ \PaperTitleB\\
Piyush Akimitsu and Ulrich Hounyo\\
November 2026\\
\PaperJEL\\
\PaperKeywords

\vspace{\baselineskip}
\begin{center}\underline{ABSTRACT}\end{center}

\noindent
\PaperAbstract

\vspace{1.5\baselineskip}
\noindent Piyush Akimitsu\\
Department of Economics\\
State University of New York at Albany\\
\href{mailto:econpiyush@icloud.com}{econpiyush@icloud.com}

\vspace{0.6\baselineskip}
\noindent Ulrich Hounyo\\
Department of Economics\\
State University of New York at Albany\\
\href{mailto:khounyo@albany.edu}{khounyo@albany.edu}
\end{singlespace}
\clearpage
\pagenumbering{arabic}
\fi

\onehalfspacing


\section{Introduction}\label{sec:intro}

Data vendors now sell firm-level social media sentiment to investors. A large branch of literature finds that text-based sentiment predicts returns \citep{antweiler2004talk, tetlock2007giving, chen2014wisdom, bartov2018twitter}. The reliability of these signals depends on who writes the posts. A sentiment index is derived from a feed to which people with information about the firm contribute, along with fads and automated content that carry no information about the firm. The index itself does not reveal the share of each contributor type, the information content of the feed, or the precision of the signal generated by such a feed. We study how much the composition of the contributor pool affects the reliability of sentiment as a signal about prices. For this, we need contributor pools that differ in a meaningful way. The literature on home bias shows that geographical proximity gives investors real informational advantages, so the investors who know a firm best are concentrated in its home market. When a firm is headquartered in a foreign country but listed in the U.S., the home-market investors who know it best are underrepresented among those who post about it on English-language Twitter.\footnote{Twitter was renamed X in July 2023. We keep the name Twitter because our primary sample starts in 2018 and the vendor's data carry that label.} Thus, comparing domestic and foreign firms listed in the U.S. allows us to compare feeds generated by crowds of different contributor compositions. We are interested in the share of each type of contributor, and geography provides variation in those shares.

Using a noisy rational expectations framework, this paper models the composition of the contributor pool and the size of the noise the pool produces, and estimates both using the generalized method of moments (GMM). Specifically, for the composition of the contributor pool, we estimate the share of repeat authors (the accounts which post about the same stock more than once on a day). Additionally, for the noise, we estimate the scale of the fad component (the transient component of the Twitter feed which carries no information about the firm). Our hypothesis is that the foreign feed is less informative than the domestic feed. The results show that both the share of the repeat authors and the scale of the fad component are larger for foreign stocks than for domestic stocks. The headline result is that the reliability of information in the social media sentiment for foreign stocks differs systematically from that of the domestic stocks. An economically coherent explanation for this difference comes from analyzing the contributor composition, which differs systematically between feeds for foreign and domestic stocks. Our structural model disciplines this interpretation. A firm's geographical location, whether domestic or foreign, is observable in the data. Thus, sorting on such geographical location lets us compare the contributor pools that differ in composition and lets us measure the difference in reliability between the two groups.

Our hypothesis about the mechanism behind this difference is the mismatch between contributors who write the tweet and people who hold information about the firm (the producer-knowledge mismatch hypothesis). Under this hypothesis, if a firm has headquarters in a foreign country (foreign firm), then most of the contributors who post about such a firm on English-language Twitter are likely not the ones who know the firm best. Even though Twitter has been a global social media platform, under our hypothesis English-language discussion of the U.S.-listed stocks is dominated disproportionately by U.S.-based users.\footnote{U.S. users accounted for approximately 20 percent of monetizable daily active users as of 2019 \citep{twitter2019annual}. \citet{paydarzarnaghi2023twitter} analyze over 9 million stock-specific tweets and show that the U.S. demographic user groups drive equity price reactions substantially more than any other group.} The U.S. investors lean more toward domestic stocks, which represents information concentration where the U.S. investors are rationally deepening their knowledge about domestic firms rather than diversifying \citep{van2009information, mondria2010home, sims2003rational}. The home-market investors of a firm with headquarters in a foreign country tend to possess an informational advantage about such a firm \citep{dvorak2005domestic, bae2008local}. The advantage is visible in prices, since home markets lead price discovery for cross-listed stocks on average \citep{eun2003cross}.\footnote{The informational advantage of the home-market investors about these foreign firms applies to both American Depositary Receipts (ADRs) and directly listed foreign shares, which form our subsample of foreign stocks. More than half of the subsample of such foreign stocks belongs to non-English-speaking countries. We call a country English-speaking when English is a first language there, following \citet{crystal2003english}.} Under our hypothesis, these home-market investors of foreign firms are underrepresented on English-language Twitter. This implies that the signal for foreign firms should have a weaker information component and more noise than the signal for domestic stocks. Investors cannot separate the signal from its error. Therefore, the prices absorb the noise carried by the signal. Since a foreign signal is noisier, the prices of foreign stocks are noisier too. Limited arbitrage lets those price effects persist \citep{shleifer1997limits}.

This matters for three main reasons. First, investors who trade using Twitter sentiment get a signal whose reliability varies with firm geography. The variation in reliability is concealed if the sentiment signals from the feed of foreign and domestic stocks are weighted equally. Second, the conventional sentiment indices are aggregates of sentiment scores across tweets for a firm's stock. This aggregate doesn't reveal either the composition of the pool by contributor type (specialists, attention-driven contributors posting information about the firm along with the day's fads, and automated accounts whose feed carries no information about the firm) or how much noise that composition adds. Third, the question bears on a large class of listings. Foreign firms list in the U.S. to reach U.S. investors. The recognition they gain lowers their cost of raising capital \citep{merton1987simple, karolyi1998companies}. A noisier feed could move the prices of the foreign listings for reasons unrelated to the true firm value.

We formalize this idea in a noisy rational expectations model in which the Twitter feed acts as a public signal \citep{hellwig1980aggregation, admati1985noisy, farboodi2022data}. We add three features to the standard setup. The first is the repeat-author share. For a given posting rate, the vendor's dispersion measure (distinct accounts per tweet) gives the share of repeat authors. The second one is the short-sale constraint \citep{miller1977risk} which allows part of the overpricing caused by optimistic noise to carry over into the next quarter. The third one is a daily dealer market \citep{kyle1985continuous} in which quoted spreads rise with informed trading. Since informed trading is not observed, we use the spread to rank stocks by it and compare domestic and foreign stocks within terciles of that rank. The model gives closed-form expressions for the moments of sentiment, returns, gains and losses, liquidity and price informativeness.

We estimate the model using GMM on the 2018 to 2024 primary sample, drawn from a panel of 471,880 stock-quarters of U.S.-listed firms covering 2010 to 2024, with firm-level Twitter sentiment from Context Analytics.\footnote{The vendor changed its sampling methodology in 2017. So, the pre-change feed is not comparable with the post-change feed. Hence, the primary sample begins in 2018 and earlier quarters are not used in the estimation.} Domestic and foreign stocks are sorted separately into terciles of the model-based informed-trading-intensity rank. Domestic-foreign comparisons are then made within each tercile of that rank. Moreover, the platform-wide changes in feed volume move measured noise for every stock. Hence, in the reduced-form regressions the domestic-foreign comparisons are made quarter by quarter. This quarter-by-quarter comparison keeps those common movements out of those regressions.

The paper has three main results. The first one is obtained from reduced-form quarterly regressions. The temporal stability across quarters of the predictive relation between measured sentiment noise and next-quarter losses, the signal's reliability, differs systematically between domestic and foreign stocks. In these regressions of next-quarter losses on sentiment noise,\footnote{The noise is the within-quarter volatility of daily per-tweet sentiment.} the mean quarterly slope is more negative for domestic stocks than for foreign stocks by 1.40 (Newey-West $t=-2.99$). In economic terms, a one-standard-deviation increase in sentiment noise increases the next-quarter losses by 0.64 percentage points of quarterly excess return for domestic stocks. The domestic-foreign difference in that response is 1.39 percentage points, which is about two thirds of the sample's mean quarterly excess return. This domestic-foreign contrast also appears as a negative quarterly slope for 78 percent of quarters for domestic stocks as opposed to 48 percent of the quarters for foreign stocks in the regressions of next-quarter losses on sentiment noise. This difference of 30 percentage points has a block-bootstrap $p$ value of 0.016 (row R14, Table~\ref{tab:battery}). If this pattern is broken down by terciles, this domestic-foreign gap in the share of quarters with a negative slope is significant in the lowest informed-trading-intensity tercile.\footnote{In the lowest intensity tercile the share of quarters with a negative slope is 0.852 for domestic stocks and 0.556 for foreign stocks (a difference of 0.296, block-bootstrap $p = 0.004$).} The gap is positive but not significant in the middle tercile and is close to zero in the top tercile where the informed trading intensity is heaviest. Additionally, the mean slope difference between domestic and foreign is strongest for ADRs at $-2.00$ (Newey-West $t=-3.72$). An ADR's U.S. price is tied to its home-market share price by arbitrage, the depositary link, which itself could weaken the loss-on-noise relation. Therefore, we look at the directly listed foreign shares, which do not have this link. The estimate of the slope difference keeps its sign at $-0.74$ but is imprecise (Newey-West $t=-1.29$, \ref{app:rob_r14d}). Thus, the mean slope difference lessens for the foreign shares without the depositary link, from 1.40 to 0.74, but keeps its sign, which suggests that the gap is not an artifact of the ADRs. In the terms of our title, the foreign signal has not disappeared but has been diluted, since its link to next-quarter losses keeps its sign in about one quarter in two rather than in nearly four in five.

The second result comes mainly from the structural estimation. The foreign feed's contributor pool has a higher share of repeat authors than that of the domestic feed, and the foreign feed also carries a larger fad component than the domestic feed. The estimated share of repeat authors is higher for foreign stocks than for domestic stocks in all terciles of informed trading intensity. Thus, the domestic-foreign gap in repeat-author share doesn't merely reflect the domestic-foreign gap in informed trading intensity. The domestic-foreign gap in repeat-author share is the widest among stocks with the most informed trading intensity. In the top tercile, the repeat-author share for foreign and domestic are 0.170 and 0.145, respectively (Table~\ref{tab:theta_hat}). The fad scale, which is a standard deviation in the vendor's per-tweet sentiment units, is 0.0293 for foreign stocks against 0.0273 for domestic stocks, so the variance of that component, which is noise by construction, is about 15 percent larger for foreign stocks. A joint Wald test rejects that the four parameters, the repeat-author share in each of the three informed-trading-intensity terciles and the fad scale, are equal for domestic and foreign stocks.\footnote{A joint Wald test rejects the null of equality of these four parameters for foreign and domestic stocks ($p = 7.2\times 10^{-10}$ with standard errors clustered by firm, $p = 1.9\times 10^{-14}$ clustered by quarter and $p = 1.9\times 10^{-5}$ double clustered by firm and quarter, Table~\ref{tab:h1_contrasts}). The foreign fad scale and the foreign top-tercile repeat-author share exceed the domestic ones in each of the 399 bootstrap replications.} The repeat-author share and the fad scale are two measures obtained from structural estimation. The third measure is the observable signal-quality index obtained directly from the observed daily feed readings. The higher the feed precision, the higher this index. The median of this index for foreign stocks is lower than that for domestic stocks in all intensity terciles, by 17 percent in the low tercile, 16 percent in the middle tercile and 7 percent in the top tercile (Table~\ref{tab:r6_dataterm}). Thus, the foreign feed carries a less precise signal than the domestic feed at every level of informed trading intensity, which shows that the precision gap is not a byproduct of the domestic-foreign gap in informed trading. The gap narrows where informed trading is heaviest.

Our third result pertains to the mechanism through which the feed's noise goes into next-quarter losses. In the model, the noise goes into the losses in two ways. The first way is through volatility. A feed with higher noise makes a day's price more volatile, which produces larger daily losses and daily gains equally. This raises losses without changing the balance between losses and gains. The second way is through overpricing. When the feed's sentiment is optimistic, the price rises above the firm's value. The investors who don't hold any stock can lower the prices back only by selling short. When short selling is closed to them, a part of the overpricing survives at the end of the quarter. In the next earnings announcement, the true value of the firm is revealed, which corrects the overpricing. This manifests as losses in the next quarter without an equivalent amount of gain to match it. This second way can make the losses exceed the gains. We ask which of the two the data support, and we use how much the price leans on the Twitter feed, which depends on feed precision. For each stock-quarter, we have a model-motivated signal-noise index that indicates how much overpricing the feed's noise would create once the price leans on the Twitter feed. This index is the product of the weight the price puts on the feed (which rises with the feed's precision) and the size of the noise the feed carries (which falls with the feed's precision). So, the index is highest at an intermediate precision level where the feed is precise enough for the market to give it real weight but still imprecise enough that it carries sizeable noise. When the price leans more on the feed than on outside knowledge (what investors know about the firm from other sources), the noise from the feed goes into the price. In the model, the more overpricing the feed's noise creates in a quarter, the larger the excess of losses over gains in the next quarter. As predicted by the overpricing correction channel, the excess of next-quarter losses over gains increases with this index ($t = 9.63$, with standard errors double clustered by firm and quarter) within each geography and informed-trading-intensity tercile. Two results prevent us from reading it as evidence for that channel. First, the estimated short-sale friction lies at its zero lower bound, so in the fitted model no overpricing survives the quarter and the correction channel is closed. Second, the channel's correction comes only from optimistic errors, so it predicts a stronger loss response after quarters with more positive-sentiment days. That stronger response does not appear, since the next quarter's loss response to noise is no steeper after quarters with more positive-sentiment days than after other quarters (row R8, Table~\ref{tab:battery}). Thus, we treat this rise as a reduced-form pattern. In the fitted model, sentiment noise reaches next-quarter losses only through return volatility. Volatility raises gains and losses alike, so it cannot produce an excess of losses over gains that rises with the index.

The model gives some additional predictions which are tested using a battery of reduced-form checks.\footnote{We re-estimate the loss-on-noise regression (row R1, Table~\ref{tab:battery}) across thirty-one subsamples and control specifications in Table~\ref{tab:additional_robustness}.} The noisier the sentiment, the larger the losses. The raw and the volume-adjusted measures of quarterly sentiment volatility, in a regression of next-quarter losses on both and tweet volume, have coefficients of $-0.317$ and $-0.612$ ($t = -6.02$ and $-2.11$, with standard errors double clustered by firm and quarter), so the loss prediction is not driven by tweet volume, to which the raw measure is tied and the volume-adjusted one is not (\ref{app:battery}). In a regression of next-quarter losses on noise and the stock's realized return volatility (the within-quarter standard deviation of its daily excess returns), the coefficients on the noise are insignificant and return volatility has a coefficient of $-14.70$ ($t = -40.91$, with standard errors double clustered by firm and quarter). Thus, noisier feeds predict larger losses because they come with more volatile returns, not because they push down the average next-quarter return. This is consistent with the short-sale friction at its zero lower bound. Next-quarter gains (the sum of a stock's positive daily excess returns over the quarter) rise with dispersion (mean ratio of distinct posting accounts to tweets within the quarter), a reduced-form regularity which the model doesn't predict. The response concentrates in the lowest intensity tercile, at 0.128 ($t = 3.15$, with standard errors double clustered by firm and quarter).

The interpretation of the structural results depends on two features. First, each stock's signal precision is recovered as its observable signal-quality index divided by a conversion constant common to all stocks. This constant absorbs the scale of the vendor's score, the serial dependence across days and the dilution of the signal by automated content, which the data do not separate (Section~\ref{sec:structural}). The estimate of this conversion constant is about 400, with a standard error of 34 (Table~\ref{tab:theta_hat}). In the domestic-foreign comparisons of signal precision, the constant cancels and the comparisons depend on the observable index alone. Second, about 85 percent of the weighted distance between model predicted and data computed moments comes from the 24 moments of the four liquidity measures, namely the quoted spread, the high-low spread, the Amihud ratio and turnover. None of the main results depend on the levels which the model recovers from those objects. The paper uses these objects mainly to rank the stocks by informed-trading-intensity terciles within which the domestic-foreign comparisons are made.

The paper makes three contributions. First, we show that the reliability of a social media sentiment signal, the stability across quarters of its link to next-quarter losses, depends on firm geography. The studies that link text-based sentiment to returns leave out of their scope who produces the sentiment and for which stocks it is reliable (Section~\ref{sec:lit}). The domestic-foreign gap in reliability holds among the stocks with the least informed trading, which shows that such a gap is not an artifact of the domestic-foreign difference in informed trading. The home-bias literature documents a home-market informational advantage in holdings and returns. We interpret this domestic-foreign gap in reliability as that advantage showing up in the feed for domestic stocks, whose home-market investors dominate English-language Twitter, and not in the feed for foreign stocks, whose home-market investors are underrepresented there. For an investor who trades on Twitter sentiment, the sentiment index is a less reliable signal for a foreign stock than for a domestic stock. Second, the paper gives the composition of the Twitter feed contributor pool and the precision of the sentiment signal from the vendor's aggregate metrics alone, without account-level data. Specifically, we get the share of repeat authors from contributor dispersion at a given posting rate (Proposition~\ref{prop:dispersion}), the fad from the part of sentiment volatility that does not fall as tweet volume rises, and the precision of the signal from that volatility as a whole (Proposition~\ref{prop:vaqsv}). Prior work backs that precision out of price informativeness, where it is a residual that cannot be traced to a source. In this paper, precision is the observable signal-quality index which is computed from the feed's daily readings, divided by one constant common to all stocks. So, the domestic-foreign comparison of precision comes from the feed alone, can be traced to the composition of the pool that produces it and can be tested against how the feed moves prices. Our hypothesis of a mismatch between who posts about a firm and who knows it is bolstered by the structural estimates which show a higher share of repeat authors and a lower signal-quality index for foreign stocks in each informed-trading-intensity tercile, and a larger fad component for the foreign feed. A part of the literature links sentiment with the feed contributors and their information. For instance, \citet{cookson2024social} relate differences in sentiment across platforms to differences in their users, while \citet{huang2024local} find that local posters on a Chinese message board are better informed than non-local ones. This paper, instead, measures contributor composition within a single platform, Twitter, and relates it to firm geography. The paper provides closed-form moment conditions which other studies and future research pertaining to other social media platforms can use to estimate the same parameters. Third, we document a regularity in next-quarter returns whose sign matches the correction of overpricing left by short-sale constraints \citep{miller1977risk}, and we find no support for that channel. Within each of the groups, domestic and foreign, and informed-trading-intensity tercile, the excess of next-quarter losses over gains rises with the model-motivated signal-noise index, which indicates how much overpricing the feed's noise would create. The estimated short-sale friction lies at its zero lower bound, and the loss response to noise is no stronger after quarters with more positive-sentiment days than after other quarters (row R8, Table~\ref{tab:battery}). Since return volatility raises gains and losses alike (Proposition~\ref{prop:vol}), neither channel of the fitted model produces the rise, and we report it as a reduced-form pattern. Thus, a loss-gain asymmetry with the sign that a mispricing channel predicts is not by itself evidence of that channel.

The rest of the paper proceeds as follows. Section 2 reviews the literature. Section 3 presents the model. Section 4 describes the data. Section 5 reports the structural estimation. Section 6 reports the reduced-form evidence. Section 7 concludes.

\section{Related Literature}\label{sec:lit}

Prior work in this area shows that investors overweigh domestic and nearby securities \citep{french1991investor}, while investment funds hold more domestic than foreign stocks \citep{lambert2026home}. This geographical proximity is more than just familiarity. It also gives real informational advantages to the investors \citep{kang1997home, ahearne2004information, coval1999home, coval2001geography}. Some other related work shows that domestic investors outperform foreign investors in the same securities \citep{dvorak2005domestic}, and that local analysts forecast more accurately in less transparent environments \citep{bae2008local}. Additionally, the investors rationally deepen initial local advantages and \citet{van2009information} provide a theoretical basis for that. In addition to geography, language shapes the attention which foreign firms receive. \citep{boulland2017news} show that investor attention rose when continental European firms began disseminating company news through an English-language electronic wire service. We study this information geography pertaining in the context of Twitter and use the domestic-foreign comparison to identify the information content of the sentiment of its feed. 

Our producer-knowledge mismatch hypothesis is based on the premise that the coverage of foreign stocks on the English-language feed by the specialists is lesser than it is for the domestic stocks. Assets which get lower recognition from the investors trade below their full-information value \citep{merton1987simple}. Investors with finite capacity direct their attention to the signals with the highest informational return \citep{sims2003rational, peng2006investor}. This leads to endogenous home bias \citep{mondria2010home, van2009information}. Retail investors are net buyers of stocks that get more attention \citep{barber2008all}. At the market level, daily aggregate retail attention predicts lower returns over the following week, while institutional attention ahead of major announcements predicts higher returns \citep{da2025market}. Evidence from surveys and experiments shows that households and finance professionals read good news about a firm as a sign of higher returns and often neglect the price adjustment that news has already caused \citep{andre2026mental}. English-language Twitter aggregates these attention decisions. Thus, the crowd-sourced signal should be more precise for domestic than for foreign stocks.

The literature establishes the connection between text-based sentiment and returns. For instance, internet message boards help predict volatility \citep{antweiler2004talk}. Pessimistic coverage in a Wall Street Journal column predicts downward price pressure which reverses later \citep{tetlock2007giving}, while the predictability from newspaper content is concentrated in recessions \citep{garcia2013sentiment}. This holds for social media as well. Seeking Alpha articles and their reader commentaries predict returns and earnings surprises \citep{chen2014wisdom}. The aggregate opinion in tweets posted just before an earnings announcement predicts the firm's earnings and its announcement return \citep{bartov2018twitter}. Similarly, a comparison between global and local news sentiment shows that the former moves returns more, and more persistently, than the latter \citep{fraiberger2021media}. \citet{allen2026squeezing} show that the discussion of the meme stocks of January 2021 on Reddit, Twitter and StockTwits strengthened the relation between retail trading and subsequent returns, and that the short squeezes which followed impaired market quality for those stocks and for their competitors. While these studies show that text-based sentiment carries price-relevant information, it is out of their scope to reflect on who produces this sentiment and for which stocks it is reliable. \citet{gomezcram2026prediction} reflect on this question for the prediction markets where prices are probabilities and contracts pay off on the outcome of the event. In their Polymarket data, about 3\% of accounts produce most of the price discovery, while the rest supply the volume of trade but little information. For social media, \citet{gilbazo2026tweeting} show one producer whose posts move money without carrying information. The posts of U.S. mutual fund families on Twitter raise the flows into their funds, and the authors reject lower search costs, higher attention and information about fund performance as the channels. We ask a similar question for social media, whether the predictive content of the feed comes from the crowd or the smaller subset of informed contributors and whether, for foreign stocks, the contributor pool tilts toward the types that carry no new information about the firm.

A recent strand of literature links this sentiment to the contributor pool's composition. \citet{cookson2024social} trace the differences in the sentiment signal of various platforms to their user population composition and platform design. \citet{huang2024local} show the informational advantage of local posters over non-local posters on a Chinese message board. Our paper focuses on English-language Twitter, links the heterogeneity in sentiment to firm geography and then estimates the contributor composition and nonfundamental component of the Twitter feed.

Our paper also relates to the literature on price informativeness. We use a noisy rational expectations model and adopt the \citet{farboodi2022data} decomposition of the \citet{bai2016informative} measure of price informativeness. In our model the precision of the sentiment signal enters that decomposition (Section~\ref{sec:model_eq}). \citet{dugast2018data} show that cheaper low-precision signals can reduce price informativeness by lowering the demand for more precise signals. The noise we study originates in the public signal and reaches prices through the weight that investors put on the feed. Its source is the part of the feed's day-to-day variation that carries no information about the firm, and that part is informative about the size of the fad and the tweet shares of the contributor types (Section~\ref{sec:model_sent}). (The related work on heterogeneous beliefs and on cross-listing is discussed in \ref{app:lit_extra}.)


\section{A Model of Sentiment, Information, and Liquidity}\label{sec:model}

Three types of contributors produce a stock's Twitter feed. (i) Specialists who post about a firm repeatedly. (ii) Attention-driven contributors who engage with any one firm occasionally. Their posts have the firm-specific signal and a transient nonfundamental component that the specialists filter out. (iii) The automated accounts which generate content mechanically. The composition of the feed based on these types determines how informative the aggregated feed is. That informativeness is given by a precision parameter in the model. A more precise public signal implies less uncertainty about firm value for the investors. Lower uncertainty implies a higher information component of price informativeness and lower return volatility. The signal noise enters prices in proportion to the weight the market rationally gives the signal. With short-sale constraints, the noise that overstates value leads to overpricing. This overpricing is corrected over the next quarter, resulting in losses. The model shows this effect is largest at an intermediate level of signal precision. A more precise signal narrows the informational advantage of informed investors over dealers, which compresses bid-ask spreads and market-impact costs. Geography, foreign and domestic, enters the model through the shares of types and the fad and bot variances. Holding the other group parameters fixed, the domestic-foreign differences in volatility, losses, informativeness, and liquidity can all be traced to differences in that signal precision.

Our framework combines three standard ingredients and a new layer: 1. A noisy rational expectations equilibrium in the tradition of \citet{hellwig1980aggregation}, \citet{admati1985noisy}, and \citet{farboodi2022data}: In such an equilibrium, prices aggregate dispersed information.\footnote{In a noisy rational expectations model investors learn from the price itself. In the supply of shares, there is a random component, without which the price would let every investor infer the private information of the others. With that random component, the price reveals that information only in part. Thus, private information still retains value. The \textit{precision} of a signal is the inverse of the variance of its noise. An investor's belief after observing the private signal, the price and the public signal has a precision equal to the precision of the belief held before observing them in addition to the individual precision of each of the three signals.} 2. The short-sale constraint in the tradition of \citet{miller1977risk}: The public signal about a firm's value has noise which makes the signal either overstate or understate that value. An understatement doesn't persist since it is removed within the quarter by ordinary buying which no investor is barred from. An overstatement raises the price and causes overpricing. Correcting it requires selling. An investor who holds no shares can sell only through a short sale. When short selling is closed to some of those investors, part of the overstatement remains in the price as overpricing. The surviving overpricing is corrected at the next earnings announcement, when the firm's realized value becomes public and the price falls to that value. Thus, overpricing today predicts low returns over the following quarter. We refer to the correction of the surviving overpricing as the drift channel (Proposition~\ref{prop:drift}). The second channel through which the signal's noise reaches gains and losses is the volatility of returns, which increases the magnitudes of gains and losses by the same amount such that the difference between those magnitudes remains unchanged (Proposition~\ref{prop:vol}). In the model, only the correction of surviving overpricing can make the magnitude of losses exceed the magnitude of gains. 3. A daily dealer market in the spirit of \citet{kyle1985continuous} and \citet{glosten1985bid}: Trading costs are the dealers' compensation for adverse selection (trading against better-informed investors). As the informed trading increases, the observed spreads and market-impact costs increase. Thus, given the feed's precision, a stock's quoted spread reveals how much informed trading it attracts. We sort stocks, within geography and quarter, into terciles of informed-trading intensity on the lagged quoted spread multiplied by the square root of prior precision plus lagged signal precision (Section~\ref{sec:model_micro}). This sort needs no records of which types of investor hold each stock. 4. The contributor model: The model links the vendor-provided dispersion metric to the concentration of repeat authors. It also splits the day-to-day variation in sentiment per tweet into two parts. For each stock and day, the data have a sentiment total which is an exponentially time-weighted sum of the sentiment scores of the tweets about the stock over the day's 24-hour window ($S_t$), and the tweet count ($N_t$). $S_t/N_t$ gives the day's sentiment per tweet. In the model, the score of each tweet contains noise specific to that tweet. An average over many tweets reduces such noise but does not eliminate it. The more tweets a day has, the less of this noise remains. This remaining noise is the sampling part of the day-to-day variation in sentiment per tweet. The fad and the automated content behave differently. The fad is a transient component of the feed for the day, so each attention-driven tweet posted that day has it. The automated content is also a component of the feed for the day, so each bot tweet posted on that day has it. Averaging the sentiment scores of the day's tweets does not reduce either of these components. The fad's contribution to a day's average sentiment score equals the fad multiplied by the fraction of the day's tweets that are attention-driven (if a tenth of the day's tweets are attention-driven then the average score carries a tenth of the day's fad, irrespective of the number of tweets). This is true also for the automated content and the fraction of the day's tweets that come from bots. These two components are the second part, the nonfundamental part, of the day-to-day variation in sentiment per tweet. The data can tell the two parts apart because the sampling part depends on the tweet count while the nonfundamental part does not (Section~\ref{sec:model_sent}). We derive how the metrics in the data link to the precision of a public signal. The nonfundamental variance is recovered from these vendor metrics. The signal precision for a quarter is recovered as the stock's observable signal-quality index (the number of days in the quarter with at least one tweet about the stock divided by the squared within-quarter volatility of daily sentiment per tweet), divided by a conversion constant that the estimation holds common across all stocks. Thus, the feed for a stock or group of stocks with a higher signal-quality index is more precise than the feed for a stock or group with a lower signal-quality index, irrespective of the value of the constant.\footnote{This index is computed for each stock-quarter. The structural estimation aggregates it to the group (domestic or foreign stocks) level for each informed-trading-intensity tercile. The headline reduced-form regressions do not use the precision, since their regressor is the within-quarter volatility of daily sentiment per tweet. Only the level of a group's precision depends on the constant. None of the findings quote that level.} The composition of contributor types is a secondary question (\ref{app:composition_check}, Section~\ref{sec:reducedform}, Table~\ref{tab:notation_map}).

\subsection{The economic environment}\label{sec:model_env}

Time is a sequence of quarters. We describe one quarter and suppress the quarter index except where needed. Within the description, periods $s = 0, 1, 2, \ldots$ count quarters from the present, so $s = 0$ is the current quarter. There is a continuum of stocks $f$. Each stock belongs to a stock group $j$, a set of stocks that share the same parameters. A stock group is defined on stocks, not on the contributors who post about them. Geography is one such grouping, splitting stocks by whether the firm is headquartered at home or abroad. The empirical work also groups stocks by the model-based informed-trading-intensity rank. The dividend that stock $f$ pays in period $s$ is $d^{*}_{fs}$. Starred symbols denote quantities in the model's value units throughout.\footnote{Assumption~\ref{assump:daily} below specifies the single positive scale that converts value units to returns.} The current dividend $d^{*}_{f0}$ is known. A share of stock $f$ is a claim to the dividend stream
\begin{equation}
d^{*}_{f1} = \upsilon_j\, d^{*}_{f0} + \bar{\epsilon}_{f1} + \epsilon_{f1}, \qquad \epsilon_{f1} \sim N(0,\Sigma_{jd}), \qquad d^{*}_{fs} = \upsilon_j^{\,s-1} d^{*}_{f1} \ \ (s \geq 2),
\end{equation}
where $\upsilon_j$ is the gross one-period persistence factor of cash flows. Let $r$ denote the gross riskless rate, one plus the net rate. We maintain $0 \leq \upsilon_j < r$ throughout, so the dividend claim has finite value. The first equality says next period's dividend equals $\upsilon_j$ times the current dividend plus two shocks. The last equality is the solved form of the recursion $d^{*}_{f,s+1} = \upsilon_j\, d^{*}_{fs}$ for $s \geq 1$ (\ref{app:proof_value}). Therefore, each dividend after period 1 is $\upsilon_j$ times the one before it. The law carries no error term, so there are no further shocks. The shock $\bar{\epsilon}_{f1}$ is a common component shared with other firms. The shock $\epsilon_{f1}$ is the firm-specific innovation, the one object investors are uncertain about. Its variance is $\Sigma_{jd}$. As in \citet{farboodi2022data}, all investors observe the common component, so all learning concerns $\epsilon_{f1}$. Trade occurs once within the quarter, at a price $P^{*}_{f1}$ set before the innovation becomes public. With no residual uncertainty after period 1, the position pays the period-1 value of the dividend stream, $V^{*}_{f1} = \sum_{s \geq 1} d^{*}_{fs}/r^{\,s-1}$, when the innovation becomes public. Substituting the dividend persistence $d^{*}_{fs} = \upsilon_j^{\,s-1} d^{*}_{f1}$ into this present-value sum gives a geometric series in $\upsilon_j/r$, which sums to $V^{*}_{f1} = \gamma_j\, d^{*}_{f1}$, where $\gamma_j \equiv r/(r-\upsilon_j)$ is the present-value multiplier of a growing perpetuity (\ref{app:proof_value}). The supply of shares is $\bar{x}_j + \tilde{x}_f$, the sum of a known mean component $\bar{x}_j$ and a random component $\tilde{x}_f \sim N(0,\Sigma_{jx})$. The random component of the quantity supplied of shares represents trades unrelated to the firm's value \citep{delong1990noise}. An investor's private signal is the firm-specific innovation and noise specific to that investor. The public signal is composed of the innovation and the public noise. The private signals factor into the prices through their average. The investor-specific noise averages out across investors. Hence, the average private signal is equal to the innovation. Thus, the price depends on---innovation, public noise and the number of shares on the market (Proposition~\ref{prop:price}). If an investor could observe the number of shares on the market then the price and the public signal could reveal the average private signal or the innovation. The random component prevents that so that the price still reflects the private signals without revealing their average. 

The model's events are spread over two quarters. Figure~\ref{fig:timeline} shows the timing of the four events, the feed, the trading, the announcement, and the correction. The feed and the trading fall in quarter $q$. The announcement and the correction fall in quarter $q+1$.

\begin{figure}[!htbp]
\centering
\caption{Timing of the mechanism and the empirical design}\label{fig:timeline}
\medskip
\resizebox{\textwidth}{!}{\begin{tikzpicture}[
  box/.style={draw, rounded corners=2pt, align=left, inner sep=5pt, text width=0.163\textwidth, minimum height=3.7cm, font=\scriptsize, execute at begin node={\hyphenpenalty=10000\relax}},
  mbox/.style={draw, dashed, rounded corners=2pt, align=left, inner sep=5pt, text width=0.56\textwidth, font=\scriptsize, execute at begin node={\hyphenpenalty=10000\relax}},
  arr/.style={->, thick, shorten >=2pt, shorten <=2pt}]
\node[box] (b1) at (0,0) {The feed is created: The contributors post about the stock on each day of quarter $q$. The quarter's daily readings form the public signal whose noise is given by the within-quarter volatility of daily sentiment per tweet, VAQSV.};
\node[box, right=0.42cm of b1] (b2) {The quarterly trade: Investors use the prior, their private signals, the price and the public signal. The market clears at the price $P^{*}_{f1}$, which incorporates the public signal with its noise.};
\node[box, right=0.42cm of b2] (b3) {The asymmetry: The pessimistic error in the signal is corrected within the quarter because of buying. The optimistic error is corrected only in part since short selling is closed to some investors. The rest of the optimistic error survives as overpricing.};
\node[box, right=0.72cm of b3] (b4) {The announcement: In quarter $q+1$, the firm reports its earnings from quarter-$q$. The innovation $\epsilon_{f1}$ becomes public.};
\node[box, right=0.42cm of b4] (b5) {The correction: The price falls to the true value revealed. The surviving overpricing is removed. As a result, the losses exceed gains in magnitude.};
\draw[arr] (b1) -- (b2);
\draw[arr] (b2) -- (b3);
\draw[arr] (b3) -- (b4);
\draw[arr] (b4) -- (b5);
\draw[decorate, decoration={brace, amplitude=6pt}]
  ([yshift=0.25cm]b3.north east -| b1.north west) -- ([yshift=0.25cm]b3.north east)
  node[midway, yshift=0.42cm, font=\footnotesize]{Quarter $q$};
\draw[decorate, decoration={brace, amplitude=6pt}]
  ([yshift=0.25cm]b4.north west) -- ([yshift=0.25cm]b5.north east)
  node[midway, yshift=0.42cm, font=\footnotesize]{Quarter $q+1$};
\node[mbox, anchor=north] (m) at ([yshift=-1.45cm]b2.south) {On each trading day of quarter $q$, a dealer takes the other side of informed and uninformed orders and sets the bid-ask spread and the market-impact cost. These two trading costs are the measurement layer from which informed-trading intensity is recovered.};
\draw[dotted, thick] (b2.south) -- (m.north);
\end{tikzpicture}}
\begin{minipage}{0.95\textwidth}\footnotesize\medskip
\textit{Notes:} A firm announces its earnings for quarter $q$ after the quarter ends, with a median lag of 35 days and 40 days for domestic and foreign firms, respectively, so the announcement typically falls in quarter $q+1$. The next-quarter gains and losses are measured over the next quarter in which the announcement falls. In the model, the investors trade once in quarter $q$, at a single price which is set before the earnings announcement. The arrows show the order of events. The dashed box shows the daily dealer market, which runs on each trading day of the quarter. The informed-trading intensity is recovered from the spread and the market-impact cost. The solid boxes show the mechanism, right from the feed to the overpricing correction.
\par\medskip\noindent Alt text: There are five boxes---three for quarter $q$ and the rest for quarter $q+1$. Box 1: The feed forms in quarter  $q$. Box 2: Investors trade at a price that incorporates the public signal and its noise. Box 3: Pessimistic noise is bought away within the quarter, while part of the optimistic noise survives because short selling is costly. Box 4: The firm announces its earnings. Box 5: The surviving overpricing is removed, so losses exceed gains. 
The dashed box under box two shows the daily dealer market, where the spread and the price impact are set and from which informed-trading intensity is measured.
\end{minipage}
\end{figure}

A continuum of investors $i$ of unit mass with mean-variance preferences and common risk-aversion coefficient $\bar{\rho}$ chooses share demands $q^i_f$ to maximize $E[W^i\,|\,\mathcal{I}^i] - \tfrac{\bar{\rho}}{2}\mathrm{Var}[W^i\,|\,\mathcal{I}^i]$. Here $\mathcal{I}^i$ is investor $i$'s information set, $W^i_0$ is initial wealth. End-of-period wealth is $W^i = rW^i_0 + \sum_f q^i_f (V^{*}_{f1} - rP^{*}_{f1})$. Investor $i$'s information set for stock $f$ contains four items. The first is the prior on next period's dividend. The cash-flow history and the common component fix its center. Its one uncertain part is the firm-specific innovation $\epsilon_{f1}$, so the prior precision is $\tau_d \equiv \Sigma_{jd}^{-1}$. The second is a private signal of precision $k^i$. A private signal is a noisy observation of the innovation that investor $i$ alone receives, for example from proprietary research or soft information. Its realized value is $\eta^i_{f1} \equiv \epsilon_{f1} + e^i_{f1}$, where the idiosyncratic noise $e^i_{f1} \sim N(0, 1/k^i)$ is independent across stocks and investors. 

A higher $k^i$ implies less idiosyncratic noise. Thus, the precision $k^i$ measures how well informed investor $i$ is. The third is the equilibrium price $P^{*}_{f1}$. The private signals enter the price through their average. Hence, observing the price gives an investor additional information about the innovation. The precision of that additional information is $\tau_p$ (equation~\eqref{eq:taup}). The fourth is a public signal extracted from the firm's Twitter feed. Section~\ref{sec:model_sent} constructs this signal from the feed and states the scope of such modeling choice. The investor's posterior is made up of a combination of these four items. Let $k^i_j$ denote investor $i$'s private-signal precision for the stocks in group $j$. The aggregate private precision in group $j$ is then given by: $K_j = \int_0^1 k^i_j\, di$, which is the integral of that precision over the unit mass of investors. $K_j$ is the total private-signal precision which the price aggregates from those investors. It measures how much private information about the firm's innovation the market holds as a whole. This aggregated total private-signal precision is high when many investors are simply informed or it could be high if each informed investor knows a lot. The model does not need these two margins separately to get the equilibrium price and the precisions behind it. $K_j$ is the informed-trading intensity of group $j$.\footnote{It is the model's counterpart of the probability of informed trading which \citet{easley1996liquidity} estimate from trade data. Reduced-form studies often use the share of a stock held by institutional investors for the same object.} The empirical analysis recovers $K_j$ from the trading costs set in a dealer market that operates on each trading day of the quarter (Section~\ref{sec:model_micro}).\footnote{In the data, the liquidity related variables are built from daily measures and averaged within the quarter, so the dealer market in the model is at the daily frequency too.} Each of the four information sources mentioned above has a precision. The prior has a precision of $\tau_d$. The price has a precision of $\tau_p$. The private signals have a precision of $K_j$. The public signal has a precision of $\tau_s$. Thus, the total precision is $T_j \equiv \tau_d + \tau_p + K_j + \tau_s$. The posterior variance of the innovation is $\overline{\Sigma}_j = 1/T_j$. The prior precision $\tau_d$, the price-signal precision $\tau_p$, and the informed-trading intensity $K_j$ take one value for each stock group $j$. The precision of the public signal $\tau_s$ is obtained for each stock-quarter (Section~\ref{sec:model_sent}). Where a group value of $\tau_s$ is needed, the estimation uses the within-group median. We keep the group subscript on $K_j$, $T_j$, and $\overline{\Sigma}_j$. We suppress it on $\tau_d$, $\tau_p$, and $\tau_s$ except when a contrast across groups is needed. In the estimation, the precisions take one value for each geography at each informed-trading-intensity tercile, six stock groups in all (Section~\ref{sec:model_micro}).

\subsection{Social media as a public signal}\label{sec:model_sent}

The Twitter feed is public, available through the vendor's aggregate metrics, and carries the public signal in the model. This paper is about the composition of the types of contributors who generate the feed and the sentiment in it. Proposition~\ref{prop:vaqsv} shows how it is constructed from the quarter's daily readings. The estimates specifically characterize the Twitter feed rather than public information in general.\footnote{This is a statement about the scope. We do not claim that the feed captures all public information relevant to pricing. We also don't claim that the composition of contributors on Twitter matches that on other platforms. Information on other venues lies outside the scope of our model. Some of the prior work in this area supports treating a single venue's sentiment as an informative public signal. For instance, \citet{Sprenger2013} find that the tweets of users who give better investment advice are retweeted more, a mechanism through which the platform aggregates information. \citet{garcia2013sentiment} and \citet{bartov2018twitter} link media and Twitter sentiment to returns and fundamentals.}

The signal is $s_{f1} = \epsilon_{f1} + u_{f1}$, where the public noise $u_{f1}$ is normal with mean zero and precision $\tau_s$. Investors condition on this signal.\footnote{The human contributors who post on the feed are investors themselves. The information they post is also the one they trade on. The predictor is the noise in the quarter $q$'s feed, and the outcomes are the gains and losses of quarter $q+1$, which rules out reverse causality. Noise that is corrected inside quarter $q$ also moves its return. We do not use that same-quarter relation for identification.} Private information doesn't show up on the feed. It reaches investors through other channels, such as proprietary research and soft information, and enters the private precision $K_j$ (Section~\ref{sec:model_env}).\footnote{Investors have finite attention. A less attentive investor extracts less from the feed and remains more uncertain about the innovation \citep{sims2003rational, peng2006investor}. In the model, the price depends on the feed's precision multiplied by the share of investors who read the feed. A more precise feed read by a few investors and a less precise feed read by many investors can move the price in the same way. Therefore, sentiment and returns alone cannot tell these two scenarios apart. The precision $\tau_s$ that we recover from the feed does not contain that share. If fewer investors read the foreign feed than the domestic feed, the gap in price-relevant precision is larger than the measured gap (footnote to \ref{hyp:mismatch}). Investor attention affects how much they trade on a signal. For instance, \citet{jiang2022attention} find that anomaly returns are higher after days with high investor attention.}

Three vendor metrics summarize the feed at the firm-quarter level. The volume-adjusted quarterly sentiment volatility $\mathrm{VAQSV}_{fq}$ is the quarter's sentiment volatility measured per tweet. A higher value maps to a less precise public signal.\footnote{The feed reports, for each stock-day, an exponentially time-weighted sentiment total $S_t$ over per-tweet scores in $[-1,1]$ and the tweet count $N_t$, with $q$ indexing quarters and $t$ days within the quarter. These are the stock-level Twitter sentiment metrics of Context Analytics, formerly Social Market Analytics, obtained through Dewey Data \citep{contextanalytics2022data}.} Quarterly sentiment volatility is the within-quarter standard deviation of the daily total, $\mathrm{QSV}_{fq} = \mathrm{SD}_{t \in q}(S_t)$. $\mathrm{VAQSV}_{fq}= \mathrm{SD}_{t \in q}(S_t/N_t)$ is the volume-adjusted quarterly sentiment volatility.\footnote{These two volatility measures have correlations of $0.32$ and $0.06$ for the full sample and the primary sample (2018Q1 onward), respectively.} As the day's tweet volume increases, so does $S_t$, which is why the volatility is measured on the per-tweet average $S_t/N_t$. The vendor's total $S_t$ weights recent tweets more than earlier ones. In the model, each per-tweet sentiment score has its own noise, $e$. The noise in a weighted average of $N_t$ such sentiment scores has a larger variance than the noise in a simple average of $N_t$ such sentiment scores, by a factor that depends on those weights. Equation~\eqref{eq:vaqsv} absorbs that factor into the constant $c_w \geq 1$. Further, $T_q$ denotes the number of days in the quarter with at least one scored tweet, weekends included, while $T_{days}$ denotes the number of trading days in the quarter with a return. The quarterly gain and loss measures are obtained by summing over $T_{days}$. For each stock and day, the vendor data gives the dispersion, $D_t = A_t/N_t$, where $A_t$ is the number of distinct accounts that post that day. Average dispersion $\overline{D}_{fq}$ is the mean of $D_t$ over $T_q$. A lower value for dispersion indicates that the feed is generated by fewer accounts. Because the formula never inspects the sentiment scores, $D_t$ measures contributor concentration rather than disagreement of opinion. Additionally, we have the average inverse tweet count, $\overline{(1/N_t)}_{fq} = \tfrac{1}{T_q}\sum_{t \in q} 1/N_t$, which is high when the daily tweet volume is low. A high value means more sampling noise in the quarter's signal.\footnote{There are two vendor methodology changes which should be noted. First, there was a dictionary update in March 2016 which changed $S_t$ across quarters. Our within-quarter moments are robust to these level shifts of $S_t$ across quarters. Second, in 2017, there was a switch from cashtag filtering to a topic-model classifier affecting the selection of tweets which would become part of the feed. To deal with this, our primary estimation sample starts with 2018Q1 (the post change period). We exclude the period prior to 2018Q1 from the estimation.}

For each stock, posts come from three contributor types. 1. The specialists $(\mathbb{S})$: Accounts that post about a firm repeatedly and whose posts carry the firm-specific signal. In the model, a specialist's tweet encompasses the firm-specific innovation and the tweet's own noise. Two forces motivate the repeated posting by the specialists. Since learning about a firm is costly, an investor who already knows a firm well gains more from learning more about it than learning about another unfamiliar firm \citep{van2009information}. An investor who has traded on private information earns profits when the price moves toward the value tied to that private information. Public posting helps such price movement, so such an investor who has private information has an incentive to post repeatedly. These two forces motivate the assumption that specialists post repeatedly. They are not evidence, and the data doesn't show how often any account or contributor type posts. Moreover, posting does not make a specialist's private signal public. A specialist's tweet consists of the firm-specific innovation and the tweet's own noise. The feed averages such tweets into the public signal, whose precision is $\tau_s$. The specialist's private signal (a separate noisy reading of the same innovation with precision $k^i$) stays with the specialist and enters the price through trading alone. 2. Attention-driven contributors $(\mathbb{A})$: Accounts which post a single tweet about a firm on a day \citep{barber2008all}. Their posts have the firm-specific signal along with the day's fad which is a transient nonfundamental component that specialists filter out and these attention-driven contributors do not. 3. Automated accounts $(\mathbb{B})$: Automated accounts, which post machine-generated content, also called bots (an informal label used for brevity). Such accounts are rare among authors but their tweets occur commonly in the feed.\footnote{Twitter's disclosures put false or spam accounts below five percent of its active users, while an independent audit from September 2022 attributed about twenty to thirty percent of U.S. content on the platform to bots. These industry measurements along with their classification error are given in \ref{app:bot_prevalence}.} The vendor's sentiment metrics are aggregated over tweets instead of accounts or contributor types. Hence, each contributor type enters these metrics through its share of tweets, and bots account for a substantial share of the tweets behind them.

$A_t$ accounts post about the stock on day $t$. Among the accounts posting about a stock in group $j$, the share of specialists is $\pi_{\mathbb{S},j}$, the share of bots is $\pi_{\mathbb{B},j}$ and the share of attention-driven contributors is $1-\pi_{\mathbb{S},j}-\pi_{\mathbb{B},j}$. Each type $g$ posts a fixed number of tweets per account per day, its posting intensity $\lambda_g$, with $\lambda_{\mathbb{S}} > 1$, $\lambda_{\mathbb{B}} > 1$ and $\lambda_{\mathbb{A}} = 1$. The two repeat intensities may differ.\footnote{Since the dispersion in the data identifies only the combined share of the two repeat types, the data cannot tell the two rates apart (Proposition~\ref{prop:dispersion}). Hence, the estimation uses a common repeat intensity $\lambda_I$ for both and fixes it at $1.5$ tweets per account per day (Section~\ref{sec:structural}).} Each tweet has a sentiment score. The sentiment scores are $\epsilon_{f1} + e$, $\beta_t + e$ and $\epsilon_{f1} + b_t + e$ for a specialist tweet, a bot tweet, and an attention-driven tweet, respectively. $b_t$ is the fad on day $t$, $\beta_t$ is the bot content on day $t$ and $e$ is the noise of the individual tweet. The firm-specific innovation, $\epsilon_{f1}$, is the part of the firm's next dividend that no investor observes without noise. Each of the model's signals, be it the private signal, the price, or the public signal, is a noisy reading of it (Section~\ref{sec:model_env}). Both a specialist tweet and an attention-driven tweet contain $\epsilon_{f1}$ in their scores. This forms the fundamental part of the score since it is news about the dividend. The fad $b_t$ and the bot content $\beta_t$ don't contain $\epsilon_{f1}$. Hence, they form the nonfundamental parts of a score. Assumption~\ref{assump:dist} states the distributions of these terms.

\refstepcounter{assump}\label{assump:dist}\textbf{Assumption \theassump\ (distribution).} The fad $b_t \sim N(0,\sigma_{b,j}^2)$ and the bot process $\beta_t \sim N(0,\sigma_{\beta,j}^2)$ are i.i.d.\ across days. The per-tweet noise $e \sim N(0,\sigma_e^2)$ is i.i.d.\ across tweets and days. The processes $b_t$ and $\beta_t$ and the noise $e$ are mutually independent and independent of $\epsilon_{f1}$.

The specialist type contains the informed specialists and institutional trading desks and investor relations accounts, since these post repeatedly with knowledge of the firm. Investor relations accounts post favorably about their own firm. If the degree of that favorable tilt is the same each day, then it increases the sentiment score by the same amount each day. QSV and VAQSV remain unchanged by it. If the degree of that favorable tilt changes from day to day, it adds day-to-day movement unrelated to the firm's innovation to the sentiment score. The model categorizes it as a nonfundamental component. The bot type contains automated accounts and also news aggregators, which repost public firm news repeatedly and add no new information about the innovation.

Dispersion depends on the shares of specialists and bots. On a day with $A_t$ posting accounts, specialists and bot accounts contribute $(\pi_{\mathbb{S},j} + \pi_{\mathbb{B},j})\lambda_I A_t$ tweets and the attention-driven type contributes $(1 - \pi_{\mathbb{S},j} - \pi_{\mathbb{B},j})A_t$ tweets. Therefore, tweets per account equal $\pi_j \lambda_I + (1 - \pi_j)$, with $\pi_j \equiv \pi_{\mathbb{S},j} + \pi_{\mathbb{B},j}$ being the share of accounts that post more than once. We call these specialists and bot accounts repeat authors.

\proposition{prop:dispersion}~\textit{(Dispersion determines the repeat-author share for a given posting intensity.) Given the feed structure, daily dispersion in group $j$ is $D = \left[\pi_j \lambda_I + (1-\pi_j)\right]^{-1}$ with $\pi_j = \pi_{\mathbb{S},j} + \pi_{\mathbb{B},j}$, strictly decreasing in $\pi_j$, and the inversion}
\begin{equation}\label{eq:pi_inversion}
\pi_j \;=\; \frac{1 - D}{D\,(\lambda_I - 1)}
\end{equation}
\textit{recovers the total repeat-author share from observed dispersion for a given $\lambda_I > 1$, with $\pi_j = 0$ at $D = 1$. Dispersion determines $\pi_j$ as a whole. It does not separate specialists from bot accounts.} \textit{(Proof in \ref{app:proof_dispersion}.)}

Proposition~\ref{prop:vaqsv} links the within-quarter volatility of the day's per-tweet sentiment to the noise in the public signal. The day's per-tweet sentiment $\bar{s}_t$ is the average of the sentiment scores of the day's $N_t$ tweets. We build it up from the three contributor types. Dispersion gives the day's tweet count as $N_t = A_t/D$. The tweet share of a type is its tweets divided by $N_t$, $\omega_{\mathbb{S},j} \equiv \pi_{\mathbb{S},j}\lambda_I D$, $\omega_{\mathbb{B},j} \equiv \pi_{\mathbb{B},j}\lambda_I D$, and $\omega_{\mathbb{A},j} \equiv (1-\pi_j)D$. The three tweet shares sum to $D\,[\pi_j \lambda_I + (1-\pi_j)]$, which equals one by Proposition~\ref{prop:dispersion}. Averaging the sentiment scores within each type gives $\epsilon_{f1} + \bar{e}^{\mathbb{S}}_t$ for specialist tweets, $\epsilon_{f1} + b_t + \bar{e}^{\mathbb{A}}_t$ for attention-driven tweets, and $\beta_t + \bar{e}^{\mathbb{B}}_t$ for bot tweets, where $\bar{e}^{g}_t$ is the average of the per-tweet noises of the $\omega_{g,j} N_t$ tweets of type $g$ and has variance $\sigma_e^2/(\omega_{g,j} N_t)$. The day's per-tweet sentiment equals the three type averages weighted by the tweet shares,
\begin{equation}\label{eq:sbar_body}
\bar{s}_t = \omega_{\mathbb{S},j}\,(\epsilon_{f1} + \bar{e}^{\mathbb{S}}_t) + \omega_{\mathbb{A},j}\,(\epsilon_{f1} + b_t + \bar{e}^{\mathbb{A}}_t) + \omega_{\mathbb{B},j}\,(\beta_t + \bar{e}^{\mathbb{B}}_t) = (1 - \omega_{\mathbb{B},j})\,\epsilon_{f1} + \omega_{\mathbb{A},j}\, b_t + \omega_{\mathbb{B},j}\,\beta_t + \bar{u}_t.
\end{equation}
The loading on the innovation is $\omega_{\mathbb{S},j} + \omega_{\mathbb{A},j} = 1 - \omega_{\mathbb{B},j}$, the sum of the tweet shares of specialists and attention-driven contributors, the two types whose tweets carry the innovation. The term $\bar{u}_t = \omega_{\mathbb{S},j}\bar{e}^{\mathbb{S}}_t + \omega_{\mathbb{A},j}\bar{e}^{\mathbb{A}}_t + \omega_{\mathbb{B},j}\bar{e}^{\mathbb{B}}_t$ is the sampling noise. It has mean zero and variance $\sigma_e^2/N_t$ (\ref{app:proof_vaqsv}). The ratio $S_t/N_t$ of the vendor's sentiment total to its tweet count is the empirical counterpart of $\bar{s}_t$.

\proposition{prop:vaqsv}~\textit{(Volume-adjusted quarterly sentiment volatility and the precision of the public signal.) Under \ref{assump:dist}, and when the factor $c_{1,t}$ defined below is the same on every day of the quarter, the conditional mean of the squared volume-adjusted quarterly sentiment volatility of a stock in group $j$, given the quarter's firm-specific innovation and the daily tweet counts, is}
\begin{equation}\label{eq:vaqsv}
E\!\left[\,\mathrm{VAQSV}_{fq}^2 \;\middle|\; \epsilon_{f1},\,\{N_t\}_{t\in q}\,\right] \;=\; \underbrace{\omega_{\mathbb{A},j}^2\,\sigma_{b,j}^2 \;+\; \omega_{\mathbb{B},j}^2\,\sigma_{\beta,j}^2}_{\;\Phi_j\;\equiv\;\text{nonfundamental}} \;+\; c_w\, \sigma_e^2\, \overline{\left(1/N_t\right)} \;\equiv\; \bar{v}_{fq},
\end{equation}
\textit{where $\Phi_j$ is the nonfundamental variance. The daily sampling variance of the per-tweet average $S_t/N_t$ is $c_{1,t}\sigma_e^2/N_t$, $c_{1,t}=1$ under equal weights and $c_{1,t}>1$ under the vendor's recency weighting, so that $c_w \geq 1$. Only the sampling-noise coefficient ($c_w\sigma_e^2$) on the average inverse tweet count ($\overline{(1/N_t)}$) enters equation~\eqref{eq:vaqsv}. $\bar{v}_{fq}$ is the average over the quarter's $T_q$ days of the daily conditional variance of the daily per-tweet sentiment. $\mathrm{VAQSV}_{fq}^2$ is unbiased for $\bar{v}_{fq}$ under \ref{assump:dist} and understates it under positive serial dependence across days (\ref{app:proof_vaqsv}). The investors condition on the public signal $s_{f1} = \epsilon_{f1} + u_{f1}$ obtained by averaging $\bar{s}_t$ over the quarter's $T_q$ days and dividing by $1-\omega_{\mathbb{B},j}$. The public noise $u_{f1}$ is the quarterly average of $\omega_{\mathbb{A},j} b_t + \omega_{\mathbb{B},j}\beta_t + \bar{u}_t$ divided by $1-\omega_{\mathbb{B},j}$. It is normally distributed with mean zero as per \ref{assump:dist}. The precision of that signal is}
\begin{equation}\label{eq:tau_s}
\tau_s \;=\; \frac{T_q}{\kappa \cdot \mathrm{VAQSV}_{fq}^2}, \qquad \kappa > 0,
\end{equation}
\textit{where $\kappa$ is the conversion constant that maps the observable signal-quality index $T_q/\mathrm{VAQSV}_{fq}^2$ into the latent precision $\tau_s$. The signal-quality index is the precision if the daily readings were independent, the day's sentiment moved one-for-one with the innovation and the vendor's score were in innovation's units. The data departs from all three of these conditions. The daily readings are serially dependent. Bot tweets don't carry any of the innovation. The day's sentiment therefore moves with the innovation by $1-\omega_{\mathbb{B},j}$ instead of one-for-one. A fixed scale converts the vendor's score in sentiment units to the innovation's cash-flow units. $\kappa$ absorbs these three departures.\footnote{The estimation treats $\kappa$ as one parameter common across firm-quarters (Section~\ref{sec:structural}). The data identify it as a whole and do not separate the three components (\ref{app:proof_vaqsv}).} The bot tweet share raises it by lowering the weight on the innovation to $1-\omega_{\mathbb{B},j}$. The squared scale that links the vendor's per-tweet score to the cash-flow innovation enters it as well. It enters because equation~\eqref{eq:sbar_body} normalizes the specialist loading to one.} \textit{(Proofs of equations \eqref{eq:vaqsv} and \eqref{eq:tau_s} in \ref{app:proof_vaqsv}.)}

The group-level precision of the public signal is $\tau_{s,j}$ where $j$ is the stock group introduced in Section~\ref{sec:model_env}. This precision of public signal differs across groups because the tweet shares ($\omega_{g,j}$) and the fad and bot variances ($\sigma_{b,j}^2$ and $\sigma_{\beta,j}^2$, respectively) that enter $\Phi_j$ are group specific (equations~\eqref{eq:vaqsv} and~\eqref{eq:tau_s}).\footnote{We write $\tau_s$ for the precision of the public signal. Where a domestic-foreign contrast is made at a fixed informed-trading-intensity tercile, the two stock groups compared are the domestic and the foreign group at that tercile (Section~\ref{sec:model_micro}).} The observable signal-quality index ($T_q/\mathrm{VAQSV}_{fq}^2$) that gives $\tau_{s,j}$ (equation~\eqref{eq:tau_s}) varies at the stock-quarter level. The estimation aggregates it to the group level (Section~\ref{sec:structural}). The linear regression in Equation~\eqref{eq:vaqsv} recovers the nonfundamental variance ($\Phi_j$) as the intercept and the sampling-noise coefficient ($c_w\sigma_e^2$) as the slope. The structural GMM estimation (Section~\ref{sec:structural}) uses the equations~\eqref{eq:vaqsv} and~\eqref{eq:tau_s} under two conventions. The first one sets the bot-content variance $\sigma_{\beta,j}^2$ to zero. That means that the nonfundamental variance $\Phi_j$ comes solely from the fad. The second one treats $\kappa$ as a parameter common to all firm-quarters (\ref{app:vaqsv_estimation}).

\subsection{Equilibrium prices and informativeness}\label{sec:model_eq}

This subsection deals with how prices respond when the public signal is added to the standard information structure. The answer is a linear equilibrium, derived by the conjecture-and-verify argument of \citet{hellwig1980aggregation} and \citet{admati1985noisy}. Our conjecture is that the price is affine in the innovation, the signal noise and the supply shock. Each investor then forms a posterior by Bayesian updating on the prior, their private signal, the public signal, and the information in the price after netting out the publicly observed signal. Demands follow from mean-variance optimization. Market clearing determines the coefficients.

\proposition{prop:price}~\textit{(Equilibrium price.) There is a linear rational expectations equilibrium in which the price $P^{*}_{f1}$ satisfies}
\begin{equation}\label{eq:price}
rP^{*}_{f1} \;=\; A_f \;+\; B_j\,\epsilon_{f1} \;+\; G_j\,u_{f1} \;+\; C_j\,\tilde{x}_f,
\end{equation}
\textit{with}
\begin{equation}\label{eq:coefs}
B_j = \gamma_j\!\left(1 - \frac{\overline{\Sigma}_j}{\Sigma_{jd}}\right)\!, \qquad
G_j = \gamma_j\, \overline{\Sigma}_j\, \tau_s, \qquad
C_j = -\,\bar{\rho}\,\gamma_j^2\, \overline{\Sigma}_j\, \frac{\tau_p + K_j}{K_j},
\end{equation}
\textit{and price-signal precision}
\begin{equation}\label{eq:taup}
\tau_p \;=\; \frac{K_j^2}{\gamma_j^2\, \bar{\rho}^2\, \Sigma_{jx}},
\end{equation}
\textit{which is independent of $\tau_s$.} \textit{(Proof in \ref{app:proof_price}.)}

Risk-averse investors pay the price $P^{*}_{f1}$ today for a persistent yet uncertain dividend stream, discounted at the gross rate $r$ (Equation~\eqref{eq:price} shows its components). A one-unit increase in the innovation ($\epsilon_{f1}$) raises the period-1 dividend ($d^{*}_{f1}$) by one. The firm's value is $V^{*}_{f1} = \gamma_j\, d^{*}_{f1}$, with $\gamma_j = r/(r-\upsilon_j)$ the present-value multiplier. The increase therefore raises $V^{*}_{f1}$ by $\gamma_j$. This multiplier is larger when the innovation is more persistent (when $\upsilon_j$ is larger). The market holds residual uncertainty about the innovation. Hence, the price reflects the innovation's value only to the degree to which the market learns about it. Investors cannot separate the feed's signal from its error. Hence, the price has the feed's signal and its noise. A part of the supply of shares is random and comes from investors who trade for reasons not related to the firm's value. This randomness keeps the price only partly revealing. The price is itself a signal, but it adds information only through the private signals that trading aggregates into it. Its response to the public signal repeats what investors already observe.\footnote{All equilibrium coefficients are derived from first principles (\ref{app:model_proofs}). At $\tau_s = 0$, $B_j$, the total precision $T_j$, and the informativeness decomposition in Proposition~\ref{prop:pinf} reduce to those in \citet{farboodi2022data}. The supply-noise coefficient $C_j$ takes the price-impact form of \citet{hellwig1980aggregation} and \citet{admati1985noisy}.} The quarterly excess return is $R \equiv V^{*}_{f1} - rP^{*}_{f1}$. Its first two moments follow from the equilibrium price (\ref{app:proof_price}). Its mean $E[R] = \bar{\rho}\gamma_j^2 \overline{\Sigma}_j \bar{x}_j$ is the premium that compensates investors for bearing the average supply $\bar{x}_j$. Its variance,
\begin{equation}\label{eq:varR}
\mathrm{Var}[R] \;=\; \gamma_j^2\, \overline{\Sigma}_j^{2} \left[\, \tau_d + \tau_s + \frac{(\tau_p + K_j)^2}{\tau_p} \,\right],
\end{equation}
falls as the feed grows more precise. A more informative feed makes the price track the firm's value more closely. Return volatility therefore falls. This is the volatility channel the loss analysis conditions on (Section~\ref{sec:model_gl}).

The second question for this subsection is how sentiment affects what the price reveals about the firm's future earnings. Following \citet{bai2016informative} and \citet{farboodi2022data}, stock-level informativeness at horizon $s$ is the covariance of horizon-$s$ earnings with the price, scaled by the price's standard deviation.\footnote{Horizon-$s$ earnings are the dividend $s$ periods after the pricing date, equal to $\upsilon_j^s\, d^{*}_{f1}$ by the dividend law (Section~\ref{sec:model_env}).}

\proposition{prop:pinf}~\textit{(Price informativeness.) Stock-level price informativeness at horizon $s$ decomposes as}
\begin{equation}\label{eq:pinf}
\mathrm{PINF}_{j,s} \;=\; \underbrace{\frac{\Sigma_{jd}}{\mathrm{SD}(P^{*}_{f1})}}_{\text{volatility}} \;\cdot\; \underbrace{\frac{\upsilon_j^{\,s}}{r - \upsilon_j}}_{\text{growth}} \;\cdot\; \underbrace{\left(1 - \frac{\overline{\Sigma}_j}{\Sigma_{jd}}\right)}_{\text{information}}, \qquad 1 - \frac{\overline{\Sigma}_j}{\Sigma_{jd}} \;=\; \frac{\tau_p + K_j + \tau_s}{\tau_d + \tau_p + K_j + \tau_s},
\end{equation}
\textit{so the information component is strictly increasing in $\tau_s$.} \textit{(Proof in \ref{app:proof_pinf}.)}

The decomposition of price informativeness has three factors. The volatility factor ($\Sigma_{jd}/\mathrm{SD}(P^{*}_{f1})$) is the innovation variance relative to the volatility of the price. A noisier price lowers measured informativeness through this factor. The growth factor ($\upsilon_j^s/(r-\upsilon_j)$) is the ratio of how strongly the innovation appears in horizon-$s$ earnings ($\upsilon_j^s$) to the rate that discounts the persistent earnings stream to present value ($r-\upsilon_j$). Today's news matters for distant earnings because of this persistence. The information factor $1-\overline{\Sigma}_j/\Sigma_{jd}$ is the fraction of the innovation variance that the price, the private signals, and the feed together remove, equal to $(\tau_p+K_j+\tau_s)/T_j$. The signal precision $\tau_s$ enters this factor directly (equation~\eqref{eq:pinf}). It reaches the volatility factor only through the price's standard deviation. In \citet{farboodi2022data}, the information factor, whose underlying precision is not observed, is backed out from measured informativeness, growth, and volatility. We do not face that limitation. Equation~\eqref{eq:tau_s} measures the signal precision $\tau_s$ from the observable signal-quality index $T_q/\mathrm{VAQSV}_{fq}^2$ up to the conversion constant $\kappa$. The model predicts that informativeness, net of the volatility and growth components, rises with $\tau_s$.

\subsection{Short-sale constraints and the asymmetry between gains and losses}\label{sec:model_gl}

The precision of the public signal should matter for losses specifically rather than through the volatility channel alone. The outcome variables are the quarter's accumulated gains and losses. The outcomes are the sums of the positive and negative parts of the daily excess return $r_{ft}$, $\mathrm{Gain}_{fq} = \sum_{t \in q} \max(r_{ft}, 0)$ and $\mathrm{Loss}_{fq} = \sum_{t \in q} \min(r_{ft}, 0)$. Connecting these outcomes to the quarterly equilibrium requires a statement about how the quarterly excess return accrues over days.

\refstepcounter{assump}\label{assump:daily}\textbf{Assumption \theassump\ (daily accrual).} The quarter contains $T_{days}$ trading days. The daily excess returns $r_{ft}$ are i.i.d.\ $N(\mu_d, \sigma_d^2)$ within the quarter. The daily parameters satisfy $T_{days}\mu_d = E[R] + \mu^{drift}$ and $T_{days}\sigma_d^2 = \mathrm{Var}[R]$, where $\mu^{drift}$ is the correction drift (Proposition~\ref{prop:drift}), a predictable component of the mean daily excess return that reverses the previous quarter's overpricing.\footnote{Returns are measured per dollar of the asset's unconditional value. A single positive scale, written $s_P$, converts the model's value units into return units and is estimated as a shared parameter (see Section~\ref{sec:structural}). A positive scale preserves the sign and the monotonicity of all predictions. Under \ref{assump:daily}, the quarterly and daily volatilities are linked by $\sigma_R = T_{days}^{1/2}\,\sigma_d$, where $\sigma_R \equiv \mathrm{Var}[R]^{1/2}$ is the standard deviation of the quarterly excess return $R$.}

As per the assumption, the quarter's excess return is the sum of independent days' excess returns with the same mean and variance. A day's mean carries two parts, the group-level risk premium and the correction drift that reverses the previous quarter's overpricing. A day's variance is the quarter's return variance divided across the days. Under \ref{assump:daily}, the expected tails are
\[
\begin{aligned}
E[\mathrm{Gain}] &= T_{days}\left[\mu_d \Phi(\mu_d/\sigma_d) + \sigma_d\,\phi(\mu_d/\sigma_d)\right],\\
E[|\mathrm{Loss}|] &= T_{days}\left[-\mu_d \Phi(-\mu_d/\sigma_d) + \sigma_d\,\phi(\mu_d/\sigma_d)\right],
\end{aligned}
\]
where $\Phi$ and $\phi$ are the standard normal distribution function and density (Proof in \ref{app:proof_vol}). Subtracting the two equations gives $E[\mathrm{Gain}] - E[|\mathrm{Loss}|] = T_{days}\,\mu_d$. The increase in daily volatility ($\sigma_d$) increases $E[\mathrm{Gain}]$ and $E[|\mathrm{Loss}|]$ equally and thus cancels in the difference. The gain-loss gap then depends on the mean daily excess return ($\mu_d$) alone. This separation between the mean and the volatility channels is the underlying mechanism for the identification in the empirical exercise. In the data, the daily volatility is the within-quarter standard deviation of daily excess returns. In a regression with the quarter's losses $\mathrm{Loss}_{fq}$ as the outcome, controlling for the daily volatility absorbs the volatility channel. Any remaining effect operates through the mean daily excess return, whose quarterly sum ($T_{days}\mu_d$) contains the correction drift ($\mu^{drift}$) (\ref{assump:daily}, Proposition~\ref{prop:drift}). The quarterly sum also contains the group-level risk premium ($E[R]$), which the group intercepts absorb. The correction drift is then identified from the variation within groups (see Section~\ref{sec:reducedform}).

\proposition{prop:vol}~\textit{(Volatility channel.) At fixed mean daily excess return $\mu_d$, $E[\mathrm{Gain}]$ and $E[|\mathrm{Loss}|]$ are increasing in the daily volatility $\sigma_d$. Under \ref{assump:daily}, $T_{days}\sigma_d^2 = \mathrm{Var}[R]$, and $\mathrm{Var}[R]$ is strictly decreasing in $\tau_s$ by equation~\eqref{eq:varR}, so a noisier public signal raises both $E[\mathrm{Gain}]$ and $E[|\mathrm{Loss}|]$ symmetrically. The drift channel is Proposition~\ref{prop:drift}. (Proof in \ref{app:proof_vol}.)}

The proposition shows separation of the volatility channel from the mean channel. A noisier feed raises return volatility (equation~\ref{eq:varR}) on a day. A day with more return volatility produces bigger swings in gains as well as losses which raises the expected gains and expected losses by the same amount so that the magnitude of the loss-gain gap remains unchanged.

\refstepcounter{assump}\label{assump:ss}\textbf{Assumption \theassump\ (short-sale constraints).} \textit{(i)} Negative realizations of the signal noise $u_{f1}$ are corrected within the quarter, meaning their price impact is removed by trading before the quarter ends. \textit{(ii)} A fraction $\delta \in [0,1]$ of the price impact of positive realizations of $u_{f1}$ survives the quarter.

A positive realization of the signal noise $u_{f1}$ holds the price above the firm's value (equation~\ref{eq:price}), an optimistic error. A negative realization holds it below, a pessimistic error. The noise $u_{f1}$ is normally distributed with mean zero. Optimistic and pessimistic errors of any given size are equally likely. Without a friction, trading would correct both errors within the quarter. The signal noise would then affect gains and losses only through the volatility channel, which leaves their gap unchanged (Proposition~\ref{prop:vol}). A loss-gain gap therefore requires a friction that blocks the correction on one side. That friction is the short-sale constraint, as in \citet{miller1977risk}. Selling short requires borrowing the stock. Mandates against short sales prohibit it for some investors \citep{almazan2004constrain}, while borrowing costs make it expensive for others \citep{davolio2002market}. Purchasing is not prohibited for any investor. This corrects the pessimistic error. An optimistic error can be corrected by a sale. The selling can come from pessimists who hold shares and from the arbitrageurs who can borrow and short. Short-selling flow is informed and predicts lower future returns \citep{boehmer2020what}. Thus, the surviving overpricing reflects the borrowing friction instead of lack of information among short sellers. The arbitrageurs have limited capital and limited horizons \citep{shleifer1997limits}. A pessimist who doesn't hold any shares and who cannot borrow cannot make any sale. Such a pessimist can abstain from purchasing but that cannot bring the price down. Thus, part of the overpricing stays in the price. The surviving fraction is $\delta$.\footnote{This pattern has empirical support. Stocks with greater analyst disagreement earn lower subsequent returns. Prices reflect the optimistic valuations when the constrained pessimists do not trade \citep{diether2002differences}.}

On the next earnings date, the firm's true value becomes public and the surviving overpricing is corrected. No buyer would be willing to pay more than the true value. Thus, the price settles at that true value. This fall in price from overpricing to the true value is a negative expected drift in the next quarter's returns. The daily excess returns scatter around a mean that carries this drift (\ref{assump:daily}). A lower mean raises the quarter's losses relative to its gains.

We now compute the expected size of the overpricing that survives the quarter. The price impact of the signal noise is $G_j u_{f1}$ by \eqref{eq:price}. Under \ref{assump:ss}, the part that survives is $\delta G_j \max(u_{f1}, 0)$, the price impact of the positive realizations scaled by $\delta$. Expected overpricing is the mean of this surviving part. The expectation $E[\max(u_{f1},0)]$ has a closed form, obtained in the same way as the expected tails with the mean set to zero (\ref{app:proof_drift}).

\proposition{prop:drift}~\textit{(Overpricing and its correction.) Under \ref{assump:ss}, expected overpricing is}
\begin{equation}\label{eq:op}
\mathrm{OP}_j \;=\; \delta\, G_j\, E[\max(u_{f1},0)] \;=\; \frac{\delta\, \gamma_j}{\sqrt{2\pi}} \cdot \frac{\sqrt{\tau_s}}{\tau_d + \tau_p + K_j + \tau_s},
\end{equation}
\textit{and under \ref{assump:daily} this is corrected as next-quarter drift $\mu^{drift} = -\mathrm{OP}_j$. $\mathrm{OP}_j$ vanishes as $\tau_s \to 0$ and as $\tau_s \to \infty$ and attains an interior maximum at}
\begin{equation}\label{eq:peak}
\tau_s^{*} \;=\; \tau_0 \;\equiv\; \tau_d + \tau_p + K_j .
\end{equation}
\textit{In measured-noise units $w \equiv \mathrm{VAQSV}$, the maximum is at $w^{*} = \sqrt{T_q/(\kappa\,\tau_0)}$. Holding $T_q$, $\kappa$, and the remaining group primitives fixed, the turning point is strictly decreasing in $K_j$.} \textit{(Proof in \ref{app:proof_drift}.)}

Expected overpricing in \eqref{eq:op} is the product of two opposing forces, scaled by the constant $\delta$. The first force is the weight the price places on the public signal, $G_j = \gamma_j\,\tau_s/(\tau_0 + \tau_s)$, from \eqref{eq:coefs} with $\overline{\Sigma}_j = 1/T_j$. This weight rises with the signal precision $\tau_s$. The second force is the mean positive noise $E[\max(u_{f1},0)]$, which equals $\sigma_u/\sqrt{2\pi}$ with $\sigma_u$ the standard deviation of $u_{f1}$. That standard deviation is $1/\sqrt{\tau_s}$, since $u_{f1}$ has precision $\tau_s$, so the second force falls as the first rises. When the feed is nearly uninformative, the market ignores it. So even if the noise is large, it cannot enter the price and there is no overpricing to correct. When the feed is highly informative, the market gives it nearly full weight, but it carries almost no noise. Expected overpricing peaks in between, at the point where the precision of the public signal equals the combined precision of all other information, $\tau_0 = \tau_d + \tau_p + K_j$ in \eqref{eq:peak}. At that point the feed provides exactly half of the market's total precision. The total is $T_j = \tau_0 + \tau_s$. $\tau_s$ is the feed's contribution to that total precision $T_j$. At $\tau_s = \tau_0$ the feed's share is $\tau_s/T_j = \tau_0/(2\tau_0) = 1/2$. When $\tau_s$ is below $\tau_0$, an increase in $\tau_s$ increases the weight more than it shrinks the noise, so the expected overpricing rises. When $\tau_s$ is above $\tau_0$, an increase in $\tau_s$ reduces the noise more than it increases the weight, so expected overpricing falls. Accordingly, the derivative of expected overpricing in $\tau_s$ is positive below $\tau_0$, zero at $\tau_0$, and negative above it.

Under \ref{assump:ss}(ii), a fraction $\delta$ of the optimistic price impact survives quarter $q$, with expected size $\mathrm{OP}_j$ in \eqref{eq:op}, and is corrected during quarter $q+1$. Accordingly, the drift specifications in Section~\ref{sec:reducedform} relate the quarter $q+1$ correction to quarter $q$ signal noise. Pessimistic noise is corrected within quarter $q$. Its price impact falls on quarter $q$ returns, not on the quarter $q+1$ correction that the drift specifications measure.

The model makes two predictions. The first is about shape. Holding the quarter's daily volatility fixed, the response of the excess of next-quarter losses over gains, $|\mathrm{Loss}_{f,q+1}| - \mathrm{Gain}_{f,q+1}$, to current-quarter measured noise first rises and then falls. The peak arrives at lower measured noise for stocks with more informed trading (Proposition~\ref{prop:drift}). For comparisons at a common $T_q$, the ordering of these peaks across groups does not depend on the conversion constant $\kappa$. The second is about sign. The correction comes only from optimistic errors. The next-quarter response is therefore stronger after quarters in which a larger share of the sentiment days were positive.

The alternative to the model is a symmetric mechanism, one in which optimistic and pessimistic sentiment errors are equally likely, move the price by the same amount, and are corrected to the same degree (\ref{app:symmetric}). Such a mechanism reaches losses only through volatility, which grows with the amount of noise and raises expected gains and expected losses by the same amount (Proposition~\ref{prop:vol}). It produces neither an interior peak nor a correction that depends on the sign of the noise. Section~\ref{sec:reducedform} tests both predictions against it. (The state dependence of the drift channel is derived in \ref{sec:model_state}.)

\subsection{The daily dealer market and the ordinal intensity index}\label{sec:model_micro}

The model shows that how much sentiment noise moves a stock's price (and hence its losses) depends on how much informed trading the stock has. If there is heavy informed trading, the price moves less. Hence, the noise matters less for the price. In such a case, a domestic-foreign gap in losses could then simply reflect a gap in informed-trading intensity rather than a gap in the precision of the public signal. Thus, a domestic-foreign comparison must hold informed-trading intensity fixed. This is achieved with the terciles.\footnote{Controlling for $\widehat{K}$ in a regression would allow comparison of the domestic and foreign stocks at the same value of $\widehat{K}$. These stock groups differ in informed-trading intensity. The regression would attribute that difference to the foreign indicator instead of removing it. As a robustness check, the three liquidity measures of Proposition~\ref{prop:micro} (the lagged quoted spread, the Amihud illiquidity ratio, and turnover) are nonetheless added as controls. The domestic-foreign difference in the loss-on-noise slope survives, at $-0.86$ (Newey-West $t=-2.21$, \ref{app:rob_r14d}).} The empirical work sorts stocks into terciles of informed-trading intensity formed inside each geography-quarter.\footnote{The headline regression (row R1, Table~\ref{tab:battery}) is one pooled regression. Its slope of next-quarter losses on measured noise differs by geography and by intensity tercile. This gives six slopes, one for each geography in each of the three terciles (equation~\eqref{eq:r1cells}). The domestic-foreign difference at a tercile is the difference between the domestic and the foreign slope in that tercile. The structural estimation reports the corresponding domestic-foreign contrasts of the repeat-author share by tercile in Table~\ref{tab:h1_contrasts}.} Thus, the empirical work uses $\widehat{K}$ only as an ordinal intensity index. For each stock-quarter, we form a ranking statistic (the previous quarter's quoted spread multiplied by $\sqrt{\tau_d + \widehat{\tau}_s}$, with $\tau_d$ the prior precision estimated for each geography and $\widehat{\tau}_s$ also from the previous quarter). Then within each geography-quarter, we rank stocks on this statistic and cut them into three terciles. Then we estimate $\widehat{K}$ for each geography and tercile, six values in all. The ranking cannot use $\widehat{K}$ itself, which is a single number per geography and tercile. $\widehat{K}$ is estimated only after the terciles are created. Unlike $\widehat{K}$, the ranking does not depend on the estimated scale constants of equation~\eqref{eq:micro}. The recovered $\widehat{K}$ increases across the three terciles (\ref{app:proof_micro}). \ref{app:dealer_detail} explains why $\widehat{K}$ can be compared across stocks within a group but not across groups. The domestic-foreign comparison conditions on the intensity tercile from the model and the foreign indicator from the data.

\subsection{Geography and the producer-knowledge mismatch hypothesis}\label{sec:model_geo}

Geography (domestic and foreign) enters the model as a grouping of stocks (Section~\ref{sec:model_env}). The hypothesis is about the contributor pool composition behind the public signal, specifically, the shares of the three types of contributors and the variances of the fad and of the bot content (equations~\ref{eq:sbar_body} and~\ref{eq:vaqsv}). Nothing else that differs between the two geographies enters the nonfundamental variance (equation~\ref{eq:vaqsv}). A domestic-foreign difference in it therefore reflects a difference in the pool composition. The hypothesis states the direction of that difference.

\refstepcounter{hyp}\label{hyp:mismatch}\textbf{Hypothesis \thehyp\ (producer-knowledge mismatch).} \textit{(a)} On English-language Twitter, foreign stocks have a larger nonfundamental variance ($\Phi_j$) than domestic stocks. \textit{(b)} Under a common conversion factor ($\kappa$), the observable signal-quality index ($T_q/\mathrm{VAQSV}_{fq}^2$) is lower for foreign stocks than for domestic stocks.

The hypothesis compares the informativeness of the feed for foreign and domestic stocks. The nonfundamental variance ($\Phi_j$) is the part of the feed's day-to-day variation that comes from the human fads and the automated content together (equation~\ref{eq:vaqsv}). It carries no information about the firm-specific innovation and does not change with the tweet count. When the measurement equation (equation~\ref{eq:vaqsv}) is fitted as a linear regression (Section~\ref{sec:model_sent}), $\Phi_j$ is the intercept. The signal precision ($\tau_s$) measures how sharply the quarter's feed narrows uncertainty about the firm-specific innovation (equation~\ref{eq:tau_s}). A larger nonfundamental variance makes the day's per-tweet sentiment average ($S_t/N_t$) more volatile across the days of the quarter. A more volatile average lowers the observable signal-quality index (holding the quarter's number of scored days and its tweet volume fixed) and delivers a less precise signal at a common conversion factor $\kappa$ (equations~\ref{eq:vaqsv} and~\ref{eq:tau_s}). Thus, a larger nonfundamental variance (the object of part (a) of \ref{hyp:mismatch}) lowers the observable signal-quality index (the object of part (b)). Part (b) of the hypothesis is a claim about the precision of the public signal. It is phrased in terms of the signal-quality index rather than $\tau_s$ for the following reason---$\tau_s$ equals the index divided by the conversion factor $\kappa$ (equation~\ref{eq:tau_s}). The conversion constant $\kappa$ is one constant shared by all stocks (Section~\ref{sec:structural}). Hence, the ordering of the precision of the public signal $\tau_s$ across stock groups is the ordering of the signal-quality index and does not depend on the estimate of $\kappa$. Both parts of \ref{hyp:mismatch} compare objects computed from the vendor's daily sentiment total and tweet count alone (Section~\ref{sec:model_sent}).

The economic intuition behind the hypothesis is that investors who understand a foreign firm's regulatory environment, the firm's management and product markets are underrepresented among its English-language contributors. Thus, the feed for foreign firms would carry a weaker signal than the feed for domestic firms. Part (a) of \ref{hyp:mismatch} relates to the nonfundamental variance as a whole, not split by contributor types. The feed has three contributor types (Section~\ref{sec:model_sent}). The empirical question is which of the contributor types produces an excess in the nonfundamental variance for foreign stocks as compared to domestic stocks. The excess in nonfundamental variance for stocks of foreign firms over those of domestic firms can reflect a larger fad variance, a larger bot-content variance, a different split of the day's tweets across the three types or a combination of these (equation~\ref{eq:vaqsv}). Although automated accounts can be identified in ground-truth samples \citep{gilani2017bots}, human fads cannot be. The data vendor's metrics are aggregated over tweets. Hence, these metrics have no account related identities (Section~\ref{sec:model_sent}). Thus, we cannot use account labels to attribute the excess in the nonfundamental variance for foreign stocks to a type of contributor. Such attribution rests on two indirect tests which are shown in the reduced-form battery (Section~\ref{sec:reducedform}). The first one uses a regression of squared measured noise on contributor dispersion (the average inverse tweet count) and the foreign indicator (Row R13a, Table~\ref{tab:battery}). Its reported object is the sign of the dispersion coefficient. The second one uses a regression of the NASDAQ trade count on an indicator for below-median dispersion and on log capitalization, on the NASDAQ subsample (Row R13b, Table~\ref{tab:battery}).\footnote{Both specifications have quarter fixed effects with standard errors double clustered by firm and quarter.}

For part (a) of \ref{hyp:mismatch}, the evidence is the foreign coefficient in the measurement equation (equation~\ref{eq:vaqsv}). We regress squared measured noise on the average inverse tweet count and the foreign indicator, with quarter fixed effects and standard errors double clustered by firm and quarter (Section~\ref{sec:model_sent}). The foreign coefficient is the difference in the intercept $\Phi_j$ between foreign and domestic stocks. We report it with and without log capitalization held fixed (Section~\ref{sec:structural}, Table~\ref{tab:size_contrasts}). Since a larger nonfundamental variance lowers the observable signal-quality index at a given number of scored days and tweet volume (equation~\ref{eq:vaqsv}), evidence for part (a) is also evidence for part (b). Part (b) is therefore not tested separately. We nonetheless report, for each informed-trading-intensity tercile, the median of the signal-quality index ($T_q/\mathrm{VAQSV}_{fq}^2$) for foreign and for domestic stocks (Table~\ref{tab:r6_dataterm}).\footnote{Both parts of the hypothesis pertain to the contributors, not to the investors who read the feed. Nonetheless, the number of investors who read the feed matters for prices. The model assumes that every investor observes the public signal. If only a fraction $\theta_j$ of investors observe it, the price responds to the signal as if its precision were $\theta_j\tau_s$ rather than $\tau_s$ (\ref{app:theta_lemma}). Home-market investors of a foreign firm, who tend to possess an informational advantage about that firm, may not read its English-language feed. Hence, a smaller $\theta_j$ for foreign stocks is plausible. Two consequences follow for the hypothesis. First, $\theta_j$ does not enter the nonfundamental variance or the signal-quality index, which are computed from the vendor's daily sentiment total and tweet count. Thus, parts (a) and (b) are unaffected. Second, let $\theta_D$ and $\theta_F$ denote the fractions for domestic and for foreign stocks, and $\tau_{s,D}$ and $\tau_{s,F}$ the corresponding precisions of the public signal. If $\theta_F \le \theta_D$, then the ratio of the domestic to the foreign signal-quality index understates the ratio of the precisions the price responds to, $\theta_D\tau_{s,D}/(\theta_F\tau_{s,F})$ (\ref{app:theta_lemma}). A smaller $\theta_F$ can then only widen the gap that part (b) states.}

\section{Data}\label{sec:data}

\subsection{Sources and panel construction}

The panel is composed of the Compustat quarterly fundamentals, the CRSP daily market data aggregated at the quarterly level, and the daily firm-level Twitter sentiment from Context Analytics (licensed through Dewey Data, \citealp{contextanalytics2022data}) aggregated at the quarterly level. The sentiment data is matched to the CRSP-Compustat merged database by ticker and quarter. The panel has 471{,}880 stock-quarters from 2010 to 2024, with the stock identified by the nine-digit CUSIP. When one ticker-quarter is attached to more than one security, we keep the larger-capitalization security, which drops the security with smaller capitalization in each of 738 ticker-quarters. The primary estimation sample starts in the first quarter of 2018 due to a change in the vendor's sampling methodology in 2017. Earlier quarters are used only in the robustness exercises. At construction, the quarter's gains and losses, the quarterly excess return, the within-quarter volatility of daily excess returns, and the residual-based gains and losses of the robustness checks are winsorized at the first and ninety-ninth percentiles of the full panel. Sentiment and liquidity measures are not winsorized. A firm-quarter enters the estimation only if it has at least twenty sentiment days, which every quantity built from the feed requires, and at least twenty trading days with a valid return, which every quantity built from returns or liquidity requires. The residual-based gains and losses of the robustness checks require at least twenty days with a residual. The primary structural sample has 98{,}308 firm-quarters. Group assignment covers 96{,}229 of them, of which 96{,}228 enter the estimation across 5{,}583 firms and 28 quarters. Foreign stocks in our sample are listed on U.S. exchanges, with foreign status classified by Compustat headquarters country. Identified by the CRSP share-level flag under a majority-of-days rule, American Depositary Receipts are 36.5\% of foreign observations. Directly listed shares, which include Canadian equities, are 63.5\% (Table~\ref{tab:composition_country}). (Calibrated constants and dependence conventions are in \ref{app:conventions}.)

\section{Structural Estimation}\label{sec:structural}

The empirical work has two parts. In the structural part, we impose the model's equations on the data and estimate the model's parameters. In the validation part, we estimate specifications that we do not derive from the model. In those specifications the model supplies only a regressor or the predicted sign of a coefficient. This section reports the four structural exercises. Section~\ref{sec:reducedform} reports the validation exercises. Table~\ref{tab:exercise_map} lists all the exercises with their classification.

\subsection{Estimation design}

We estimate the model in one step by the generalized method of moments on ninety-one closed-form moments.\footnote{Since the model moments have closed forms, a simulated method of moments would add simulation error without any gain. Maximum likelihood would require a joint density for every data object behind the moments and would score each daily return against the normal distribution of Assumption~\ref{assump:daily}. The method of moments uses this normality through expected gain and expected loss and leaves the other moment blocks distribution free.} The ninety-one moments are the model-implied values of the data moments (\ref{sec:model_to_estimation}), which cover the liquidity measures, the firm's cash flows, price informativeness, dispersion, squared measured noise, and the quarter's gains and losses. Using all of them has the same two advantages as in the microstructure inversion. The first advantage is that the measurement error in any one moment does not pass into the estimates because each parameter is estimated from every moment block jointly. The second advantage is that it allows testing of the relations which the model imposes upon the moments. A parameter vector that reproduces all 91 moments can exist only if the data satisfies each one of the relations which the model imposes upon the moments. Those relations hold for the population moments. The data moments are sample estimates of the population moments and, in a finite sample, carry sampling error. Hence, some amount of misfit remains even when the model is correctly specified. A test of over-identifying restrictions checks whether the misfit that remains at the estimates is larger than what the sampling error alone would produce. It shows that the misfit is indeed larger. The test (forty-six degrees of freedom for forty-five free parameters) rejects the null that all ninety-one relations hold (\ref{app:fit_detail}). Thus, the model is misspecified in at least one relation, which matters for the choice of weighting matrix. Under correct specification the probability limit of the estimate is the true parameter vector, irrespective of the weighting matrix. Under misspecification it is the parameter vector that minimizes the weighted distance between the model moments and the data moments, which depends on the weighting matrix. Thus, the weighting must be fixed in advance rather than chosen from the fit \citep{hall2003large}. We pre-specify a diagonal weighting matrix. Each diagonal entry is the inverse of the variance of that moment across 999 firm-block bootstrap replications of the data moments. Standard errors on the parameters come from the sandwich covariance for a fixed weighting matrix, which combines the Jacobian of the model moments, the weighting matrix, and the covariance of the data moments. For that covariance we use the firm-block bootstrap covariance of the moments, shrunk toward its diagonal by the Ledoit-Wolf rule. We set the posting intensity at 1.5. Hence, it is treated as a known constant in the covariance, so that the standard errors cover the forty-five estimated parameters. The assignment of stocks to informed-trading-intensity terciles is itself computed from the sample (Section~\ref{sec:model_micro}), while the sandwich covariance takes it as given. To account for the sampling variation in the tercile assignment, we run a firm-block bootstrap with 399 replications. Each replication draws firms with replacement (keeping each firm's complete time series) and repeats the tercile assignment, the data moments and the estimation, with the weighting matrix held at its point-estimate value (\ref{app:estdetail}).

\subsection{Headline estimates and the joint Wald test}

Table~\ref{tab:theta_hat} shows the parameter estimates. The GMM objective function (the weighted distance between the data moments and the model moments) has more than one local minimum. The headline estimate is the local minimum at which that distance is the smallest among the minima that satisfy the pre-specified saturation criterion and the acceptance checks (\ref{app:basin_detail}). The repeat-author shares are identified from contributor dispersion (Proposition~\ref{prop:dispersion}) and the fad scale from the nonfundamental variance (equation~\ref{eq:vaqsv}). Thus, the domestic-foreign contrasts in these estimates rest on two contrasts in the data. Contributor dispersion (the number of distinct accounts per unit of posting volume) is lower for the foreign feed than for the domestic feed. The nonfundamental variance (the part of sentiment variance not coming from firm fundamentals) is larger for the foreign feed. In order to check whether these two contrasts reflect geography and not firm size, we re-estimate both with log capitalization held fixed. Contributor dispersion for foreign stocks remains lower than that for domestic stocks ($t=-4.34$). The nonfundamental variance for foreign stocks remains higher than that of domestic stocks ($t=6.12$) (Table~\ref{tab:size_contrasts}). To translate the dispersion into repeat-author shares (Proposition~\ref{prop:dispersion}) requires a value for the posting intensity (the number of tweets per account per day that a repeat author posts). The moments identify the posting intensity weakly, because dispersion identifies the repeat-author shares only jointly with the posting intensity. A higher intensity paired with lower shares produces the same dispersion. When the posting intensity is estimated freely, the estimate is 1,102.98 and every repeat-author share is at its zero lower bound (\ref{app:headline_detail}). The freely estimated shares reached a bound, so we set the posting intensity at 1.5. This value lies inside the range of 1.2 to 1.85 implied by the external posting rates of \ref{app:composition_check}. At the set posting intensity, the feed for foreign stocks has the higher repeat-author share in each of the three informed-trading-intensity terciles. The fad scale (estimated for the feed of each geography and held equal across its three terciles, \ref{app:vaqsv_estimation}) is higher for the foreign feed too. A joint Wald test rejects the null that the three repeat-author shares and the fad scale are equal across geography. The statistic (four degrees of freedom) is 48.56 with standard errors clustered by firm ($p = 7.2\times 10^{-10}$), 70.38 with standard errors clustered by quarter ($p = 1.9\times 10^{-14}$) and 27.05 with standard errors double clustered by firm and quarter ($p = 1.9\times 10^{-5}$) (Table~\ref{tab:h1_contrasts}). To make sure that the value 1.5 does not drive these findings, we repeat the estimation at eight different values of the posting intensity from 1.2 to 1.85. The feed for foreign stocks maintains the higher share of repeat-authors and fad scale at every such value of posting intensity.\footnote{For each of these eight values of posting intensity, the joint Wald test rejects the null that the three shares of repeat-authors and the fad scale are equal across geography, when standard errors are clustered by firm, when standard errors are clustered by quarter and when the standard errors are double clustered by firm and quarter. The largest value among these $p$ values is $1.0\times 10^{-4}$ (Table~\ref{tab:lambda_profile}). This shows that the result that the foreign feed has the higher repeat-author shares and the higher fad scale is statistically significant at every value of the posting intensity in this range and robust to clustering variations. \ref{app:headline_detail} reports the identification diagnostics.}

\begin{table}[!htbp]\centering
\begin{small}
\caption{Headline structural parameter estimates with asymptotic sandwich standard errors and firm-block bootstrap intervals}\label{tab:theta_hat}
\begin{tabular}{lcc >{\raggedright\arraybackslash}p{6.2cm}}
\hline\hline
Parameter & Estimate & Sandwich SE & Bootstrap summary \\
\hline
$\pi$, Domestic T1 & 0.1744 & 0.0020 & 95\% interval: [0.1712, 0.1775] \\
$\pi$, Domestic T2 & 0.1573 & 0.0019 & 95\% interval: [0.1542, 0.1604] \\
$\pi$, Domestic T3 & 0.1447 & 0.0018 & 95\% interval: [0.1417, 0.1472] \\
$\pi$, Foreign T1 & 0.1820 & 0.0050 & 95\% interval: [0.1744, 0.1895] \\
$\pi$, Foreign T2 & 0.1637 & 0.0045 & 95\% interval: [0.1566, 0.1710] \\
$\pi$, Foreign T3 & 0.1698 & 0.0047 & 95\% interval: [0.1616, 0.1769] \\
$\sigma_b$, domestic & 0.02732 & 0.00031 & 95\% interval: [0.02686, 0.02775] \\
$\sigma_b$, foreign & 0.02927 & 0.00041 & 95\% interval: [0.02858, 0.02985] \\
$\sigma_e^2$ & 0.0001642 & 1.70e-05 & 95\% interval: [0.0001377, 0.0001878] \\
$s_P$ & 1.564 & 0.0574 & 95\% interval: [1.479, 1.721] \\
$\kappa$ & 394.2 & 34.2 & 95\% interval: [294, 490] \\
\hline
\addlinespace
$\delta$, at its zero lower bound & 0.0003728 & -- & Same value in each of the 399 replications \\
\hline\hline
\end{tabular}
\begin{minipage}{0.95\textwidth}\footnotesize\medskip
\textit{Notes:} 45 parameters are estimated using a one-step GMM with 91 moments while the posting intensity is set at 1.5 (Section~\ref{sec:structural}). A diagonal weighting matrix is used where there is an inverse of the bootstrap variance of each moment on the diagonal. Standard errors are computed from the sandwich estimator of the asymptotic covariance matrix of the GMM estimator (Section~\ref{sec:structural}). The covariance matrix of the data moments in the sandwich estimator is estimated by a firm-block bootstrap with 999 replications and shrunk toward its diagonal by the Ledoit-Wolf method. The 95\% intervals in the last column are the 2.5th and 97.5th percentiles of the estimates over 399 firm-block bootstrap replications. Each replication resamples firms with replacement and re-estimates the parameters on the resampled panel with the same weighting matrix as the point estimates. The surviving overpricing fraction ($\delta = 1/(1+e^{-z})$ with $z$ unrestricted) is strictly between 0 and 1. It approaches 0 only as $z$ decreases without bound. At the reported value ($\delta = 0.0003728, z = -7.89$), the GMM objective still increases in $\delta$. Thus, the estimate lies at its zero lower bound. The reported value is where the optimizer stopped along the transform. Each of the 399 replications starts from the point estimate and stops at the same value. Since the delta method does not apply to an estimate at the edge of its parameter space, no standard error is reported for $\delta$. Treating the posting intensity as free in the covariance (at the same point estimates) multiplies the standard errors of the repeat-author shares by 7.9 to 18.7. Dispersion identifies the shares only jointly with the posting intensity (Section~\ref{sec:structural}).
\end{minipage}
\end{small}
\end{table}

All four domestic-foreign contrasts in Table~\ref{tab:h1_contrasts} are negative, which affirms that the feed for foreign stocks has the higher repeat-author share in each of the three informed-trading-intensity terciles and the higher fad scale. The higher foreign fad scale and the higher foreign top-tercile share hold in each of the 399 bootstrap replications. These contrasts provide supporting evidence for our producer-knowledge mismatch hypothesis.

\begin{table}[!htbp]\centering
\begin{small}
\caption{Domestic-foreign contrasts in the repeat-author shares and the fad scale, with firm-block bootstrap confidence intervals and the joint Wald test of their equality across geography}\label{tab:h1_contrasts}
\begin{footnotesize}\setlength{\tabcolsep}{4pt}
\begin{tabular}{lcccc}
\hline\hline
Object & Estimate & Bootstrap median & \shortstack{95\% interval\\or $p$ value} & \shortstack{Share of\\replications $<0$} \\
\hline
$\pi$ contrast, tercile 1 & -0.00766 & -0.00778 & [-0.01565, 0.00032] & 0.972 \\
$\pi$ contrast, tercile 2 & -0.00638 & -0.00647 & [-0.01361, 0.00127] & 0.950 \\
$\pi$ contrast, tercile 3 & -0.02512 & -0.02454 & [-0.03260, -0.01723] & 1.000 \\
$\sigma_b$ contrast & -0.00195 & -0.00195 & [-0.00247, -0.00139] & 1.000 \\
Joint Wald, clustered by firm & 48.56 & -- & $p = 7.2\times 10^{-10}$ & -- \\
Joint Wald, clustered by quarter & 70.38 & -- & $p = 1.9\times 10^{-14}$ & -- \\
Joint Wald, double clustered & 27.05 & -- & $p = 1.9\times 10^{-5}$ & -- \\
\hline\hline
\end{tabular}
\end{footnotesize}
\begin{minipage}{0.95\textwidth}\footnotesize\medskip
\textit{Notes:} The table shows repeat-author shares in the three terciles (of informed trading intensity) and the fad scale for domestic and foreign stocks, with the posting intensity set at 1.5. The 95\% confidence intervals are bootstrap (sampling firms with replacement) percentile intervals from 399 replications. The last column is the share of those replications in which the contrast is negative (implying that foreign stocks have the higher repeat-author share in each of the three terciles and the higher fad scale). The joint Wald test examines the null that all four contrasts are zero (with four degrees of freedom) using standard errors clustered by firm, clustered by quarter and double clustered by firm and quarter \citep{cameron2011robust}. 
\end{minipage}
\end{small}
\end{table}

The two data contrasts of this subsection, the lower contributor dispersion and the higher nonfundamental variance of the foreign feed, also hold within size terciles (Table~\ref{tab:size_contrasts}, row R16, \ref{app:rob_size}). The dispersion gap keeps its sign in all three size terciles and is significant in the two smaller ones ($t=-2.49$ and $-3.54$). The variance gap is significant in all three ($t=3.12$, $5.38$ and $5.21$). The next task is showing who the repeat authors are, whether it is the specialists who also trade or automated accounts that post without trading. We regress the NASDAQ trade count of a stock-quarter on an indicator for below-median dispersion (a narrower contributor pool), with log capitalization as a control (Row R13b, Table~\ref{tab:battery}). Quarters with a narrower contributor pool have 8,121 more trades ($t=10.5$). Thus, repeat posting goes together with trading, which is a feature of the specialists not the automated accounts. Then we regress squared measured noise on dispersion, with the inverse tweet volume and the foreign indicator as controls (Row R13a, Table~\ref{tab:battery}). If repeat posting by automated accounts added noise to the feed, a narrower pool would show more noise and the coefficient on dispersion would be negative. The coefficient is $1.6\times 10^{-5}$ ($t=0.12$) (95\% CI = [$-2.5\times 10^{-4}$, $2.8\times 10^{-4}$]). Thus, there is no evidence of the automated-content channel differing between the domestic and the foreign feed. This result also does not show that the channel is small, since the interval is wide relative to the coefficient. The data helps identify the repeat-author share as a whole and does not split it into specialists and automated accounts (Proposition~\ref{prop:dispersion}). Hence, we put no number on the automated share. Classifying an account as automated or human from its tweets is correct for about 77 percent of accounts at best (\ref{app:bot_prevalence}), so a count of automated accounts in our feeds would mislabel about one account in four.

\subsection{The conversion constant}

The conversion constant ($\kappa$) converts the observable signal-quality index into the precision of the public signal ($\tau_s$). The estimate of $\kappa$  is 394 (se=34 and 95\% CI=[294, 490], Table~\ref{tab:theta_hat}, \ref{app:kappa_detail}). When we fix $\kappa$ below that confidence interval at 240 or above that confidence interval at 565, and re-estimate the other parameters, the GMM objective rises by only 0.7 and 0.4 percent (Table~\ref{tab:kappa_profile}). Thus, the fit of the model barely depends on the value of $\kappa$. $\kappa$ absorbs the three departures of the data from the conditions under which the signal-quality index would equal $\tau_s$, namely serial dependence of the daily readings, the attenuation of the day's sentiment by automated content, and the scale of the vendor's per-tweet sentiment score relative to the firm's cash-flow innovation (Section~\ref{sec:model_sent}). The serial dependence can be obtained from the autocorrelation of the daily per-tweet series. The median lag-one autocorrelation is near 0.04 in both geographies with a dependence multiplier near 1.1 (\ref{app:serial_dependence}). The remainder of $\kappa$ reflects the scale of the vendor's score and the attenuation by automated content, which the data do not separate (\ref{app:proof_vaqsv}). Thus, we don't calibrate an automated-account level. The sampling uncertainty in the estimate of $\kappa$ affects none of the domestic-foreign comparisons of signal quality. Those comparisons are ratios of $\tau_s$ across groups, in which $\kappa$ (common to all groups) cancels and leaves a ratio of the signal-quality index. The slopes of gains and of losses on the lagged signal-quality index are estimated with the index in logarithms, where $\kappa$ enters only as an intercept. Only the level of $\tau_s$ depends on $\kappa$. None of our findings uses that level.

\subsection{The short-sale friction}

In the model, a fraction ($\delta$) of the overpricing from an optimistic error in the public signal survives the quarter because of the short-sale friction. This surviving fraction of overpricing is corrected in the next quarter and it manifests as an excess of losses over gains (Proposition~\ref{prop:drift}). The estimate of $\delta$ lies at its zero lower bound (Table~\ref{tab:theta_hat}). With $\delta$ at its zero lower bound, the model is able to explain the next-quarter gains and losses without the mechanism involving correction of surviving overpricing. The model moments for the means of next-quarter gains and losses and for their slopes on the signal-quality index depend on the quarter's return variance (the volatility channel) and on the mean daily excess return (the group-level risk premium plus the correction drift, \ref{assump:daily}). The correction drift is proportional to $\delta$. With $\delta$ at its zero lower bound, just the return variance and the risk premium match these moments to their data counterparts. The zero lower bound doesn't affect the concentration and fad contrasts, which are identified from contributor dispersion and the nonfundamental variance (Section~\ref{sec:structural}), or the reliability gap, which is reduced-form evidence (row R14, Table~\ref{tab:battery}).

\subsection{Fit, misfit, and the informativeness block}

The GMM objective is the sum over the 91 moments (45 free parameters) of the squared distance between each data moment and its model-implied counterpart (the misfit of that moment), each distance weighted by the inverse of that moment's bootstrap variance (Section~\ref{sec:structural}). At the headline estimate (Table~\ref{tab:theta_hat}), the GMM objective is equal to 4,338. The model imposes 46 restrictions on the data moments beyond those needed to estimate the parameters (the overidentifying restrictions). The Hansen $J$ test rejects the null hypothesis that these 46 restrictions hold at the 5 percent level (\ref{app:fit_detail}). Table~\ref{tab:perblock_j} is an attempt to locate the misfit by splitting the objective at the headline estimate across the nine blocks of moments and reporting (for each block) its share of the objective and its largest moment-level $t$ statistic (the misfit of a moment divided by the moment's bootstrap standard error). The liquidity block holds 24 moments. Each is the mean of one of the four liquidity measures of \ref{sec:model_to_estimation} (the quoted spread, the high-low spread, the Amihud ratio, and turnover) in one of the six geography-tercile groups (domestic and foreign stocks at each of the three informed-trading-intensity terciles). The weighted squared misfit of these 24 moments makes up 84.8 percent of the objective. The largest $t$ statistic among them is 28.5 (a miss of 28.5 bootstrap standard errors on one moment). Most of the liquidity block's misfit is due to the restriction that a single scale constant ($a_s$) links the model's spread to the quoted spread in all six groups (equation~\eqref{eq:micro}). If we let $a_s$ differ across the six groups (which adds five free parameters) the objective falls from 4,338 to 1,588. The liquidity block misfit doesn't affect the assignment of stocks to informed-trading-intensity terciles through the estimated scale constants ($a_s$, $a_{cs}$ and $a_\lambda$). This is because such assignment is done before the joint estimation on a ranking statistic (Section~\ref{sec:model_micro}) which doesn't contain the scale constants. When the terciles are formed using the high-low spread or using the quoted spread (capped at its adverse-selection bounds), the domestic-foreign difference in the slope of next-quarter losses on measured noise in the lowest intensity tercile ($\beta_{D1}-\beta_{F1}$ in equation~\eqref{eq:r1cells}) keeps its sign. When the terciles are formed on the lagged quoted spread alone, the reliability gap (the domestic-foreign difference in the share of the 27 quarters in which the quarterly slope of next-quarter losses on measured noise is negative) is positive in each of the three spread terciles (\ref{app:fit_detail}).

The misfit of the liquidity block also doesn't affect the parameters which justify the main results (the repeat-author shares, $\pi_j$, and the fad scales, $\sigma_{b,j}$). This is because excluding the liquidity block from the moment system leaves the domestic-foreign contrasts in the repeat-author share by tercile and in the fad scale unchanged up to five decimals. If we let $a_s$ differ across groups, the domestic-foreign contrasts change in magnitude by less than 12 percent and don't change sign (Table~\ref{tab:liq_robust}). The repeat-author share in each tercile and the fad scale remain higher for the foreign stocks under the Ledoit-Wolf-shrunk full-matrix weighting (Table~\ref{tab:variants}). The misfit of the liquidity block also shows up in the regressions of the quoted spread and the Amihud illiquidity ratio on the previous quarter's measured noise and dispersion (log capitalization and per-day turnover as controls and quarter fixed effects, row R5, Table~\ref{tab:battery}). In the quoted spread regression, the coefficient on lagged measured noise is $-0.034$ ($t=-5.41$). In the Amihud ratio regression, it is $-6.30$ ($t=-3.88$). Both signs are the opposite of the prediction of the adverse-selection mechanism in the dealer market that the spread and the price-impact ratio widen as the public signal grows noisier (Proposition~\ref{prop:micro}). The informativeness block (the 12 price-informativeness moments, Table~\ref{tab:perblock_j}) has a misfit of its own. Price informativeness (the covariance of the price with earnings one to four quarters ahead, scaled by the standard deviation of the price, Proposition~\ref{prop:pinf}) cannot be negative in the model, since none of the three factors in equation~\eqref{eq:pinf} is negative. In the data it is negative for foreign stocks in informed-trading-intensity terciles 2 and 3 at each of the four horizons (Table~\ref{tab:r6_dataterm}). The removal of the 12 informativeness moments from the moment system changes the repeat-author shares and the fad scales by less than $10^{-5}$ (Table~\ref{tab:variants}). Thus, neither the misfit of the liquidity block nor the misfit of the informativeness block changes the domestic-foreign contrasts in the repeat-author share and in the fad scale (\ref{app:fit_detail}).

\begin{table}[!htbp]\centering
\caption{Concentration and fad contrasts under liquidity-block robustness}\label{tab:liq_robust}
\begin{tabular}{@{}lccc@{}}
\hline\hline
Domestic-foreign contrast & Headline & \shortstack{Liquidity block\\excluded} & \shortstack{Common spread\\scale relaxed} \\
\hline
$\pi$ contrast, tercile 1 & $-0.00766$ & $-0.00766$ & $-0.00678$ \\
$\pi$ contrast, tercile 2 & $-0.00638$ & $-0.00638$ & $-0.00701$ \\
$\pi$ contrast, tercile 3 & $-0.02512$ & $-0.02512$ & $-0.02594$ \\
$\sigma_b$ contrast & $-0.00195$ & $-0.00195$ & $-0.00198$ \\
\hline\hline
\end{tabular}
\begin{minipage}{0.95\textwidth}\footnotesize\medskip
\textit{Notes:} The first three rows show the domestic-foreign contrasts in the repeat-author share ($\pi$) for each informed-trading-intensity tercile in that order. The last row is the domestic-foreign contrast in the fad scale ($\sigma_b$). When a contrast is negative, foreign stocks have the higher value. The headline column is the same as the estimate column of Table~\ref{tab:h1_contrasts}. The middle column shows the contrasts re-estimated with the 24 liquidity moments (the quoted spread, the high-low spread, the Amihud ratio, and turnover for each of the six geography-tercile groups) excluded from the moment system, with the free parameters re-estimated on the remaining 67 moments. They equal the headline contrasts up to five decimal places. The parameters that the liquidity moments identify (the three scale constants, $a_s$, $a_{cs}$, $a_\lambda$, and the six noise-trading standard deviations, $\sigma_{x,j}$), and the conversion constant $\kappa$, are held at their headline values. In the last column, the single scale constant $a_s$ that links the model's spread to the quoted spread in all six groups (equation~\eqref{eq:micro}) is replaced by one $a_s$ per group and all parameters are re-estimated. Both variants start from the headline estimate with the posting intensity set at 1.5. The variants that remove the informativeness block or use the Ledoit-Wolf-shrunk full-matrix weighting are in Table~\ref{tab:variants}.
\end{minipage}
\end{table}

\section{Reduced-Form Evidence and Falsification}\label{sec:reducedform}

The parameter estimates in Section~\ref{sec:structural} are obtained by minimizing the weighted distance between the 91 data moments and their model-implied counterparts. This section is about whether the model's implications hold in regressions that do not impose such structure. The regressions use quarter fixed effects, standard errors double clustered by firm and quarter, the primary sample from 2018 onward, and the screens of Section~\ref{sec:data}.\footnote{Since the standard errors are double clustered by quarter and firm for a panel which spans only 27 quarters, the cluster-robust standard errors can understate sampling variation and the $t$ tests can over-reject. Therefore, for five of the battery's headline coefficients, we also show the $p$ values from the wild-cluster bootstrap with Rademacher weights \citep{cameron2008bootstrap} (which remains valid with few clusters) with stars (Table~\ref{tab:battery}). Each of the five values is below 0.05.} Wherever the lagged sentiment enters the regressions, the lag is formed after the sample is cut at 2018Q1, which ensures that no sentiment value from before the vendor's methodology change enters any regression. The specification used is,

\begin{equation}\label{eq:rfspec}
Y_{f,q+h} \;=\; \beta\, X_{fq} \;+\; \Gamma' Z_{fq} \;+\; \alpha_q \;+\; \eta_{fq},
\end{equation}
where $X_{fq}$ is the row's sentiment regressor measured in quarter $q$, $Y_{f,q+h}$ is the row's outcome measured $h$ quarters later, the vector $Z_{fq}$ collects the row's stated controls and conditioning, and $\alpha_q$ are quarter fixed effects. The horizon is $h=1$ for every row in which the outcome is measured one quarter after the sentiment regressor (the prediction rows and the two liquidity rows, row R5, Table~\ref{tab:battery}). The horizon is $h=0$ for the two rows in which the outcome and the regressor are measured in the same quarter (rows R13a and R13b, Table~\ref{tab:battery}). For the rows that use the announcement-aligned outcome, the outcome quarter is the calendar quarter of the earnings announcement that follows the quarter for which sentiment is measured (the next quarter for most firm-quarters, Table~\ref{tab:seams}).\footnote{Two rows depart from this specification. Row R14 (Table~\ref{tab:battery}) runs a cross-sectional regression of next-quarter losses on measured noise per quarter and geography, and tests the domestic-foreign difference in the share of quarters with a negative slope by a circular block bootstrap over quarters. Row R12 (Table~\ref{tab:battery}) is not a regression but a rank correlation between two stock rankings, described in \ref{app:battery}. Each row's specification is in Table~\ref{tab:battery_specs}.}

The slope of next-quarter total losses on measured noise differs across the six geography-tercile cells (row R1, Table~\ref{tab:battery}). The regressor is interacted with the indicators for geography and for intensity tercile, and each cell has its own intercept,
\begin{equation}\label{eq:r1cells}
Y_{f,q+1} \;=\; \sum_{g\in\{D,F\}}\;\sum_{k=1}^{3}\left(\beta_{gk}\,X_{fq} + \delta_{gk}\right)\mathbf{1}\{f\in gk\} \;+\; \Gamma' Z_{fq} \;+\; \alpha_q \;+\; \eta_{fq},
\end{equation}
where cell $gk$ collects the stock-quarters of geography $g$ in intensity tercile $k$ and $\delta_{gk}$ is its intercept, with one cell intercept normalized. The terciles are formed within each geography-quarter. Thus, cell membership is reassigned every quarter. The vector $Z_{fq}$ consists of the controls only (the FF3 betas, MAX, log capitalization, and buzz). The implied slope $\beta_{gk}$ is the response of next-quarter total losses to measured noise in cell $gk$. The domestic-foreign difference at intensity tercile $k$ is given by $\beta_{Dk} - \beta_{Fk}$.

\begin{table}[!htbp]\centering
\caption{Headline reduced-form results}\label{tab:battery}
\begin{tabularx}{\textwidth}{@{}>{\RaggedRight}X>{\RaggedRight}p{0.30\textwidth}>{\RaggedRight\arraybackslash}p{0.20\textwidth}@{}}
\hline\hline
Row & Headline estimate & Statistic \\
\hline
\multicolumn{3}{@{}l}{\textit{Panel A. Main reduced-form evidence}} \\
\addlinespace
Reliability by geography (R14) & Mean quarterly domestic-foreign slope difference, $-1.40$. Domestic share of negative-beta quarters 0.778, foreign share 0.481 & Newey-West $t = -2.99$, share difference $p = 0.016$ \\
Volatility and losses, unconditional levels (R1) & Implied slopes negative in all six cells & Largest $|t| = 1.60$ \\
Dispersion and gains by tercile (R4) & $+0.128^{***}$ in the low tercile & $t = 3.15$ \\
Volume-contamination horse race (R9) & Raw signal $-0.317^{***}$, adjusted $-0.612^{**}$ & $t = -6.02$ and $-2.11$ \\
Consolidated joint regression, losses (R11) & Noise $-1.64^{***}$, dispersion $+1.10^{***}$ & $t = -4.88$ and $+12.7$ \\
\addlinespace
\multicolumn{3}{@{}l}{\textit{Panel B. Mechanism and diagnostic evidence}} \\
\addlinespace
Fad-filtered noise diagnostic (R3FAD) & Squared term negative in all three terciles & Largest $|t| = 0.31$ \\
Drift-isolated level, announcement aligned (R2) & $+0.211$ & $t = 1.15$ \\
State dependence, announcement aligned (R7) & $-23.8$ & $t = -1.98$ \\
Correction after optimism, announcement aligned (R8) & $+0.214$ & $t = 0.21$ \\
Spread on lagged noise (R5) & $-0.034$ & $t = -5.41$ \\
Price impact on lagged noise (R5) & $-6.30$ & $t = -3.88$ \\
Attribution sign test (R13a) & $+1.6\times 10^{-5}$ & $t = 0.12$ \\
Trading coincidence (R13b) & $+8{,}121$ trades & $t = 10.5$ \\
Similarity-index bridge (R12) & Rank correlation $-0.379$ & Diagonal share $0.227$ \\
\hline\hline
\end{tabularx}
\begin{minipage}{0.95\textwidth}\footnotesize\medskip
\textit{Notes:} Each row's specification is in Table~\ref{tab:battery_specs}. Stars mark the wild-cluster bootstrap $p$ value for the five coefficients of rows R4, R9, and R11, with three stars for $p$ below 0.01, two for $p$ below 0.05 and one for $p$ below 0.10.
\end{minipage}
\end{table}

\subsection{Reliability differs by geography}\label{sec:rf_reliability}

\begin{figure}[!htbp]\centering
\caption{Quarterly reliability of the sentiment-loss relation by geography}\label{fig:r14_reliability}
\includegraphics[width=0.85\textwidth]{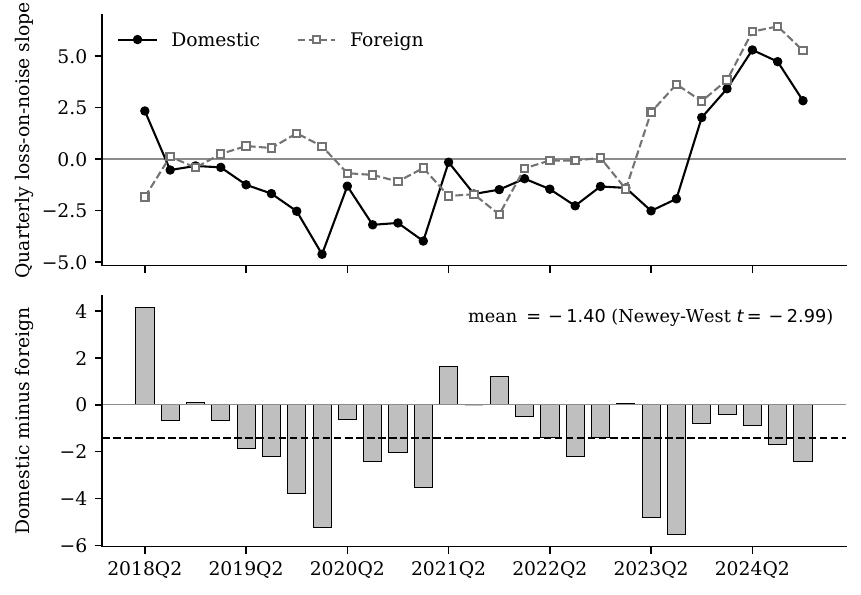}
\begin{minipage}{0.85\textwidth}\footnotesize\medskip
\textit{Notes:} The top panel of the figure shows the quarterly slopes of next-quarter total losses on measured noise, estimated separately by geography with log capitalization and the FF3 betas as controls, over the 27 usable quarters of the primary sample. Each geography-quarter cell requires at least 15 stocks. The bottom panel plots the domestic-foreign difference in these slopes. The dashed line is its mean ($-1.40$, Newey-West $t=-2.99$) at four lags (row R14, Table~\ref{tab:battery}).
\par\medskip\noindent Alt text: The two panels cover the period from 2018 to 2024. The top panel plots the quarterly loss-on-noise slope for domestic stocks (solid line) and for foreign stocks (dashed line). The solid line is negative in more quarters. The bottom panel shows bars for the domestic-foreign difference by quarter. The mean (indicated by the dashed line) is $-1.40$ (Newey-West $t=-2.99$). 
\end{minipage}
\end{figure}

For each geography-quarter, we regress next-quarter losses on measured noise across stocks using log capitalization and the FF3 betas as controls. Then, we compare foreign and domestic on two statistics (row R14, Table~\ref{tab:battery}). The first one is the mean of the slope over quarters, which allows us to compare the size of the next-quarter losses-noise relation. The second one is the share of quarters in which the slope is negative, which allows us to compare the stability of the sign of the next-quarter losses-noise relation across quarters. The results show that the mean quarterly slope is more negative for domestic stocks than for foreign stocks by 1.40 (Newey-West $t=-2.99$).\footnote{27 of the 28 quarters of the primary sample are used as the first quarter has no in-sample predecessor to supply the regressor. The Newey-West standard error of the mean slope difference allows autocorrelation in the quarterly slope differences up to four quarters. The difference in shares is tested with a circular block bootstrap over quarters, with blocks of four consecutive quarters and 999 draws (the same resampled quarters for domestic and foreign). The mean slope difference is negative and significant at 5 percent when any single sample year is left out (\ref{app:rob_r14e}).} The slope is negative in about four of five quarters for domestic stocks as opposed to one in two for foreign stocks (a domestic-foreign difference in shares of quarters with a negative slope of 0.296, block-bootstrap $p=0.016$). A one-standard-deviation increase in measured noise (0.0099 in the estimation sample) deepens next-quarter losses by 0.64 percentage points of quarterly excess return for domestic stocks. The mean domestic-foreign slope difference of $-1.40$ times the standard deviation of 0.0099 is $-0.0139$ in the loss variable (a next-quarter loss deeper by 1.39 percentage points for domestic stocks relative to foreign stocks). The mean domestic-foreign slope difference survives in sign after we add a richer control set consisting of the lagged quoted spread, the Amihud illiquidity ratio, and turnover to the size and factor-beta controls ($-0.86$, Newey-West $t=-2.21$, \ref{app:rob_r14d}). Figure~\ref{fig:r14_reliability} shows the quarterly slopes and their difference. This instability for the foreign stocks suggests that the next-quarter losses-noise slope for foreign stocks does not have a stable sign from quarter to quarter, implying that the foreign signal becomes diluted.

The reliability gap could reflect a domestic-foreign difference in informed-trading intensity rather than a difference in the feeds (Section~\ref{sec:model_micro}). To hold informed-trading intensity fixed, we rerun the quarterly regressions with the same controls within each informed-trading-intensity tercile and compare the share of quarters with a negative slope for foreign and domestic stocks. The domestic-foreign difference in the share of quarters with a negative slope survives in sign in tercile 1 with a difference of 0.296 (share for domestic stocks is 0.852 and for foreign stocks is 0.556, block-bootstrap $p=0.004$). The difference for tercile 2 is 0.148 and is not significant. In tercile 3, the difference closes to $-0.037$ and is not significant. Hence, informed-trading intensity does not explain the gap away, but the gap is confined to tercile 1. Across terciles 1 to 3, the share of quarters with a negative slope is 0.852, 0.889, and 0.741 for domestic stocks (23, 24, and 20 of 27 quarters), and 0.556, 0.741, and 0.778 for foreign stocks (15, 20, and 21 of 27 quarters). We do not test the differences between adjacent terciles within a geography, since the model gives no prediction for how the gap should change from one tercile to the next. The gap closes in tercile 3. The model has an explanation for a closing gap but without giving the tercile in which the gap should close. When investors know a lot about a stock from sources other than the feed, the investors put lesser weight on the feed so that sentiment in the feed affects the price the least. The loss-on-noise slope is then close to zero, and its sign in any one quarter is a matter of sampling error. Which tercile has the most such outside knowledge depends on which part of it varies across terciles. If informed trading is the part that varies, tercile 3 has the most. If the prior precision is the part that varies, as it is at the headline estimates, tercile 1 has the most. The two readings predict opposite orderings of the gap across terciles. The domestic shares rise from tercile 1 to 2 and fall in tercile 3, which fits neither ordering. Hence, we give the tercile pattern no structural interpretation (\ref{app:tercile_pattern}).

We classify the foreign stocks using three categories---by listing type (ADRs and directly listed shares), by the language of the headquarters country (English-speaking and non-English-speaking), and by country (Canada and other foreign). The separation by listing type shows a significant difference between ADRs and directly listed foreign shares. The share of quarters with a negative slope is 0.296 (8 of 27 quarters) for ADRs against 0.593 (16 of 27 quarters) for directly listed foreign shares (a difference of $-0.296$, block-bootstrap $p=0.022$). The price of an ADR listing in the U.S. depends on its home-market share price through the depositary link, which can weaken the relation between the feed's noise and next-quarter losses because of reasons unrelated to information.\footnote{The difference in the share of quarters with a negative slope between stocks of foreign firms headquartered in English-speaking countries and those headquartered in non-English-speaking countries is $-0.074$ ($p=0.561$). The same difference between Canadian and other foreign stocks is also $-0.074$ ($p=0.597$). Neither of these two differences is significant (Table~\ref{tab:reliability_splits}, \ref{app:reliability_splits}). The mean slopes by listing type are in \ref{app:rob_r14d}.}\textsuperscript{,}\footnote{There are other alternatives to informed-trading intensity as the conditioning variable, such as analyst coverage, institutional ownership, and media coverage. Analyst coverage requires IBES data, institutional ownership requires 13F holdings records and the media coverage requires a news archive. Since the licensed access to this data is not available, we cannot test these alternatives directly.} The reliability gap survives within informed-trading-intensity tercile 1 and with log capitalization controlled in every cross-section. These two results leave less room for a domestic-foreign difference in analyst coverage, institutional ownership or media coverage to explain the gap, but we do not test these three.

\subsection{The volatility channel in reduced form}

Higher sentiment noise in the present quarter implies larger losses in the next quarter. The slope of next-quarter losses on noise in the headline loss regression is negative in all six geography-tercile cells. However, none of them are individually statistically significant (row R1, Table~\ref{tab:battery}). The slope of next-quarter losses on noise, in a pooled specification with dispersion and the full set of control variables, is negative and significant (row R11, Table~\ref{tab:battery}). After controlling for realized volatility in the regression, the cell slopes move toward zero or turn positive but none of them is statistically distinguishable from zero (row R2, Table~\ref{tab:battery}). Thus, the response of next-quarter losses to noise works through return volatility instead of through a lower mean return. A less precise public signal leads to more fluctuations in price, which show up as larger losses. The coefficients on measured noise and dispersion have opposite signs in the joint regression (row R11, Table~\ref{tab:battery}). Thus, while a signal with higher noise predicts larger next-quarter losses, a wider or more dispersed contributor pool predicts smaller losses, with each effect surviving in the presence of the other. Therefore, the two measures carry different information about the feed. One is about the precision of the feed's signal and the other is about the span of the feed's contributors. The paper treats them as distinct channels.

\subsection{The correction channel in reduced form}\label{sec:rf_correction}

In the model, the expected overpricing that survives a quarter is proportional to $\sqrt{\tau_s}/(\tau_0+\tau_s)$ (equation~\eqref{eq:op}), which varies with the precision of the public signal (equation~\eqref{eq:peak}). For each stock-quarter we evaluate this factor with $\tau_0$ at the headline estimate of the group's precisions ($\tau_d$, $\tau_p$ and $K_j$) and with $\tau_s$ at the stock's estimated precision of the public signal in the previous quarter (which is equal to the previous quarter's signal-quality index divided by the headline estimate of the conversion constant, equation~\eqref{eq:tau_s}). We refer to this as the model-motivated signal-noise index. We then project the excess of next-quarter losses over gains on this index,
\begin{equation}\label{eq:driftproj}
\bigl|\mathrm{Loss}_{f,q+1}\bigr| - \mathrm{Gain}_{f,q+1} \;=\; b_3\,\mathrm{Index}_{fq} \;+\; \Gamma' Z_{fq} \;+\; \alpha_q \;+\; \gamma_g \;+\; \eta_{fq},
\end{equation}
where $\mathrm{Index}_{fq}$ is the model-motivated signal-noise index, $Z_{fq}$ consists of the factor betas, the lottery control and log capitalization. $\alpha_q$ are quarter fixed effects and $\gamma_g$ are geography-tercile group effects which confine the comparison to stock-quarters within the same geography-tercile group. The excess of losses over gains in the quarter of the next earnings announcement rises with the index ($t=9.63$). If the group effects are removed, the slope turns negative, since the comparison then also runs across groups and mixes the differences in group averages with the relation within groups (\ref{app:drift_detail}). Since the structural friction parameter lies at its zero lower bound (Table~\ref{tab:theta_hat}) and the correction after optimistic quarters is not detectable (row R8, Table~\ref{tab:battery}), the rise of the loss-gain excess with the index has the sign which the overpricing correction mechanism predicts but the fitted model does not produce. Thus, the rise is reported as a reduced-form pattern instead of a structural identification of an overpricing correction caused by short-sale constraints.

\subsection{Gains and the span of the contributor pool}\label{sec:rf_gains}

The results also show that the next-quarter gains rise with the span of the contributor pool as measured by dispersion (the number of distinct posting accounts per tweet), mainly where informed trading intensity is low. The slope of next-quarter gains on dispersion is positive and significant in informed-trading-intensity tercile 1 (row R4, Table~\ref{tab:battery}), while in terciles 2 and 3 it is statistically indistinguishable from zero ($t = 0.11$ and $0.34$). Holding the other group primitives fixed, the model implies that the price responds more to the feed where informed trading is lighter (equation~\eqref{eq:coefs}). This direction is seen in the slope of gains on dispersion, which is positive in tercile 1 and indistinguishable from zero in terciles 2 and 3. Holding cash-flow predictability fixed does not remove the pattern, but within tercile 1 it is strongest for the firms with the most predictable cash flows, the opposite of what the model implies (\ref{app:rob_r4_innov}, Table~\ref{tab:r4_innovvar}). A wider contributor pool predicts larger next-quarter gains (row R4, Table~\ref{tab:battery}) and smaller next-quarter losses in the joint regression (row R11, Table~\ref{tab:battery}). Therefore, a wider contributor pool raises the mean of the next-quarter returns. The volatility channel cannot produce such an effect, since it raises expected gains and expected losses by the same amount and leaves the mean return unchanged (Proposition~\ref{prop:vol}). The correction channel cannot produce that pattern either, since it is closed by the zero lower bound on the short-sale friction. We therefore report the relation between the span of the contributor pool and next-quarter gains as a reduced-form regularity which the model doesn't predict.

The tercile orderings in Section~\ref{sec:rf_gains} and in \ref{app:drift_detail} are subject to the same limitation. The estimated informed-trading intensities ($K_j$) are also smallest in tercile 3 for both geographies. Thus, the tercile orderings in Sections~\ref{sec:rf_reliability} and \ref{sec:rf_gains} and in \ref{app:drift_detail} agree with the model's comparative statics only when the other group primitives are held fixed. They do not follow from the estimated values of $\tau_0$ and $K_j$ across the groups.

\section{Conclusion}\label{sec:conclusion}

This paper compares the Twitter feed sentiment of foreign and domestic stocks. Our hypothesis is that the feed for foreign stocks is less informative than the feed for domestic stocks. The main results support the hypothesis. Twitter sentiment provides a more reliable signal for stocks of domestic firms than for stocks of foreign firms. The predictive relation between measured sentiment noise and next-quarter losses is stronger for domestic stocks than for foreign stocks. This relation maintains its sign in four out of five quarters for domestic stocks and about one out of two quarters for foreign stocks. This difference could reflect a gap in informed-trading intensity between domestic and foreign stocks rather than a gap in their feeds, since heavier informed trading makes the price respond less to the feed's noise. However, the difference survives when domestic and foreign stocks are compared within informed-trading-intensity tercile 1, which has the lowest informed trading.

The composition of contributors provides an explanation for this difference. A sentiment index constructed from the Twitter feed of a U.S.-listed stock consists of two components. The first one consists of the beliefs of investors who hold information about the firm. The second one consists of fads and automated content that carry no information about the firm. The share of each component in the sentiment index depends on the composition of the contributor pool by type. The estimates show that the contributor pool for the feed for foreign stocks has a higher share of repeat authors than that of the feed for domestic stocks, in each of the three informed-trading-intensity terciles. The estimates also show that the foreign feed carries a larger fad component. Thus, the feed for foreign stocks carries a larger nonfundamental component and gives a less precise public signal. Hence, the composition difference between the two feeds explains why the sentiment index is a less reliable signal for foreign stocks than for domestic stocks. Our interpretation of these differences is the producer-knowledge mismatch, which is the mismatch between the contributors who post about the firm and the people who hold information about the firm. For a foreign firm, the investors who understand its regulatory environment, management, and product markets are underrepresented among its English-language contributors. The structural model disciplines this interpretation. We do not observe the geography or the information sets of the individual contributors.

The model also predicts an asymmetry between next-quarter gains and losses. Due to the optimistic error in the public signal, there is overpricing which survives the quarter and manifests as a loss in the next quarter because of the short-sale friction. The results also show that the excess of next-quarter losses over gains rises with the model-motivated signal-noise index within each geography-tercile group. The mechanism of correction of overpricing from the model predicts the direction of this relation. However, as the estimate of the short-sale friction lies at its zero lower bound, the fitted model is not able to produce this rise. Therefore, we report the rise as a reduced-form pattern.

The results from this paper are limited to English-language Twitter and to the stocks listed on the U.S. exchanges. An open question that remains for future research is whether the mismatch between who posts about a firm and who knows about a firm also shows up on the local-language social media platforms of the home markets of the foreign firms. A measurement of the sentiment index on those platforms would allow the testing of a mechanism similar to ours but with the two geographical areas reversed. On a local-language platform in the home market of a foreign firm, investors who know the firm best would be a dominant part of the contributors who post about that firm, while the investors who know a U.S. firm best would be a small part of the contributors who post about a U.S. firm. A mechanism along similar lines to ours would imply that the feed for the firm headquartered in that market is more informative than that for the U.S.-headquartered firm. Such a mechanism would also imply that the relation between measured sentiment noise and next-quarter losses is more stable across quarters for the firm headquartered in that market than for the U.S.-headquartered firm. Our model provides the closed-form moment conditions that such studies on other social media platforms would use to estimate their parameters.

\bibliographystyle{aer}
\bibliography{Sentiment_stocks}

\clearpage
\addtocontents{toc}{\protect\setcounter{tocdepth}{2}}
\thispagestyle{empty}
\begin{center}
\vspace*{0.04\textheight}
{\LARGE\bfseries Internet Appendix\par}
\bigskip
{\Large to\par}
\bigskip
{\Large\bfseries Where Has the Foreign Signal Gone? Contributor Composition and the Reliability of Social Media Sentiment\par}
\end{center}
\vspace{1em}
{\small\setstretch{1.15}
\makeatletter\def\tmptocnumberline#1{\hb@xt@ 12em{#1\hfill}}\makeatother
\tableofcontents
}
\clearpage
\appendix
\renewcommand{\appendixname}{Internet Appendix}
\setcounter{table}{0}


\section{Additional Estimation Detail and Robustness}\label{app:estdetail}

This section reports the estimation details and the robustness results that the paper refers to. Table~\ref{tab:notation_map} shows each of the model's objects with its empirical counterpart.

\begin{footnotesize}
\begin{longtable}{@{}>{\RaggedRight}p{0.14\textwidth}>{\RaggedRight}p{0.26\textwidth}>{\RaggedRight\arraybackslash}p{0.54\textwidth}@{}}
\caption{Model objects and their empirical counterparts}\label{tab:notation_map}\\
\toprule
Symbol & Object & Empirical counterpart \\
\midrule
\endfirsthead
\multicolumn{3}{@{}l}{\textit{Table~\ref{tab:notation_map}, continued}}\\
\toprule
Symbol & Object & Empirical counterpart \\
\midrule
\endhead
\midrule
\endfoot
\bottomrule
\endlastfoot
$D$ & contributor dispersion & quarterly mean of the vendor's daily dispersion, unique accounts per tweet \\
$\pi_j$ & repeat-author share (auxiliary) & inversion \eqref{eq:pi_inversion} applied to $\overline{D}$, used in \ref{app:composition_check} \\
$\lambda_I$ & posting intensity of repeat authors & estimated in the joint estimation when identified, with a pre-specified weak-identification rule (\ref{app:headline_detail}), and fixed at literature posting rates for the inversion (\ref{app:composition_check}) \\
$\omega_{\mathbb{B},j}$ & bot tweet share & not separately identified (tested in Section~\ref{sec:reducedform}) \\
$\mathrm{VAQSV}$ & per-tweet sentiment volatility & within-quarter standard deviation of daily per-tweet sentiment $S_t/N_t$ \\
$\Phi_j$ & nonfundamental variance & intercept of \eqref{eq:vaqsv} by geography \\
$\overline{1/N_t}$ & sampling-noise regressor & quarterly mean of daily $1/N_t$ over positive-volume days \\
$T_q$ & positive-volume sentiment days & number of positive-volume sentiment days in the quarter \\
$\tau_s$ & public signal precision & $T_q / (\kappa\,\mathrm{VAQSV}^2)$ \\
$r$ & gross riskless rate & one plus the mean quarterly risk-free rate (calibrated) \\
$\bar{\rho}$ & risk-aversion coefficient & calibrated (sensitivity reported) \\
$\mathrm{spread}_j$ & adverse-selection cost & quarterly mean of the daily quoted relative spread (the \citet{corwin2012simple} spread enters the estimation as a second measure with its own scale constant, and the quarterly median quoted spread is a reported companion) \\
$\mathrm{impact}_j$ & price-impact ratio & \citet{amihud2002illiquidity} illiquidity ratio \\
$\sigma_{x,j}$ & noise trading & quarterly turnover divided by the number of return days, a per-day rate (the zero-return share as an auxiliary check) \\
$\upsilon_j,\ \Sigma_{jd}$ & cash-flow persistence, prior variance & AR(1) of operating income before depreciation over total assets (net income over total assets as a robustness alternative) with fiscal-quarter-of-year effects (year-over-year variant in robustness) \\
$\mathrm{PINF}$ inputs & informativeness regression & operating income before depreciation over total assets at horizon $q+s$ on market capitalization over total assets at $q$ (CRSP market value, with Compustat market value as a robustness alternative) \\
$\mathrm{SD}(P^{*}_{f1})$ & price dispersion in the informativeness factor & standard deviation of market capitalization over total assets residualized on the same controls (CRSP, the same price source as all return objects) \\
$T_{days}$ & return days in the quarter & number of return days in the quarter (group mean in estimation) \\
$\sigma_d$ & daily return volatility (fitted) & within-quarter standard deviation of daily excess returns (residual idiosyncratic volatility as a robustness alternative) \\
Gain, $|$Loss$|$ & signed daily sums & within-quarter sums of positive and of negative daily excess returns \\
 & model-exact versions & the same sums of daily out-of-sample three-factor residuals \\
positive-sentiment share & sign of realized fads & share of positive-sentiment days among the classified days of quarter $q$ \\
$cay$ & consumption-wealth ratio & excluded, since the published series ends in 2019Q3 (footnote to Assumption~\ref{assump:tvprior}) \\
$\mathrm{FU}_q$ & financial uncertainty & the financial-uncertainty index of \citet{ludvigson2021uncertainty} \\
geography & domestic or foreign & domestic if the Compustat headquarters location is the United States, foreign otherwise \\
\end{longtable}
\end{footnotesize}

\begin{table}[!htbp]\centering
\footnotesize
\caption{The empirical exercises and their classification}\label{tab:exercise_map}
\begin{tabularx}{\textwidth}{@{}>{\raggedright}p{0.135\textwidth}>{\raggedright}p{0.21\textwidth}>{\raggedright}p{0.175\textwidth}>{\raggedright}p{0.22\textwidth}>{\raggedright\arraybackslash}X@{}}
\toprule
Exercise & Equation source & Delivers & Used for & Reported in \\
\midrule
\multicolumn{5}{@{}l}{\textit{Structural, model equations imposed}} \\
\addlinespace
Measurement regression & Proposition~\ref{prop:vaqsv}. & $\sigma_e^2 c_w$ and $\Phi_j$. & The sampling-noise acceptance gate and the geography contrast in nonfundamental variation. & Section~\ref{sec:structural}, \ref{app:vaqsv_estimation}. \\
\addlinespace
Microstructure inversion & Proposition~\ref{prop:micro}. & $\sigma_{I,j}$, $\sigma_{x,j}$, the scale constants, and $\widehat{K}$. & The ordinal intensity ranking behind all tercile splits. & Section~\ref{sec:structural}, \ref{app:estdetail}. \\
\addlinespace
Joint moment system & The 91 model-implied moments. & $\pi_j$, $\sigma_{b,j}$, and the conversion constant $\kappa$. & Supporting evidence for \ref{hyp:mismatch} when the estimated contrasts have the hypothesized signs. & Tables~\ref{tab:theta_hat} and~\ref{tab:h1_contrasts}. \\
\addlinespace
Short-sale friction & The model's functional form. & $\delta$, at its zero lower bound. & Assigns the excess of next-quarter losses over gains to the volatility channel. & Table~\ref{tab:theta_hat}. \\
\addlinespace
\multicolumn{5}{@{}l}{\textit{Empirical validation, model-motivated}} \\
\addlinespace
Drift projection & Equation~\eqref{eq:driftproj}. The model supplies the index. & $b_3$ and its $t$ statistic. & Validates the correction prediction within groups. & Section~\ref{sec:reducedform}, Table~\ref{tab:b3_drift}. \\
\addlinespace
Reduced-form battery & Equation~\eqref{eq:rfspec}, with each row's specification in Table~\ref{tab:battery_specs}. & The headline estimates of Table~\ref{tab:battery}. & Tests the mechanism's implications without the model. & Table~\ref{tab:battery}, \ref{app:battery}. \\
\bottomrule
\end{tabularx}
\end{table}

\textit{Outcome variables and the common component.}\label{app:outcome_construction} In the data, the common component that all investors observe (Section~\ref{sec:model_env}) is measured by the three Fama-French factors (FF3). Excess returns are measured against the riskless rate. The residual from a regression of the stock's daily excess return on the three factors removes the common component. The quarterly factor realizations are common to all stocks. The quarter fixed effects of the regressions absorb them. The residual-based gains and losses of the robustness checks (Section~\ref{sec:data}) are the sums over the quarter of the positive and of the negative daily residuals, with each stock's FF3 betas fixed at their estimates from the end of the previous calendar quarter and with no intercept subtracted. This construction keeps the mean daily excess return, which carries the correction drift that the model predicts (\ref{assump:daily}). Table~\ref{tab:b3_drift} reports the drift projection on these residual-based outcomes.

\subsection{From the measurement propositions to the estimation}\label{app:vaqsv_estimation}

\label{sec:model_to_estimation}We estimate the model's parameters by the generalized method of moments on ninety-one closed-form moments (Section~\ref{sec:structural}). The estimator minimizes a weighted distance between the data moments and their model-implied counterparts, with a pre-specified diagonal weighting matrix. For each geography and informed-trading-intensity tercile, we estimate the informed-trading intensity ($K_j$), the noise-trading standard deviation ($\sigma_{x,j}$), the cash-flow persistence ($\upsilon_j$), the innovation variance ($\Sigma_{jd}$), the repeat-author share ($\pi_j$), and the group-level risk premium ($\mu_{0,j}$), which is given by the mean daily excess return net of the correction drift (\ref{assump:daily}). The mean supply of shares ($\bar{x}_j$) enters through this risk premium (Section~\ref{sec:model_eq}). For each geography, we estimate the fad scale ($\sigma_{b,j}$), with the bot-content variance ($\sigma_{\beta,j}^2$) set to zero (Section~\ref{sec:model_sent}). For all stocks together, we estimate the conversion constant ($\kappa$), the surviving overpricing fraction ($\delta$), the scale that converts value units to return units ($s_P$), the sampling-noise coefficient ($c_w\sigma_e^2$) and the scale constants that map the model's spread and price-impact ratio to the quoted spread ($a_s$), the high-low spread ($a_{cs}$), and the Amihud ratio ($a_\lambda$), with the turnover scale normalized to one (Proposition~\ref{prop:micro}). The posting intensity ($\lambda_I$) is set at 1.5 (Section~\ref{sec:model_sent}). The gross riskless rate ($r$) and the risk-aversion coefficient ($\bar{\rho}$) are calibrated (Section~\ref{sec:data}). Within each geography and intensity tercile, the data moments are the means of the four liquidity measures, namely the quoted spread (the daily ask minus bid over the quote midpoint, averaged within the quarter), the high-low spread (the \citet{corwin2012simple} estimator of the same spread from daily high and low prices), the Amihud ratio (daily absolute return over dollar volume, averaged within the quarter) and turnover (daily share volume over shares outstanding, entered per day as the quarterly sum divided by the number of return days). The moments also include the persistence and the innovation variance of the firm's cash flows, price informativeness at horizons of one and four quarters (Proposition~\ref{prop:pinf}), the mean of dispersion, and the mean of squared measured noise. The last within-group moments are the variance of the quarterly excess return, the means of the quarter's gains and of its losses, and the slopes of gains and of losses on the lagged signal-quality index. The remaining moment is the slope of squared measured noise on the average inverse tweet count in the pooled sample (\ref{app:estdetail}). We conduct two exercises which precede the GMM and supply tercile assignment and starting values for it. The first is the measurement regression, which fits the measurement equation (equation~\ref{eq:vaqsv}) by ordinary least squares across firm-quarters. The second is the microstructure inversion of Proposition~\ref{prop:micro}. Both use estimates of $\upsilon_j$ and $\Sigma_{jd}$ by group obtained from the firm's cash flows before the GMM. The measurement regression gives the nonfundamental variance ($\Phi_j$) by geography and the sampling-noise coefficient ($c_w\sigma_e^2$), which serve as starting values for the GMM. The inversion fits the group means of the quoted spread, the high-low spread, the Amihud ratio, and turnover to their model counterparts (equation~\ref{eq:micro}). The quoted spread and the Amihud ratio alone would give two equations in the two unknowns ($\sigma_{I,j}$ and $\sigma_{x,j}$) for each group (Section~\ref{sec:model_micro}). We have two equations in two unknowns. We can solve them for exact values of $\sigma_{I,j}$ and $\sigma_{x,j}$, whatever values the two data moments take. So any measurement error in either moment passes straight into the estimates. We also cannot test the relations that the model imposes among the liquidity measures. Hence, we over-identify the inversion on purpose. We fit the group means of the quoted spread, the high-low spread, the Amihud ratio, and turnover together, four equations in the same two unknowns. We pick $\sigma_{I,j}$ and $\sigma_{x,j}$ to minimize the weighted sum of squared differences between the four data moments and their model counterparts. We weight each difference by the inverse of that moment's bootstrap variance. Thus, a moment measured with more sampling noise gets a smaller weight and moves the estimates less. In the model, the bid-ask spread is proportional to $\sigma_{I,j}$ (equation~\ref{eq:micro}). The quoted spread and the high-low spread both measure that spread. Hence, each is proportional to $\sigma_{I,j}$, with its own scale constant, $a_s$ for the quoted spread and $a_{cs}$ for the high-low spread. So the estimate of $\sigma_{I,j}$ draws on both spreads. A systematic error in the quoted spread then moves it less than if the quoted spread were the only measure of $\sigma_{I,j}$. The over-identification is also what makes the model testable. The exactly identified system is not. At the estimated $\sigma_{I,j}$ and $\sigma_{x,j}$, we compare each data moment with its model counterpart and scale the distance by the moment's bootstrap standard error. This tests whether the relations of equation~\ref{eq:micro} hold in the data. We report these over-identifying checks (\ref{app:liq_kyle}).

We fit the linear regression of equation~\eqref{eq:vaqsv} across firm-quarters. We read the nonfundamental variance $\Phi_j$ off its intercept and the sampling-noise coefficient $c_w\sigma_e^2$ off its slope. We measure the quarterly sentiment volatility on the per-tweet average $S_t/N_t$, and give our reason for this in Section~\ref{sec:model_sent}. The conversion constant $\kappa$ absorbs three departures, and the data identify $\kappa$ as a whole.

We map Proposition~\ref{prop:vaqsv} to the data with two conventions (Section~\ref{sec:model_sent}). The first is that we set the bot-content variance ($\sigma_{\beta,j}^2$) to zero. Under it, the nonfundamental variance ($\Phi_j$) comes from the fad alone and equals $(1-\omega_j)^2\,\sigma_{b,j}^2$, where $\omega_j$ is the combined tweet share of the two repeat types, the specialists and the automated accounts (\ref{app:proof_dispersion}). This attribution of the nonfundamental variance to the fad is an assumption. Dispersion identifies the repeat-author share ($\pi_j$), the combined account share of the two repeat types. It does not separate specialists from automated accounts (Proposition~\ref{prop:dispersion}). An automated share that differs between domestic and foreign stocks is the case that would move the domestic-foreign comparison of signal quality (\ref{app:kappa_cases}). The second is that we treat the conversion constant $\kappa$ as one parameter common to all firm-quarters. The moment system identifies its level (\ref{app:kappa_detail}). In the joint estimation, we hold the fad scale ($\sigma_{b,j}$) equal across each geography's three informed-trading-intensity terciles (Section~\ref{sec:structural}). With the fad scale held equal, the differences in the nonfundamental variance across a geography's three terciles are what identifies the posting intensity ($\lambda_I$). The moments identify it weakly, so we set it at 1.5 (Section~\ref{sec:structural}).

\subsection{Conventions that discipline the moments}\label{app:conventions}

The gross quarterly riskless rate is 1.0056 (one plus the mean quarterly riskless rate over the primary sample). The risk-aversion coefficient is 0.02 from \citet{farboodi2022data}. Standard errors on the ordinary least squares estimates are clustered two ways, by firm and by quarter, with the conservative rule that sets degrees of freedom to the smaller number of clusters minus one \citep{cameron2011robust,petersen2009estimating}. Estimates that depend on generated regressors use a firm-block bootstrap that repeats the generating steps inside each replication.\footnote{Section~\ref{sec:structural} describes the bootstrap and reports the number of replications.}

\textit{Conventions for the liquidity moments.}\label{app:liq_measures} \ref{sec:model_to_estimation} defines the four liquidity measures. For the moments on the quoted spread and on the high-low spread, we use the firm-quarters whose quarter-end price is at least five dollars. The reason is that the one-cent minimum tick puts a floor under the relative spread of a low-priced stock. The zero-return share, the auxiliary over-identifying check of Proposition~\ref{prop:micro}, is the fraction of the quarter's trading days with a return of exactly zero, the trading-cost indicator of \citet{lesmond1999new}. The Amihud ratio in our data has a scale that differs from the values published elsewhere. Therefore, we treat the three scale constants ($a_s$, $a_{cs}$ and $a_\lambda$) as parameters and do not compare the level of any liquidity measure with another study's. Equation~\eqref{eq:micro} does not fix the scale of the noise-trading standard deviations on its own. Multiplying every $\sigma_{x,j}$ by one constant, multiplying $a_\lambda$ by the same constant and dividing $a_T$ by it leaves the model's price-impact ratio and turnover unchanged. We fix that scale by normalizing $a_T$ to one.

\textit{Model dependence of the liquidity identification.}\label{app:liq_kyle} \ref{app:dealer_detail} states that the spread does not depend on the amount of noise trading. That property belongs to the Kyle batch market. In a sequential market of the \citet{glosten1985bid} type, the dealer posts quotes for a fixed trade size. There the spread depends on the ratio of informed trades to noise trades as well as on the informed investor's advantage ($\sigma_{I,j}$). The spread would then depend on the amount of noise trading ($\sigma_{x,j}$) as well. It would no longer identify the informed investor's advantage on its own. The over-identifying checks that \ref{app:dealer_detail} refers to are the high-low spread, turnover, the zero-return share and the number of registered market makers in the stock, which is available for the NASDAQ subsample. The high-low spread and turnover are among the moments of the joint estimation (\ref{sec:model_to_estimation}). The zero-return share and the market-maker count are not.

The model moment for the variance of the quarterly excess return is the average over the firm-quarters of the group of the closed-form return variance, with each firm-quarter at its own signal precision ($\tau_s$) from equation~\eqref{eq:tau_s}. The model moments for the means of the quarter's gains and losses are the expected gain and the expected loss of Section~\ref{sec:model_gl}, computed once per group at the group averages over those firm-quarters of the expected overpricing and of the return variance. The slopes of gains and of losses on the lagged signal-quality index are estimated with quarter fixed effects, coded to sum to zero, in the data moments and in the model moments alike. The quarter fixed effects remove what moves the index for every stock in a quarter at once. The pooled slope of squared measured noise on the average inverse tweet count is estimated with the same quarter fixed effects and a foreign indicator. Without the quarter fixed effects that pooled slope has the opposite sign in the data. That slope is reported for comparison and is not among the 91 moments. The group-level precision of the public signal is the median of the signal-quality index across the group's firm-quarters, converted through equation~\eqref{eq:tau_s}, because the index is heavy tailed. We assign stocks to informed-trading-intensity terciles in two steps. In the first step, we rank stocks within each geography and quarter on the lagged quoted spread alone. In the second, we rank them on the ranking statistic of Section~\ref{sec:model_micro}, the lagged quoted spread multiplied by the square root of the prior precision plus the lagged signal precision. The slopes are estimated with the index lagged one quarter, an approximation to the model's timing, in which the quarter's own signal precision moves the quarter's price (Section~\ref{sec:model_eq}). Within a firm, the logarithm of the index has a correlation of 0.32 with its value one quarter earlier in this sample. The group intercepts of the slope regressions absorb the group-level risk premium ($\mu_{0,j}$), which the model derives (\ref{assump:daily}). For that reason the mean of the quarterly excess return is not a moment of the system. This is an assumption.

\subsection{The optimizer, the basin table, and the saturation criterion}\label{app:basin_detail}

We start the optimizer from two families of starting values. Table~\ref{tab:basin} shows, for each family, the number of starting values, the range of the objective at which they end, how many of them end at a finite objective and whether the lowest point of the family satisfies the saturation criterion. The first family is one starting value built from the two exercises that precede the GMM (\ref{sec:model_to_estimation}), with the conversion constant started at one, together with ten Latin-hypercube draws around it. From these eleven starting values, the optimizer ends at objectives between 76,842.0 and 83,033.8. When we continue the optimization from the lowest of them, the objective falls to 76,752.4. At that point six of the model's gain and loss moments underflow to exactly zero, so that the objective is flat in the parameters behind the underflow and the optimizer stops there. The saturation criterion excludes any candidate estimate at which one of the model's gain, loss or variance moments underflows to exactly zero, unless no other candidate ends at a finite objective. Thus, it excludes the point reached from the first family. The second family consists of fourteen starting values set from the data moments at seven values of the conversion constant from 1.05 to 400 and two scales of the informed-trading intensity. From these the optimizer ends at objectives between 4,267.9 and 4,400.2. Continuing the optimization from the lowest of them gives the interior optimum, with an objective of 4,250.3 and no moment at zero. At the interior optimum the posting intensity is free and every repeat-author share is at its zero lower bound (Section~\ref{sec:structural}). Hence, six of the twenty acceptance checks are not met there, one for each share. We accept an estimate only if the conversion constant is at least one, the posting intensity exceeds one and, in each of the six groups, the cash-flow persistence lies below the gross riskless rate, the microstructure inversion has a positive denominator and the repeat-author share lies strictly between zero and one. These are the twenty acceptance checks. We set the posting intensity at 1.5 by the rule stated in \ref{app:headline_detail} and re-estimate the other forty-five parameters starting from the interior optimum. The result is the headline estimate. It satisfies the saturation criterion and the twenty acceptance checks. At the headline estimate the largest component of the gradient with respect to the forty-five free parameters is 0.016. When we restart the optimizer from the headline estimate, the objective changes by less than $10^{-10}$.

\begin{table}[!htbp]\centering\footnotesize\setlength{\tabcolsep}{4pt}
\caption{Starting values for the GMM estimation and the headline estimate}\label{tab:basin}
\begin{tabular}{lccccc}
\hline\hline
Start family & Starts & Objective range & Finite & Criterion & Headline \\
\hline
Pre-specified value and Latin hypercube & 11 & 76842.0--83033.8 & 11 & -- & -- \\
\quad lowest objective in the saturated region & -- & 76752.4 & -- & not satisfied & no \\
Set from the data, spanning the noise scale & 14 & 4267.9--4400.2 & 14 & -- & -- \\
\quad interior optimum, posting intensity free & -- & 4250.3 & -- & satisfied & no \\
\quad interior optimum, posting intensity set at 1.5 & -- & 4337.7 & -- & satisfied & yes \\
\hline\hline
\end{tabular}
\begin{minipage}{0.95\textwidth}\footnotesize\medskip
\textit{Notes:} This table shows the two families of starting values for the GMM estimation. For each family it shows the number of starting values, the range of the objective at which they end, the number that end at a finite objective, whether the lowest point of the family satisfies the saturation criterion and whether that point is the headline estimate. The indented rows show the point we reach when we continue the optimization from the lowest starting value of each family, which is why their objectives lie below the range of the family. The point reached from the first family has six of the model's gain and loss moments at exactly zero and its six variance moments above zero but tiny. The saturation criterion excludes it. The point reached from the second family (the interior optimum) satisfies the criterion. Its posting intensity is free and every repeat-author share is at its zero lower bound. Hence, six of the twenty acceptance checks are not met (Section~\ref{sec:structural}). In the last row we re-estimate the other forty-five parameters with the posting intensity set at 1.5, starting from the interior optimum (\ref{app:headline_detail}). Its objective is higher than that of the interior optimum, because holding one parameter fixed does not lower the minimum. Section~\ref{sec:structural} states the rule by which we pick the headline estimate among the candidates.
\end{minipage}
\end{table}

The weighting matrix is a further convention. Since the estimate depends on the weighting matrix under misspecification, we fix the weighting in advance rather than choosing it from the fit \citep{hall2003large}. Section~\ref{sec:structural} states the reason. For the headline estimate we use the diagonal weighting matrix described there. We report the Ledoit-Wolf-shrunk full-matrix weighting as a variant (Table~\ref{tab:variants}).

\subsection{Headline estimates, identification, and the posting-intensity profile}\label{app:headline_detail}

Table~\ref{tab:theta_hat} shows the headline estimates and Section~\ref{sec:structural} states the reading. Three diagnostics bear on the identification. The first diagnostic is a comparison of two estimates of the sampling-noise coefficient. The GMM gives $1.64\times 10^{-4}$ and the measurement regression of \ref{sec:model_to_estimation} gives $1.95\times 10^{-4}$. The second diagnostic concerns the posting intensity, which the moments identify weakly (Section~\ref{sec:structural}). We fixed a rule for weak identification in advance. Whenever a freely estimated repeat-author share reaches a bound, or the Jacobian of the model moments has a near-zero singular value in the direction of the posting intensity, we set the posting intensity at 1.5 and estimate the other parameters at that value. When we estimate the posting intensity freely, the estimate is 1,102.98 and every repeat-author share is below $10^{-4}$, at its zero lower bound. Thus, the first condition holds. The third diagnostic is the lowest posting intensity at which the repeat-author share lies below one in each of the six groups. By Proposition~\ref{prop:dispersion}, the repeat-author share in a group is a function of the mean of dispersion in that group and the posting intensity. For the share to lie below one, the posting intensity must exceed the inverse of the mean of dispersion. The largest of the six inverses is 1.0974, for foreign stocks in the first tercile. The value 1.5 lies above it. Table~\ref{tab:lambda_profile} shows the joint Wald test of Table~\ref{tab:h1_contrasts} and the signs of its four domestic-foreign contrasts at each of the eight values of the posting intensity from 1.2 to 1.85.

Table~\ref{tab:theta_hat} reports twelve of the forty-five estimated parameters. Table~\ref{tab:theta_hat_rest} reports the remaining thirty-three.

\begin{table}[!htbp]\centering
\begin{small}
\caption{Remaining structural parameter estimates}\label{tab:theta_hat_rest}
\begin{tabular}{lccc}
\hline\hline
Parameter & Estimate & Sandwich SE & 95\% bootstrap interval \\
\hline
$K_j$, Domestic T1 & 0.0006961 & 0.0328 & [0.0006079, 0.0007872] \\
$K_j$, Domestic T2 & 0.0008342 & 0.0392 & [0.0007689, 0.0009368] \\
$K_j$, Domestic T3 & 0.0005565 & 0.0262 & [0.0004759, 0.0006699] \\
$K_j$, Foreign T1 & 0.001008 & 0.0474 & [0.0008557, 0.001271] \\
$K_j$, Foreign T2 & 0.0009730 & 0.0455 & [0.0007576, 0.001189] \\
$K_j$, Foreign T3 & 0.0006285 & 0.0296 & [0.0005184, 0.0008492] \\
\addlinespace
$\sigma_{x,j}$, Domestic T1 & 0.01042 & 0.000153 & [0.01019, 0.01070] \\
$\sigma_{x,j}$, Domestic T2 & 0.01041 & 0.000220 & [0.009982, 0.01084] \\
$\sigma_{x,j}$, Domestic T3 & 0.02170 & 0.00115 & [0.02014, 0.02323] \\
$\sigma_{x,j}$, Foreign T1 & 0.01103 & 0.000706 & [0.009609, 0.01268] \\
$\sigma_{x,j}$, Foreign T2 & 0.002912 & 0.000229 & [0.002451, 0.003574] \\
$\sigma_{x,j}$, Foreign T3 & 0.03750 & 0.00584 & [0.02946, 0.04492] \\
\addlinespace
$\upsilon_j$, Domestic T1 & 0.8302 & 0.00775 & [0.8131, 0.8403] \\
$\upsilon_j$, Domestic T2 & 0.8446 & 0.00671 & [0.8310, 0.8512] \\
$\upsilon_j$, Domestic T3 & 0.8828 & 0.00535 & [0.8694, 0.8909] \\
$\upsilon_j$, Foreign T1 & 0.8642 & 0.0118 & [0.8412, 0.8804] \\
$\upsilon_j$, Foreign T2 & 0.8729 & 0.0142 & [0.8469, 0.8837] \\
$\upsilon_j$, Foreign T3 & 0.8951 & 0.00840 & [0.8794, 0.9102] \\
\addlinespace
$\Sigma_{jd}$, Domestic T1 & 0.0003889 & 2.24e-05 & [0.0003613, 0.0004254] \\
$\Sigma_{jd}$, Domestic T2 & 0.0006308 & 4.07e-05 & [0.0005556, 0.0007152] \\
$\Sigma_{jd}$, Domestic T3 & 0.0007201 & 6.47e-05 & [0.0005397, 0.0009509] \\
$\Sigma_{jd}$, Foreign T1 & 0.0002448 & 3.98e-05 & [0.0001885, 0.0003100] \\
$\Sigma_{jd}$, Foreign T2 & 0.0004454 & 9.04e-05 & [0.0003367, 0.0006438] \\
$\Sigma_{jd}$, Foreign T3 & 0.0006021 & 0.000103 & [0.0003825, 0.0008149] \\
\addlinespace
$\mu_{0,j}$, Domestic T1 & 0.0004797 & 7.88e-05 & [0.0004557, 0.0005027] \\
$\mu_{0,j}$, Domestic T2 & 0.0003342 & 0.000117 & [0.0003039, 0.0003747] \\
$\mu_{0,j}$, Domestic T3 & 0.0001347 & 0.000169 & [0.00008719, 0.0001827] \\
$\mu_{0,j}$, Foreign T1 & 0.0003640 & 0.000123 & [0.0003127, 0.0004264] \\
$\mu_{0,j}$, Foreign T2 & 0.0002689 & 0.000150 & [0.0001693, 0.0003413] \\
$\mu_{0,j}$, Foreign T3 & $-0.00004670$ & 0.000219 & [$-0.0001595$, 0.00008915] \\
\addlinespace
$a_s$ & 10.35 & 243 & [9.226, 12.00] \\
$a_{cs}$ & 108.8 & 2559 & [97.66, 126.5] \\
$a_\lambda$ & 0.1670 & 3.93 & [0.1475, 0.1957] \\
\hline\hline
\end{tabular}
\begin{minipage}{0.95\textwidth}\footnotesize\medskip
\textit{Notes:} This table shows the thirty-three parameters estimated jointly with those in Table~\ref{tab:theta_hat}, at the headline estimate with the posting intensity set at 1.5. For each of the six geography-tercile groups, it gives the informed-trading intensity ($K_j$), the noise-trading standard deviation ($\sigma_{x,j}$), the cash-flow persistence ($\upsilon_j$), the innovation variance ($\Sigma_{jd}$) and the risk premium ($\mu_{0,j}$). The last three rows give the scale constants for the quoted spread ($a_s$), the high-low spread ($a_{cs}$) and the Amihud ratio ($a_\lambda$). We compute the sandwich standard errors and the 95\% intervals over the 399 firm-block bootstrap replications as in Table~\ref{tab:theta_hat}. The sandwich standard errors of the six informed-trading intensities and of the three scale constants exceed their estimates by more than twenty times. The Jacobian of the model moments has a near-zero singular value, $1.1\times 10^{-7}$ of the largest, in a direction that raises the six intensities together and lowers the three scale constants together. Hence, the moments identify these nine parameters weakly. Each bootstrap replication starts from the point estimate, and the intervals for these nine parameters are narrow.
\end{minipage}
\end{small}
\end{table}

\begin{table}[!htbp]\centering
\caption{Joint Wald test of the equality of the repeat-author shares and the fad scale across geography at eight values of the posting intensity}\label{tab:lambda_profile}
\begin{footnotesize}
\setlength{\tabcolsep}{3pt}
\begin{tabular}{lccccc}
\hline\hline
Posting intensity & Objective & Firm Wald ($p$) & Quarter Wald ($p$) & Two-way Wald ($p$) & Contrast signs \\
\hline
1.20 & 4442.6 & 62.34 ($9.4\times10^{-13}$) & 86.06 ($9.1\times10^{-18}$) & 37.31 ($1.6\times10^{-7}$) & all negative \\
1.30 & 4396.0 & 56.09 ($1.9\times10^{-11}$) & 80.01 ($1.7\times10^{-16}$) & 32.78 ($1.3\times10^{-6}$) & all negative \\
1.40 & 4361.0 & 51.37 ($1.9\times10^{-10}$) & 74.44 ($2.6\times10^{-15}$) & 29.31 ($6.7\times10^{-6}$) & all negative \\
1.50 & 4337.7 & 48.56 ($7.2\times10^{-10}$) & 70.38 ($1.9\times10^{-14}$) & 27.05 ($1.9\times10^{-5}$) & all negative \\
1.60 & 4321.9 & 46.86 ($1.6\times10^{-9}$) & 67.43 ($7.9\times10^{-14}$) & 25.54 ($3.9\times10^{-5}$) & all negative \\
1.70 & 4310.6 & 45.78 ($2.7\times10^{-9}$) & 65.23 ($2.3\times10^{-13}$) & 24.51 ($6.3\times10^{-5}$) & all negative \\
1.80 & 4302.2 & 45.06 ($3.9\times10^{-9}$) & 63.55 ($5.2\times10^{-13}$) & 23.76 ($8.9\times10^{-5}$) & all negative \\
1.85 & 4298.8 & 44.79 ($4.4\times10^{-9}$) & 62.85 ($7.3\times10^{-13}$) & 23.47 ($1.0\times10^{-4}$) & all negative \\
\hline\hline
\end{tabular}
\end{footnotesize}
\begin{minipage}{0.95\textwidth}\footnotesize\medskip
\textit{Notes:} This table shows the joint Wald test of Table~\ref{tab:h1_contrasts} at eight values of the posting intensity from 1.2 to 1.85. At each value we re-estimate the other forty-five parameters and compute the four domestic-foreign contrasts (in the three repeat-author shares and the fad scale). We test the null that the four contrasts are zero with standard errors clustered by firm, clustered by quarter and double clustered by firm and quarter, as in Table~\ref{tab:h1_contrasts}. At each value we treat the posting intensity as a known constant in the covariance, as we do at 1.5 (Section~\ref{sec:structural}). The second column is the GMM objective at each value. The last column shows that all four contrasts are negative at each value, so that foreign stocks have the higher repeat-author share in each tercile and the higher fad scale. We reject the null at each value under each clustering. The largest $p$ value is $1.0\times 10^{-4}$.
\end{minipage}
\end{table}

\subsection{The conversion constant, profile and shape check}\label{app:kappa_detail}

Section~\ref{sec:structural} reports the estimate of the conversion constant ($\kappa$) and the three departures of the data that $\kappa$ absorbs. Table~\ref{tab:kappa_profile} shows the profile of the GMM objective function in $\kappa$ (the objective at twenty-five values of $\kappa$, with the other parameters re-estimated at each value). We report this profile and the sandwich standard error of $\kappa$ rather than a confidence set obtained by inverting the GMM objective function, because the Hansen $J$ test rejects the null hypothesis that the 46 overidentifying restrictions hold (\ref{app:fit_detail}) and such a set is not valid under misspecification. Across the 399 firm-block bootstrap replications, the 2.5th and 97.5th percentiles of the estimate of $\kappa$ are 294 and 490. The standard deviation of the estimate across the replications is 47.5.

\begin{table}[!htbp]\centering
\caption{GMM objective function at twenty-five values of the conversion constant}\label{tab:kappa_profile}
\begin{tabular}{rrr@{\hspace{2.5em}}rrr}
\hline\hline
$\kappa$ & Objective & \shortstack{Ratio to\\minimum} & $\kappa$ & Objective & \shortstack{Ratio to\\minimum} \\
\hline
1.1 & 30,401.5 & 7.007 & 43.1 & 4,789.0 & 1.104 \\
1.4 & 22,904.5 & 5.279 & 57.4 & 4,656.6 & 1.073 \\
1.9 & 8,992.3 & 2.073 & 76.4 & 4,598.9 & 1.060 \\
2.5 & 8,959.8 & 2.065 & 101.6 & 4,535.4 & 1.045 \\
3.3 & 8,953.5 & 2.064 & 135.3 & 4,471.3 & 1.031 \\
4.4 & 7,782.7 & 1.794 & 180.0 & 4,413.9 & 1.017 \\
5.8 & 5,034.2 & 1.160 & 239.6 & 4,369.7 & 1.007 \\
7.8 & 4,979.3 & 1.148 & 318.8 & 4,343.7 & 1.001 \\
10.3 & 4,938.3 & 1.138 & 424.3 & 4,338.5 & 1.000 \\
13.7 & 4,915.3 & 1.133 & 564.6 & 4,355.3 & 1.004 \\
18.3 & 4,895.9 & 1.128 & 751.4 & 4,394.9 & 1.013 \\
24.3 & 4,870.9 & 1.123 & 1,000.0 & 4,457.0 & 1.027 \\
32.4 & 4,838.1 & 1.115 & & & \\
\hline\hline
\end{tabular}
\begin{minipage}{0.95\textwidth}\footnotesize\medskip
\textit{Notes:} This table shows the GMM objective function at twenty-five values of the conversion constant ($\kappa$), equally spaced in $\log\kappa$ from 1.05 to 1,000. At each value, we fix $\kappa$ at that value, set the posting intensity at 1.5 and re-estimate the other parameters, taking the estimates at the neighboring value as starting values. We do this twice across the twenty-five values, once starting from the interior optimum of Table~\ref{tab:basin} and once from the point reached from the first family of starting values in that table. The run from the interior optimum has the lower objective at each of the twenty-five values and is the one shown. The ratio is the objective divided by its minimum over the twenty-five values, 4,338.5 at $\kappa$ of 424.3. At the headline estimate ($\kappa$ of 394.2), the objective is 4,337.7, which is 0.02 percent below that minimum.
\end{minipage}
\end{table}

We pre-specified a check of the shape of the loss and variance responses to the signal-quality index. The check is an attempt to bound the conversion constant from the data alone. We look for an interior peak in the excess of losses over gains and for a kink in the response of the return variance (a level of the index around which the slope of the logarithm of the return variance on the logarithm of the index falls from zero toward minus one). Neither is found between the 5th and the 95th percentile of the index in any of the six groups. The slope of losses on the index is negative in the low and middle terciles of the index in each of the six groups. In the top tercile of the index, the slope is smaller in magnitude than in the middle tercile of the index and its sign differs across the six groups. This pattern is consistent with the estimates, in which the surviving fraction of overpricing ($\delta$) is at its zero lower bound (Table~\ref{tab:theta_hat}), so that the response of next-quarter losses to the index runs through the volatility channel alone. The bound from the check ($3.0\times 10^{4}$ on the product of the conversion constant and the precision of outside knowledge) rests on the premise that the correction drift dominates. The estimates do not satisfy that premise. Hence, the bound does not apply and the check is uninformative about $\kappa$. Thus, the evidence on the level of $\kappa$ comes from the moment system. At the headline estimate, the precision of the public signal is 0.14 to 0.24 of the precision of outside knowledge in five of the six groups. For foreign stocks in the first tercile, the ratio is 0.082. The moments for the means of that group's gains and losses carry the least information about $\kappa$ of the six groups. These ratios are consistent with Table~\ref{tab:kappa_profile}, in which the objective lies within 1 percent of its minimum from $\kappa$ of 240 to 565. This does not affect the joint Wald test of Table~\ref{tab:h1_contrasts}, because that group's repeat-author share is identified from contributor dispersion (Proposition~\ref{prop:dispersion}), which does not depend on $\kappa$.

\subsection{Serial dependence of the daily per-tweet series}\label{app:serial_dependence}

One of the three components which is absorbed by the conversion constant ($\kappa$) is the serial dependence of the daily per-tweet sentiment ($S_t/N_t$) (equation~\eqref{eq:tau_s}, Sections~\ref{sec:model_sent} and \ref{sec:structural}). Positive serial dependence raises the dependence multiplier ($\kappa_{0,fq}$) above one. Because of that, $\mathrm{VAQSV}_{fq}^2$ understates the average daily variance $\bar{v}_{fq}$ (equation~\eqref{eq:vaqsv_bias}, \ref{app:proof_vaqsv}). In order to deal with this, we order the days with at least one scored tweet by date and subtract the within-quarter mean (for each firm-quarter), which gives the daily deviations. Then we summarize the dependence across days in these deviations three ways. First, we take the lag-one autocorrelation over the full sequence of these scored days (the first-order piece of the covariance sum in $\kappa_{0,fq}$). Second, since the days with no scored tweet can lie between two scored days, we take the same autocorrelation over days which are calendar-adjacent alone (pairs of days that are consecutive on the calendar). Third, we divide the long-run variance of the deviations by their sample variance. This ratio is $\kappa_{0,fq}$ itself, the multiplier built from the dependence at every lag rather than the first lag alone. Then we estimate the long-run variance with the Bartlett kernel at the Newey-West automatic bandwidth. This ratio equals one when the days are independent and exceeds one when there is positive serial dependence. Table~\ref{tab:serial_dependence} shows the three statistics for 98,290 firm-quarters.

\begin{table}[!htbp]\centering
\footnotesize\setlength{\tabcolsep}{3pt}
\caption{Serial dependence of the daily per-tweet sentiment series by geography}\label{tab:serial_dependence}
\begin{tabular}{lcccccc}
\hline\hline
 & \multicolumn{2}{c}{Domestic} & \multicolumn{2}{c}{Foreign} & \multicolumn{2}{c}{Domestic-foreign} \\
Statistic & Mean & Median & Mean & Median & Gap & (SE) \\
\hline
Lag-one autocorrelation, all days & 0.049 & 0.043 & 0.044 & 0.040 & 0.005 & (0.002) \\
Lag-one autocorrelation, calendar-adjacent days & 0.069 & 0.056 & 0.063 & 0.052 & 0.006 & (0.002) \\
Long-run variance ratio & 1.091 & 1.061 & 1.082 & 1.047 & 0.010 & (0.003) \\
\hline
Firm-quarters & \multicolumn{2}{c}{82,701} & \multicolumn{2}{c}{15,589} & & \\
\hline\hline
\end{tabular}
\begin{minipage}{0.95\textwidth}\footnotesize\medskip
\textit{Notes:} This table shows three statistics of serial dependence in the daily per-tweet sentiment series, for domestic and for foreign stocks. For each firm-quarter, we order the series over the days with at least one scored tweet and subtract the within-quarter mean of the series. The first row is the lag-one autocorrelation over the full sequence of those days. The second row is the lag-one autocorrelation over pairs of calendar-adjacent days (days that are consecutive on the calendar) alone. For the third row, we divide the long-run variance, estimated with the Bartlett kernel at the Newey-West automatic bandwidth $\lfloor 4(T_q/100)^{2/9}\rfloor$, by the sample variance. The sample is the primary sample (2018Q1 onward) with at least twenty sentiment days per firm-quarter. The gap is the domestic-foreign difference in the statistic. The standard error of the gap is double clustered by firm and quarter, from a regression of the statistic on a foreign indicator. The second statistic is missing for 287 domestic and 56 foreign firm-quarters with fewer than five pairs of calendar-adjacent days.
\end{minipage}
\end{table}

The median lag-one autocorrelation over the days with at least one scored tweet is 0.043 for domestic stocks and 0.040 for foreign stocks (first row, Table~\ref{tab:serial_dependence}). Thus, the serial dependence across days is small. The domestic-foreign gaps are between 0.005 and 0.010 ($t$ between 2.4 and 2.9). A difference in serial dependence of this size moves the ratio of the dependence multipliers of the two geographies by about one percent, against domestic-foreign gaps of 7 to 17 percent in the median signal-quality index across the informed-trading-intensity terciles (Table~\ref{tab:r6_dataterm}). The sample autocorrelation is biased downward by about $1/T_q$, which is about 0.02 at the median $T_q$ of 54 days. The lag-one autocorrelation corrected for this bias is about 0.06 in both geographies, which is still small. The same dependence enters the conditional mean of $\mathrm{VAQSV}_{fq}^2$ through the factor $(T_q-\kappa_{0,fq})/(T_q-1)$ in equation~\eqref{eq:vaqsv_bias}, which is 0.998 at the median $T_q$ for a dependence multiplier of 1.1. Thus, the estimates of the nonfundamental variance ($\Phi_j$) move by a fraction of one percent. The dependence multiplier is near 1.1 (third row, Table~\ref{tab:serial_dependence}), which is a small part of the estimated conversion constant of 394 (Table~\ref{tab:theta_hat}). Hence, with $\kappa$ held common across the two geographies (Section~\ref{sec:structural}), the serial dependence overstates the domestic-foreign ratio of $\tau_s$ by about one percent.

\subsection{The drift projection}\label{app:drift_detail}

Section~\ref{sec:rf_correction} states how we read the projection of the excess of next-quarter losses over gains on the model-motivated signal-noise index (equation~\eqref{eq:driftproj}). Table~\ref{tab:b3_drift} shows the slope $b_3$ under both the announcement and the calendar alignment, with the gains and losses built from the daily excess returns and with the residual-based gains and losses of \ref{app:outcome_construction}. The last row shows the slope without the group effects. Since the short-sale friction parameter lies at its zero lower bound (Table~\ref{tab:theta_hat}), the fitted model does not produce this slope (Section~\ref{sec:rf_correction}). The gain and loss block holds twelve moments, whose weighted squared misfit makes up 0.4 percent of the objective (Table~\ref{tab:perblock_j}).

\begin{table}[!htbp]\centering
\caption{Slope of the loss-gain excess on the signal-noise index under the announcement and the calendar alignment, with and without the group effects}\label{tab:b3_drift}
\begin{tabular}{lcccc}
\hline\hline
Specification & $b_3$ & SE & $t$ & $N$ \\
\hline
Announcement, excess returns & +10.13 & 1.05 & 9.63 & 87{,}301 \\
Announcement, residual-based & +10.68 & 0.894 & 11.95 & 86{,}900 \\
Calendar, excess returns & +10.98 & 1.18 & 9.33 & 89{,}939 \\
Calendar, residual-based & +11.80 & 1.01 & 11.72 & 89{,}357 \\
Announcement, excess returns, without group effects & $-2.67$ & 0.64 & $-4.18$ & 81{,}301 \\
\hline\hline
\end{tabular}
\begin{minipage}{0.95\textwidth}\footnotesize\medskip
\textit{Notes:} This table shows the slope $b_3$ of equation~\eqref{eq:driftproj} (the projection of the excess of next-quarter losses over gains on the model-motivated signal-noise index), with the index evaluated at the headline estimate (Section~\ref{sec:rf_correction}). We estimate this projection with quarter fixed effects, geography-tercile group effects, and the controls of equation~\eqref{eq:driftproj} (the factor betas, the lottery control, and log capitalization). The standard errors are clustered by firm and by quarter. The first four rows differ in two choices. We measure the gains and losses either in the quarter of the next earnings announcement (the announcement alignment) or in the next calendar quarter (the calendar alignment). We construct the gains and losses from the daily excess returns or from the residual-based gains and losses of \ref{app:outcome_construction}. The last row is a different composition run where we drop the group effects (81,301 stock-quarters). We construct the gains and losses from the daily excess returns under the announcement alignment, keep the quarter fixed effects but date them at the outcome quarter, and lag the index one quarter more than in the first four rows. The slope becomes negative. We interpret this reversal as an effect of composition (Section~\ref{sec:rf_correction}). Since this index has no natural scale, we report only the sign of $b_3$ and its $t$ statistic.
\end{minipage}
\end{table}

The model also predicts that the loss-gain excess rises with current-quarter measured noise only up to a certain level of measured noise and falls beyond it (Proposition~\ref{prop:drift}). This peak level is lower in groups with more informed trading. To check this, we regress the loss-gain excess on current-quarter measured noise and its square, separately in each informed-trading-intensity tercile. The fitted curve peaks at a lower level of measured noise in the middle tercile than in the low tercile (whether the excess is measured in the next calendar quarter or in the quarter of the next earnings announcement). This is the direction predicted by the model, but the peak levels for the low and middle terciles differ by less than one standard error (Table~\ref{tab:r3_turning}). The peak level for the top tercile is too imprecise and is left out of the ordering.

\subsection{Fit and misfit, full diagnostics}\label{app:fit_detail}

The Hansen $J$ statistic \citep{hansen1982large} is 5,503 on 46 degrees of freedom at the estimate under the Ledoit-Wolf-shrunk full-matrix weighting (the last column of Table~\ref{tab:variants}), against a 5 percent critical value of 62.8. Thus, the test rejects the null that the 46 overidentifying restrictions hold. Section~\ref{sec:structural} splits the objective across the nine blocks of moments to locate the misfit (Table~\ref{tab:perblock_j}). The concentration and fad contrasts keep their signs when we exclude the liquidity block and when we let the scale constant $a_s$ differ across the six groups (Table~\ref{tab:liq_robust}).

The liquidity block misfit also appears in the reduced form regressions of the quoted spread and the Amihud illiquidity ratio of quarter $q+1$, each on the measured noise and the dispersion of quarter $q$ (log capitalization and per-day turnover as controls, quarter fixed effects, and standard errors double clustered by firm and quarter). The spread regression uses the firm-quarters with a share price of at least five dollars (81,420 firm-quarters) and the Amihud regression uses 98,773 firm-quarters. In the quoted spread regression, the coefficient on measured noise is $-0.034$ ($t=-5.41$). In the Amihud ratio regression, it is $-6.30$ ($t=-3.88$). In the dealer market of \ref{app:dealer_detail}, the spread ($a_s\sigma_{I,j}$) and the price-impact ratio ($a_\lambda\sigma_{I,j}/\sigma_{x,j}$) both fall as the precision of the public signal rises (Proposition~\ref{prop:micro}), and higher measured noise means lower precision. Hence, the model predicts a positive coefficient on measured noise in both regressions, and the data give a negative one in both. The per-day turnover control doesn't affect this sign (coefficient $-1.67\times10^{-5}$ with $t=-0.01$ in the spread regression and $-0.062$ with $t=-1.25$ in the Amihud regression).

Section~\ref{sec:structural} states that the liquidity block misfit doesn't affect the assignment of stocks to informed-trading-intensity terciles. We check this in two ways (rows R10 and R14C). In row R10, we form the terciles using the high-low spread instead of the quoted spread in the ranking statistic, or using the quoted spread capped at each of its four adverse-selection bounds, and re-run the headline regression of row R1 (equation~\eqref{eq:r1cells}). The domestic-foreign difference in the tercile 1 slope, $\beta_{D1}-\beta_{F1}$, is $+0.77$ ($t=1.41$) under the headline assignment, $+0.74$ ($t=0.91$) under the high-low spread and between $+0.66$ and $+1.14$ under the four caps. In row R14C, we form the terciles on the lagged quoted spread alone within each geography-quarter and re-run the quarterly regressions of row R14 within each spread tercile. The reliability gap is $+0.037$ ($p=0.901$) in the tercile with the narrowest spreads, $+0.333$ ($p=0.004$) in the middle tercile and $+0.111$ ($p=0.482$) in the tercile with the widest spreads, with $p$ values from the circular block bootstrap over quarters of row R14 (Table~\ref{tab:spread_sort}).

Price informativeness for foreign stocks as a whole is negative in the primary sample under both price sources (CRSP market value and, as the robustness alternative, Compustat market value, Table~\ref{tab:seams}). Within the sample of the joint estimation, the estimate at the one-quarter horizon is imprecise for both American Depositary Receipts and directly listed foreign shares. The estimate is more negative at the four-quarter horizon than at the one-quarter horizon. If we take the price one quarter earlier, the four-quarter estimate is more negative for both listing types. Its $t$ statistic is $-2.3$ for foreign stocks as a whole and $-2.0$ for directly listed foreign shares. For American Depositary Receipts, the $t$ statistic is $-0.8$. Price informativeness for domestic stocks at the one-quarter horizon is positive in the same sample ($t=3.2$). Thus, the domestic-foreign contrast in price informativeness keeps its sign in each of these splits. On its own, the foreign estimate is significant in some splits and imprecise in others. Table~\ref{tab:variants} shows the pre-specified estimation variants. The largest misfit is in the second moments of foreign stocks in the third informed-trading-intensity tercile (the illiquid-foreign group of Table~\ref{tab:variants}, whose second moments are the variance of the quarterly excess return and the innovation variance of the firm's cash flows). That misfit does not pass into the repeat-author shares or the fad scales. We confirm this by dropping those second moments and then the group itself (the two illiquid-foreign columns of Table~\ref{tab:variants}). The repeat-author shares then move by 0.0066 or less and the fad scales by 0.001 or less. The four domestic-foreign contrasts keep their signs.

Table~\ref{tab:r6_dataterm} shows the implied price-informativeness data term by group and horizon. We build the term at each horizon as the signed price-informativeness estimate divided by the volatility factor and the growth factor of Proposition~\ref{prop:pinf}, both taken from the first-stage estimates that precede the GMM. In the model this term equals the information factor $1-\overline{\Sigma}_j/\Sigma_{jd}$ (equation~\eqref{eq:pinf}). This information factor is at most one. For domestic stocks in the first tercile the term is positive at all four horizons and lies above one, above that model bound. In the domestic second and third terciles it lies between $-0.35$ and 0.26. For foreign stocks in the first tercile it is positive at two horizons and negative at the other two. In the foreign second and third terciles it is negative at every horizon and most negative at the four-quarter horizon. We interpret the negative sign as evidence that home markets lead price discovery for cross-listed foreign stocks. The U.S. price of such foreign stocks carries less information about their future earnings. If we remove the informativeness block from the moment system, it moves the repeat-author shares and the fad scales by less than $10^{-5}$ (Table~\ref{tab:variants}). Thus, the concentration and fad contrasts do not depend on this block.

Table~\ref{tab:psi} shows, for each group, the ratio of the variance of the quarterly excess return across the group's firm-quarters to its value under independent days with the same daily variance (the independence benchmark). If the daily excess returns do not correlate across days, the variance of their quarterly sum equals the sum of the daily variances, even when the daily variance changes from day to day. Thus, day-to-day movement in volatility alone leaves the ratio at one. The ratios instead lie between 0.766 and 0.916, from 8 to 23 percent below one. We check two explanations for the shortfall in the notes to Table~\ref{tab:psi}. The unwinsorized ratio is lower in each of the six groups. Hence, winsorization asymmetry is not the cause. The measured lag-one autocorrelation lowers the ratio by less than six percentage points. That leaves within-quarter reversal over more than one lag as the remaining explanation. Therefore, we take the data moment for the variance of the quarterly excess return to be that variance itself rather than the independence benchmark.

\subsection{Dependence and the acceptance gate, full detail}\label{app:dependence_detail}

Two kinds of dependence enter the inference of Section~\ref{sec:structural}. The first is the dependence within a firm across its quarters, which enters the standard errors and the confidence intervals. The second is the dependence within a quarter across its firms, which enters the joint Wald test. Section~\ref{sec:structural} describes the firm-block bootstrap, whose replications keep each firm's complete time series and repeat the generating steps. Hence, the dependence across a firm's quarters and the sampling error of the generated regressors both pass into the confidence intervals. For the dependence across the firms of one quarter, Section~\ref{sec:structural} reports the joint Wald test with two further clusterings of the standard errors, by quarter and two-way by firm and quarter (Table~\ref{tab:h1_contrasts}). We build these tests from three covariances of the data moments. The firm covariance is the covariance of the data moments across the 999 firm-block bootstrap replications of Section~\ref{sec:structural}. For the quarter covariance we instead draw quarters with replacement, again over 999 replications. The two-way covariance (the covariance under both kinds of dependence at once) is the firm covariance plus the quarter covariance minus the covariance across 999 replications that draw firm-quarters with replacement. We shrink each of these three covariances toward its diagonal by the Ledoit-Wolf rule of Section~\ref{sec:structural} before we form the two-way covariance. The measurement regression that precedes the GMM (\ref{sec:model_to_estimation}) has one acceptance gate. We accept the regression only when the sampling-noise coefficient ($c_w\sigma_e^2$) is positive and its $t$ statistic exceeds 3 with standard errors clustered by firm. Since the measurement regression carries quarter fixed effects, the sampling-noise coefficient is identified mainly from differences across firms within a quarter. Hence, we cluster by firm in the gate. The coefficient is $1.95\times 10^{-4}$ ($t=11.7$ clustered by firm, $t=2.2$ double clustered by firm and quarter). We apply the gate to the measurement regression only and not to the moments of the joint estimation.

\subsection{Reduced-form battery, specifications and full results}\label{app:battery}

\begin{table}[!htbp]\centering
\footnotesize
\caption{Specifications of the reduced-form rows}\label{tab:battery_specs}
\begin{tabularx}{\textwidth}{@{}>{\raggedright}p{0.055\textwidth}>{\raggedright}p{0.23\textwidth}>{\raggedright}p{0.25\textwidth}>{\raggedright\arraybackslash}X@{}}
\toprule
Row & Outcome & Predictor & Conditioning, controls, and design \\
\midrule
R1 & Next-quarter total losses. & Measured noise. & Foreign-by-intensity-tercile interactions. Controls are the FF3 betas, MAX, log capitalization, and buzz. Volatility is excluded by design. Announcement aligned. \\
\addlinespace
R3FAD & Next-quarter excess of losses over gains. & Fad-filtered noise and its square. & By intensity tercile. Turning points carry delta-method intervals. Announcement aligned. \\
\addlinespace
R4 & Next-quarter total gains. & Dispersion. & By intensity tercile. Controls as in R2. \\
\addlinespace
R5 & Quoted spread, next quarter. & Measured noise and dispersion. & Controls are log capitalization and turnover. The spread outcome uses the five-dollar price screen. \\
\addlinespace
R5 & Amihud ratio, next quarter. & Measured noise and dispersion. & Controls are log capitalization and turnover. \\
\addlinespace
R9 & Next-quarter total losses. & Raw and volume-adjusted signals jointly, with tweet volume. & Controls as in R1. \\
\addlinespace
R11 & Next-quarter losses and gains, one regression each. & Measured noise and dispersion jointly. & Full controls, the FF3 betas, MAX, log capitalization, company age, spread, Amihud ratio, and turnover. \\
\bottomrule
\end{tabularx}
\end{table}

\begin{table}[!htbp]\centering
\footnotesize
\ContinuedFloat
\caption{Specifications of the reduced-form rows (continued)}
\begin{tabularx}{\textwidth}{@{}>{\raggedright}p{0.055\textwidth}>{\raggedright}p{0.23\textwidth}>{\raggedright}p{0.25\textwidth}>{\raggedright\arraybackslash}X@{}}
\toprule
Row & Outcome & Predictor & Conditioning, controls, and design \\
\midrule
R12 & Not a regression. & Similarity index against $\widehat{K}$. & Spearman rank correlation and the within-geography-quarter tercile cross-tabulation. \\
\addlinespace
R13a & Squared measured noise, same quarter. & Dispersion, with inverse tweet volume and the foreign indicator. & The reported object is the sign of the dispersion coefficient. \\
\addlinespace
R13b & NASDAQ trade count, same quarter. & Below-median-dispersion indicator. & Log capitalization as control, on the NASDAQ subsample. \\
\addlinespace
R14 & Quarterly cross-sections of losses. & Measured noise, per quarter and geography. & Log capitalization and the FF3 betas as controls. The statistic is the share of quarters with a negative slope, with a circular block bootstrap over quarters. \\
\addlinespace
R7 & Next-quarter total losses. & Measured noise interacted with demeaned financial uncertainty. & The R2 design, announcement aligned. The uncertainty main effect is absorbed by the quarter fixed effects. \\
\addlinespace
R8 & Next-quarter total losses. & Measured noise interacted with the optimism share. & The R2 design, announcement aligned. \\
\addlinespace
R2 & Next-quarter total losses. & Measured noise. & The R1 design plus the volatility control, announcement aligned. The drop from R1 isolates the volatility channel. \\
\bottomrule
\end{tabularx}
\begin{minipage}{0.95\textwidth}\footnotesize\medskip
\textit{Notes:} This table shows the specification used for each of the rows for the battery of reduced form regressions. Unless stated otherwise, the specification used (equation~\eqref{eq:rfspec}) uses quarter fixed effects, double-clustered standard errors (firm and quarter), primary sample (from 2018 onward) and the usual screens (Section~\ref{sec:data}). In order to test whether sentiment in a quarter predicts an outcome in the next quarter, we take the lag of a regressor by one quarter (true for most of the rows in the table). When we need a within-quarter relation, we don't take lags (Rows R13a and R13b). Conditioning on the informed-trading-intensity terciles implies interacting the sentiment regressor with the tercile indicators which gives each tercile its own intercept. In some rows, the outcome is taken from the quarter of the firm's next earnings announcement instead of the next calendar quarter (Section~\ref{sec:reducedform} and \ref{app:battery}).
\end{minipage}
\end{table}

We describe the design of the battery of reduced form regressions in Section~\ref{sec:reducedform}. For five of the rows (R1, R2, R3FAD, R7 and R8, Table~\ref{tab:battery}), we form the outcome under two alignments. In order to align the outcome with the earnings news, we take it from the quarter of the firm's next earnings announcement (the announcement alignment, which we report as the headline). To check that this timing does not drive the result, we also take the outcome from the next calendar quarter and repeat these five regressions (the calendar alignment). The outcome quarter is the same under both alignments for 92.4 percent of the firm-quarters that have an announcement-aligned outcome. We have the firm's next announcement date for 99.0 percent of the primary sample.

\textit{Losses and volatility (R1, R2, R3FAD).} We give the reading of the loss regression without the volatility control in Section~\ref{sec:reducedform} (row R1, Table~\ref{tab:battery}). In order to isolate the volatility channel, we add realized volatility as a control (row R2, Table~\ref{tab:battery}). In the base cell (domestic stocks in the lowest tercile), the coefficient on measured noise is not statistically distinguishable from zero whether we take the outcome from the firm's next earnings announcement quarter or the next calendar quarter (0.211 with $t=1.15$ and 0.301 with $t=1.60$, respectively). In the announcement-aligned run, realized volatility itself enters with a coefficient of $-14.70$ ($t=-40.91$). The other five cell slopes are also indistinguishable from zero in both cases. To check that sampling noise does not drive the result, we subtract the estimated sampling-noise component from measured noise. Within each tercile, we fit the excess of next-quarter losses over gains on the filtered noise and its square (row R3FAD, Table~\ref{tab:battery}). The coefficient on the squared filtered noise is negative in all three terciles in both cases. Thus, the fitted curve is concave in each tercile. The turning points of these fits are imprecise. Hence, we report the turning points from the fits on raw measured noise instead (Table~\ref{tab:r3_turning}). Before we truncate it at zero, the filtered variance is negative for 8.2 percent of the firm-quarters when we take the outcome from the announcement quarter (7.9 percent from the next calendar quarter).

\textit{Gains and dispersion (R4).} Section~\ref{sec:rf_gains} states the reading.

\textit{Volume horse race (R9).} We enter the raw quarterly signal and the volume-adjusted signal together in one regression of next-quarter total losses. The coefficient on each is negative and statistically significant ($-0.317$ with $t=-6.02$ on the raw signal and $-0.612$ with $t=-2.11$ on the adjusted signal, row R9, Table~\ref{tab:battery}). Thus, the volume adjustment is informative about next-quarter losses in a way the raw signal is not.

\textit{Consolidated regression (R11).} Table~\ref{tab:battery} shows the headline estimates of the loss regression in row R11. Section~\ref{sec:reducedform} states the reading. In the gain regression, the coefficient on measured noise is positive and the coefficient on dispersion is negative. By design, we also control for the quoted spread in this specification (Table~\ref{tab:battery_specs}), which is one of the inputs of the ranking statistic by which we sort stocks into informed-trading-intensity terciles (Section~\ref{sec:model_micro}).

\textit{Similarity-index bridge (R12).} To see whether a sort on firm characteristics alone would reproduce our sort into informed-trading-intensity terciles, we compare the two (row R12, Table~\ref{tab:battery}). We rank the stocks by a similarity index built from firm characteristics alone, and again by the informed-precision index behind that sort, which we build from the lagged quoted spread and the lagged signal precision. The rank correlation between the two rankings is negative, at $-0.379$. Only 0.227 of the stocks are in the same tercile under both rankings (\ref{app:similarity_check}). In the same appendix, we also sort the stocks by the similarity index and ask how next-quarter losses relate to measured noise and how next-quarter gains relate to dispersion.

\textit{Liquidity (R5).} Section~\ref{sec:structural} states the reading.

\textit{Attribution (R13a, R13b).} We check whether automated accounts affect the feeds for domestic and foreign stocks in a different manner (rows R13a and R13b, Table~\ref{tab:battery} and Section~\ref{sec:structural}) and we find that it does not. However, the result could be because the difference is too small. To get clarity on that, we would need the variance of the automated content, which is not in the data. Instead, we show that the feed for foreign stocks would be the more precise of the two only if it carried 8 percent more weight on the firm's true signal than the one for domestic stocks (\ref{app:kappa_cases}). We also check whether the average dispersion shifts suddenly at any one quarter due to change in characteristics of the social media platform. We treat each quarter as a possible break date, but not the first or last 15 percent of the 28 quarters, since a break that close to the start or end would leave too few quarters on one side to estimate the mean there. The largest Chow statistic is 2.71 (fourth quarter of 2023). The second largest is 2.47 (third quarter of 2023). This is a descriptive scan since we don't compare the Chow statistics with the critical values. At that break, dispersion decreases more for stocks priced below five dollars or in the smallest size tercile of their quarter than for the others ($t=-8.9$ on the interaction of an indicator for the quarters after the break with an indicator for these stocks, with quarter fixed effects and standard errors double-clustered by firm and quarter). Automated posting is most likely to be in these stocks. Several documented changes to the platform could cause such a break.\footnote{We take these dates from the platform's own posts and filings and from news reports at the time. In 2018, Twitter removed large numbers of fake accounts (reported by the Washington Post on July 6, 2018 and announced on the company blog on July 11, 2018) through May and June 2018. The platform changed owners on October 27, 2022. It restricted outside access to its data for 2023 (announced on February 2, 2023, with paid access from March 29, 2023). It began paying creators and capping how much users could read on July 1, July 13, and July 28, 2023. It tested a fee for new accounts on October 17, 2023. We find no larger platform-wide event inside the fourth quarter of 2023 itself.} The foreign coefficient in the measurement equation (equation~\eqref{eq:vaqsv}) is the domestic-foreign difference in the nonfundamental variance. It has the same sign and stays significant when we drop the quarters around these events, one at a time and then together (the 2018 removal of fake accounts, the 2022 change of ownership, and the 2023 limits on data access, with $t=6.3$, 6.5, 6.4 and 6.5 on 82{,}814 firm-quarters). We interpret the detected quarter as coming after the platform's 2023 changes, not a causal effect, since a change in the platform's data access can also change which tweets the vendor collects, and with it the dispersion which we observe can change too.

\textit{Reliability (R14).} Section~\ref{sec:reducedform} states the design and the reading. In Sections~\ref{sec:intro} and \ref{sec:reducedform}, we scale the slopes by the standard deviation of measured noise in the estimation sample, 0.0099. Rounded to four decimals, this standard deviation is the same for domestic and for foreign stocks. The sample's mean quarterly excess return in Section~\ref{sec:intro} is the pooled mean of the quarterly cumulative excess return in the estimation sample, 0.0211.

\textit{State dependence and correction after optimism (R7, R8).} In row R7 (Table~\ref{tab:battery}), the coefficient on measured noise interacted with demeaned financial uncertainty is $-23.8$ ($t=-1.98$) under the announcement alignment and $-11.3$ ($t=-1.63$) under the calendar alignment. Section~\ref{sec:structural} states the reading of the correction after optimism (row R8, Table~\ref{tab:battery}). The coefficient on measured noise interacted with the optimism share is 0.214 ($t=0.21$) under the announcement alignment and 0.292 ($t=0.29$) under the calendar alignment.

\subsection{Cross-sectional structure on a characteristics-only sort}\label{app:similarity_check}

We wish to know whether our two patterns (gains-dispersion and loss-noise relations) still show up if we sort stocks in different ways. Apart from the sorting of stocks using the informed-trading-intensity (Section~\ref{sec:model_micro}), we also sort them using a similarity index built from firm characteristics alone (without the sentiment feed, the quoted spread or the structural model). This similarity index is weighted cosine similarity between a stock's characteristics and a benchmark stock formed by taking median of all stock features using weights from a random-forest \citep{hobergphillips2016, koffi2025innovative}. We refer to the stocks closest to the benchmark as typical and those farthest from the benchmark as atypical. The results show that only 0.227 of the stocks are in the same tercile under both types of sorting as opposed to one third which we would get if the two sortings were unrelated. Their rank correlation is $-0.379$. Thus, a stock's representativeness is a weak indication of its intensity tercile. The paper's model uses sorting by intensity and makes no use of sorting by similarity.

Table~\ref{tab:similarity_check} shows the two patterns by similarity tercile and geography. The gains-dispersion relation is positive for the atypical stocks and near zero for the typical ones. This positive response is marginally significant for domestic stocks and imprecise for foreign stocks. If the similarity sort ran opposite to the intensity sort, the gain-dispersion response would show more in the typical stocks. It shows up in atypical stocks instead. The loss-noise relation doesn't show any pattern across the similarity terciles. Its sign varies across the cells. The loss-noise relation is weak even when the sorting is by intensity (row R1, Table~\ref{tab:battery}). Such a weak relation would not survive on a sorting such as by similarity which barely overlaps with intensity.

\begin{table}[!htbp]\centering
\caption{Cross-sectional structure of the two channels on the characteristics-only similarity sort}\label{tab:similarity_check}
\begin{tabularx}{\textwidth}{@{}>{\RaggedRight}X cccc@{}}
\hline\hline
Similarity tercile & Domestic slope & Domestic $t$ & Foreign slope & Foreign $t$ \\
\hline
\multicolumn{5}{@{}l}{\textit{Panel A. Next-quarter gains on dispersion}} \\
Tercile 1 (atypical) & $+0.069$ & 1.91 & $+0.034$ & 0.46 \\
Tercile 2 & $+0.056$ & 1.71 & $+0.058$ & 1.20 \\
Tercile 3 (typical) & $-0.006$ & $-0.19$ & $-0.022$ & $-0.45$ \\
\multicolumn{5}{@{}l}{\textit{Panel B. Next-quarter losses on measured noise}} \\
Tercile 1 (atypical) & $+0.72$ & 1.30 & $+1.07$ & 1.45 \\
Tercile 2 & $-0.89$ & $-2.09$ & $-1.02$ & $-1.42$ \\
Tercile 3 (typical) & $-0.52$ & $-1.70$ & $-0.29$ & $-0.60$ \\
\hline\hline
\end{tabularx}
\begin{minipage}{0.95\textwidth}\footnotesize\medskip
\textit{Notes:} This table shows the two reduced-form channels by similarity tercile and geography, each as the per-cell slope from a foreign-by-tercile interaction with standard errors double clustered by firm and quarter. We form the similarity terciles within geography and quarter. Tercile 1 is the least similar to the benchmark, the atypical end. Tercile 3 is the most similar, the typical end. Panel A is next-quarter total gains on dispersion. Panel B is next-quarter total losses on measured noise. The loss channel has 89{,}978 firm-quarters in each cell.
\end{minipage}
\end{table}

\subsection{Robustness}\label{app:robustness}

\textit{Reliability splits (R14B).}\label{app:reliability_splits} Table~\ref{tab:reliability_splits} reports all cells and contrasts. Section~\ref{sec:reducedform} states the reading.

\begin{table}[!htbp]\centering
\footnotesize
\caption{Reliability splits, sign stability across quarters}\label{tab:reliability_splits}
\begin{tabular}{lcccc}
\hline\hline
Split & Share, first subset & Share, second subset & Difference & $p$ \\
\hline
\multicolumn{5}{@{}l}{\textit{Domestic versus foreign, within intensity terciles}} \\
\addlinespace
Tercile 1 (low) & 0.852 & 0.556 & $+0.296$ & 0.004 \\
Tercile 2 & 0.889 & 0.741 & $+0.148$ & 0.224 \\
Tercile 3 (high) & 0.741 & 0.778 & $-0.037$ & 0.787 \\
\addlinespace
\multicolumn{5}{@{}l}{\textit{Within foreign stocks}} \\
\addlinespace
English versus non-English headquarters & 0.444 & 0.519 & $-0.074$ & 0.561 \\
Canada versus other foreign & 0.407 & 0.481 & $-0.074$ & 0.597 \\
ADR versus directly listed & 0.296 & 0.593 & $-0.296$ & 0.022 \\
Domestic versus directly listed foreign & 0.778 & 0.593 & $+0.185$ & 0.116 \\
\hline\hline
\end{tabular}
\begin{minipage}{0.95\textwidth}\footnotesize\medskip
\textit{Notes:} This table shows the share of quarters in which the slope of next-quarter total losses on measured noise is negative (row R14B; 27 usable quarters). We run this regression for each subset in each quarter, with log capitalization and the FF3 betas as controls. We skip any quarter in which a subset has fewer than 15 firms. We classify a headquarters country as English-speaking when English is its first language \citep{crystal2003english}. We classify the stock as an ADR when the CRSP ADR flag holds on a majority of its days. We test each domestic-foreign difference with the R14 circular block bootstrap over quarters (block length four, 999 replications, paired across the two subsets). The shares count only the sign of each slope, not its size. The median number of firms per quarter is 979, 975, and 918 for the domestic terciles, 173, 171, and 153 for the foreign terciles, 231 and 257 for the English and non-English splits, 116 and 371 for Canada and other foreign stocks, and 197 and 306 for ADRs and directly listed shares.
\end{minipage}
\end{table}

\textit{Mean slopes by listing type and richer controls (R14D).}\label{app:rob_r14d} We run two more specifications, with the R14 per-quarter design, screens, and quarters unchanged. In the first, we report the plain difference in mean quarterly slopes for ADRs and for directly listed foreign shares. The difference in the mean quarterly slope between domestic stocks and ADRs is $-2.00$ (Newey-West $t=-3.72$). The difference between domestic stocks and directly listed foreign firms is $-0.74$ ($t=-1.29$). These two differences agree with the negative-slope shares in Table~\ref{tab:reliability_splits}. Most ADRs are non-Canadian. So the split by listing type and the split by country overlap. In the second, we add the lagged quoted spread, the Amihud illiquidity ratio, and turnover to the R14 controls on the full sample. The domestic share of negative-slope quarters stays at 0.778. The foreign share is 0.556. We leave out the firm's age since its IPO, which we have for only 60.9 percent of the sample, 60.1 percent of the domestic firm-quarters, and 65.7 percent of the foreign firm-quarters. On the subsample where that age exists, with age and the full control set, the difference is $-0.11$ ($t=-0.25$). We attribute this weakening to both the smaller subsample and the added controls.

\textit{Headline reliability across sample years (R14E).}\label{app:rob_r14e} The headline reliability estimate is the mean across quarters of the domestic-foreign slope of next-quarter total losses on measured noise ($-1.40$, Newey-West $t=-2.99$, row R14, Table~\ref{tab:battery}). We re-estimate it seven times, each time dropping one sample year. It stays negative and significant at the 5 percent level every time. The estimate ranges from $-1.17$ without 2023 to $-1.73$ without 2018 and the Newey-West $t$ statistic from $-2.34$ to $-4.26$. Leaving out 2018 restricts the window to 2019 through 2024 and removes the three 2018 quarters, which strengthens the estimate to $-1.73$ ($t=-4.26$). This cuts against the pull of 2018 on the slope for foreign stocks in the low tercile, which the three 2018 quarters inflate (\ref{app:rob_loyo}).

\begin{sloppypar}
\textit{Additional robustness (R15).}\label{app:additional_robustness} We re-estimate row R1, the regression of next-quarter total losses on measured noise (Table~\ref{tab:battery}), on thirty-one alternative samples and specifications (Table~\ref{tab:additional_robustness}). The foreign contrast stays negative in all thirty-one re-estimations.
\end{sloppypar}

\textit{Leave-one-year-out.}\label{app:rob_loyo} There is a caveat for the leave-one-year-out re-estimations (Table~\ref{tab:additional_robustness}). When we drop 2018, the slope for foreign stocks in the first informed-trading-intensity tercile flattens to $-0.080$, while dropping any other year leaves it between $-0.75$ and $-1.28$. So much of that slope in the first tercile comes from the 2018 quarters.

\begin{table}[!htbp]\centering
\footnotesize\setlength{\tabcolsep}{3pt}
\caption{Additional robustness of the unconditional loss regression}\label{tab:additional_robustness}
\begin{tabular}{lrcccc}
\hline\hline
Sample or specification & $N$ & Foreign contrast & $p$ & Foreign slope, Tercile 1 & $p$ \\
\hline
Baseline. & 89,011 & $-0.771$ & 0.171 & $-0.914$ & 0.151 \\
\addlinespace
\multicolumn{6}{@{}l}{\textit{Foreign listing, language, and country splits}} \\
\addlinespace
Excluding foreign ADRs. & 83,769 & $-0.989$ & 0.101 & $-1.118$ & 0.108 \\
Excluding directly listed foreign firms. & 80,831 & $-0.783$ & 0.398 & $-0.959$ & 0.332 \\
Foreign restricted to English-speaking. & 81,444 & $-0.780$ & 0.148 & $-0.916$ & 0.168 \\
Foreign restricted to non-English. & 83,156 & $-0.792$ & 0.358 & $-0.976$ & 0.278 \\
Excluding Canada. & 86,080 & $-0.641$ & 0.301 & $-0.799$ & 0.236 \\
\addlinespace
\multicolumn{6}{@{}l}{\textit{Leave-one-year-out}} \\
\addlinespace
Excluding 2018. & 80,165 & $-0.370$ & 0.531 & $-0.080$ & 0.886 \\
Excluding 2019. & 76,910 & $-0.758$ & 0.287 & $-0.984$ & 0.242 \\
Excluding 2020. & 74,506 & $-1.154$ & 0.030 & $-0.877$ & 0.155 \\
Excluding 2021. & 73,576 & $-0.672$ & 0.295 & $-0.753$ & 0.277 \\
Excluding 2022. & 73,047 & $-1.152$ & 0.057 & $-1.049$ & 0.128 \\
Excluding 2023. & 76,453 & $-0.523$ & 0.372 & $-0.935$ & 0.165 \\
Excluding 2024. & 79,409 & $-0.781$ & 0.179 & $-1.277$ & 0.045 \\
\hline\hline
\end{tabular}
\end{table}

\begin{table}[!htbp]\centering
\footnotesize\setlength{\tabcolsep}{3pt}
\ContinuedFloat
\caption{Additional robustness of the unconditional loss regression (continued)}
\begin{tabular}{lrcccc}
\hline\hline
Sample or specification & $N$ & Foreign contrast & $p$ & Foreign slope, Tercile 1 & $p$ \\
\hline
\multicolumn{6}{@{}l}{\textit{Leave-one-country-out}} \\
\addlinespace
Excluding China. & 87,401 & $-0.947$ & 0.084 & $-1.115$ & 0.066 \\
Excluding the United Kingdom. & 87,842 & $-1.079$ & 0.082 & $-1.235$ & 0.082 \\
Excluding Israel. & 88,090 & $-0.760$ & 0.181 & $-0.909$ & 0.162 \\
Excluding Bermuda. & 88,242 & $-0.681$ & 0.257 & $-0.839$ & 0.204 \\
Excluding Ireland. & 88,452 & $-0.783$ & 0.169 & $-0.925$ & 0.144 \\
Excluding Switzerland. & 88,597 & $-0.821$ & 0.138 & $-0.958$ & 0.132 \\
Excluding Brazil. & 88,640 & $-0.709$ & 0.208 & $-0.843$ & 0.181 \\
Excluding the Netherlands. & 88,674 & $-0.763$ & 0.181 & $-0.902$ & 0.159 \\
Excluding Luxembourg. & 88,702 & $-0.702$ & 0.219 & $-0.839$ & 0.195 \\
Excluding Argentina. & 88,738 & $-0.741$ & 0.189 & $-0.898$ & 0.164 \\
Excluding Singapore. & 88,742 & $-0.777$ & 0.165 & $-0.921$ & 0.152 \\
Excluding the Cayman Islands. & 88,758 & $-0.760$ & 0.177 & $-0.898$ & 0.162 \\
Excluding Germany. & 88,766 & $-0.786$ & 0.159 & $-0.915$ & 0.151 \\
Excluding Greece. & 88,774 & $-0.785$ & 0.163 & $-0.928$ & 0.143 \\
Excluding Japan. & 88,780 & $-0.893$ & 0.115 & $-1.037$ & 0.115 \\
Excluding Mexico. & 88,785 & $-0.745$ & 0.191 & $-0.891$ & 0.168 \\
\addlinespace
\multicolumn{6}{@{}l}{\textit{Alternative tweet-volume controls}} \\
\addlinespace
Volume control, log total tweets. & 89,011 & $-0.620$ & 0.284 & $-1.420$ & 0.033 \\
Volume control, mean inverse tweet count. & 89,011 & $-0.519$ & 0.381 & $-1.394$ & 0.040 \\
Volume controls, buzz and log tweets. & 89,011 & $-0.630$ & 0.275 & $-1.406$ & 0.034 \\
\hline\hline
\end{tabular}
\begin{minipage}{0.95\textwidth}\footnotesize\medskip
\textit{Notes:} In each row we re-estimate the announcement-aligned specification of row R1 (Table~\ref{tab:battery}), next-quarter total losses on measured noise interacted with the foreign indicator and the informed-trading-intensity tercile (full factorial), with the FF3 betas, MAX, log capitalization, and buzz as controls, quarter fixed effects, and standard errors double clustered by firm and quarter. The foreign contrast is the coefficient on noise interacted with the foreign indicator. The foreign slope, Tercile 1, is the implied foreign loss-noise slope in the lowest informed-trading tercile. The English split classifies headquarters countries by English as a first language following \citet{crystal2003english}; the ADR split uses the CRSP share-level ADR flag under a majority-of-days rule. Leave-one-country-out covers the seventeen foreign countries with at least 200 estimation-sample observations, with Canada's exclusion shown in the splits panel; the thirty-four smaller countries, the largest being India with 199 observations, are not dropped individually. The volume rows replace or augment the buzz control with alternatives measured over the outcome window, as the buzz control is (row R15).
\end{minipage}
\end{table}

\begin{table}[!htbp]\centering
\footnotesize
\caption{Sample composition by headquarters country}\label{tab:composition_country}
\begin{tabular}{lrrrr}
\hline\hline
Country & Firm-quarters & Firms & Share of sample (\%) & Share of foreign (\%) \\
\hline
United States. & 75,589 & 4,486 & 84.9 & -- \\
Canada. & 2,931 & 176 & 3.3 & 21.8 \\
China. & 1,610 & 136 & 1.8 & 12.0 \\
United Kingdom. & 1,169 & 95 & 1.3 & 8.7 \\
Israel. & 921 & 74 & 1.0 & 6.9 \\
Bermuda. & 769 & 44 & 0.9 & 5.7 \\
Ireland. & 559 & 31 & 0.6 & 4.2 \\
Switzerland. & 414 & 26 & 0.5 & 3.1 \\
Brazil. & 371 & 30 & 0.4 & 2.8 \\
Netherlands. & 337 & 21 & 0.4 & 2.5 \\
Luxembourg. & 309 & 19 & 0.3 & 2.3 \\
Argentina. & 273 & 15 & 0.3 & 2.0 \\
Singapore. & 269 & 25 & 0.3 & 2.0 \\
Cayman Islands. & 253 & 17 & 0.3 & 1.9 \\
Germany. & 245 & 20 & 0.3 & 1.8 \\
Greece. & 237 & 18 & 0.3 & 1.8 \\
Japan. & 231 & 11 & 0.3 & 1.7 \\
Mexico. & 226 & 12 & 0.3 & 1.7 \\
Other (34 countries). & 2,298 & 160 & 2.6 & 17.1 \\
Total. & 89,011 & 5,416 & 100.0 & -- \\
\hline\hline
\end{tabular}
\begin{minipage}{0.95\textwidth}\footnotesize\medskip
\textit{Notes:} This table shows how the stocks in the baseline estimation sample split across their Compustat headquarters countries (the announcement-aligned R1 screens, Table~\ref{tab:additional_robustness}; row R15). The foreign stocks make up 13,422 firm-quarters from 51 countries. Each country's count is its number of distinct firms. The total row shows the number of distinct firms over all countries.
\end{minipage}
\end{table}

\textit{Size-conditioned measurement contrasts (R16).}\label{app:rob_size} Table~\ref{tab:size_contrasts} shows all the cells. Section~\ref{sec:structural} states the reading.

\begin{table}[!htbp]\centering
\footnotesize
\caption{Size-conditioned measurement contrasts, foreign coefficients}\label{tab:size_contrasts}
\begin{tabular}{lrcccc}
\hline\hline
Specification & $N$ & Foreign coefficient & SE & $t$ & $p$ \\
\hline
\multicolumn{6}{@{}l}{\textit{Panel A. Contributor dispersion}} \\
\addlinespace
No size control. & 98,291 & $-0.0057$ & 0.0018 & -3.09 & 0.0046 \\
Log capitalization. & 98,291 & $-0.0074$ & 0.0017 & -4.34 & 0.0002 \\
Size tercile 1. & 32,771 & $-0.0051$ & 0.0021 & -2.49 & 0.0194 \\
Size tercile 2. & 32,757 & $-0.0096$ & 0.0027 & -3.54 & 0.0015 \\
Size tercile 3. & 32,763 & $-0.0040$ & 0.0026 & -1.53 & 0.1386 \\
\addlinespace
\multicolumn{6}{@{}l}{\textit{Panel B. Nonfundamental variance}} \\
\addlinespace
No size control. & 98,291 & $+7.42$ & 1.17 & 6.35 & 0.0000 \\
Log capitalization. & 98,291 & $+7.15$ & 1.17 & 6.12 & 0.0000 \\
Size tercile 1. & 32,771 & $+5.39$ & 1.73 & 3.12 & 0.0043 \\
Size tercile 2. & 32,757 & $+7.08$ & 1.32 & 5.38 & 0.0000 \\
Size tercile 3. & 32,763 & $+8.91$ & 1.71 & 5.21 & 0.0000 \\
\hline\hline
\end{tabular}
\begin{minipage}{0.95\textwidth}\footnotesize\medskip
\textit{Notes:} In Panel A, the contributor dispersion is regressed on the foreign indicator, while in Panel B, the squared measured noise is regressed on the average inverse tweet count and the foreign indicator, both with quarter fixed effects and standard errors clustered by firm and quarter, and with and without controls for size. We construct terciles using size within each quarter on log capitalization. Domestic and foreign stocks share the same bins. Lower dispersion means fewer distinct accounts per tweet. A negative coefficient in Panel A means foreign stocks have lower dispersion than domestic stocks, the concentration contrast. A positive coefficient in Panel B means foreign stocks have more nonfundamental variance than domestic stocks, the fad contrast. The Panel B coefficients are small in level. We scale them and their standard errors by $10^{5}$ for the table. We keep the firm-quarters that pass the sentiment screens and have non-missing dispersion and log capitalization (row R16).
\end{minipage}
\end{table}

\textit{Reliability under a model-free sort (R14C).}\label{app:rob_spreadsort} We give the design and reading of this exercise in Section~\ref{sec:structural} (row R14C, Table~\ref{tab:spread_sort}). The lagged quoted spread is only one term of the ranking statistic (Section~\ref{sec:model_micro}). It is a noisy proxy for informed-trading intensity. So the gap sits in a different tercile under the two sorts, in spread tercile 2 here and in intensity tercile 1 under the intensity sort. Either way, we draw the same conclusion, that the reliability gap survives the tercile split and does not come from how we build the sort. Table~\ref{tab:spread_sort} shows all the cells and the contrasts.

\begin{table}[!htbp]\centering
\footnotesize
\caption{Reliability under a model-free spread sort}\label{tab:spread_sort}
\begin{tabular}{lcccc}
\hline\hline
Spread tercile & Share, domestic & Share, foreign & Difference & $p$ \\
\hline
Tercile 1 (narrowest spread) & 0.407 & 0.370 & $+0.037$ & 0.901 \\
Tercile 2 & 0.741 & 0.407 & $+0.333$ & 0.004 \\
Tercile 3 (widest spread) & 0.778 & 0.667 & $+0.111$ & 0.482 \\
\hline\hline
\end{tabular}
\begin{minipage}{0.95\textwidth}\footnotesize\medskip
\textit{Notes:} This table shows the share of quarters for which the slope of next-quarter losses on measured noise is negative (row R14C; 27 usable quarters). This regression is run for domestic and the foreign stocks separately, within each tercile of the spread, with log capitalization and the FF3 betas as controls. We construct the terciles using the lagged quoted spread for each geography-quarter which is a sorting not related to the model. We test each domestic-foreign difference with the R14 circular block bootstrap over quarters (block length four, 999 replications, paired across the two subsets).
\end{minipage}
\end{table}

\textit{Gains on dispersion holding cash-flow predictability fixed (R4B).}\label{app:rob_r4_innov} The informed-trading-intensity terciles also differ in the estimated cash-flow innovation variance, which is smallest in tercile 1 and rises across the terciles. So the least informed-traded stocks also have the most predictable cash flows. We ask whether the pattern of row R4 (Table~\ref{tab:battery}), in which dispersion predicts next-quarter gains in tercile 1 only, survives when we hold cash-flow predictability fixed. Table~\ref{tab:r4_innovvar} shows three results. First, when we add the log innovation variance to the row R4 design as a control, alone or interacted with dispersion, the slope in intensity tercile 1 stays positive and significant, while the slopes in terciles 2 and 3 are indistinguishable from zero. Second, within each cash-flow-predictability tercile, the slope is largest in intensity tercile 1, where it is significant in predictability terciles 1 and 2. Third, within intensity tercile 1 the slope is concentrated in predictability tercile 1, the firms with the most predictable cash flows, and is indistinguishable from zero in predictability tercile 3. The model's outside-knowledge channel implies a reverse pattern. Firms with more predictable cash flows have a smaller cash-flow innovation variance. Therefore, the investors know more about such a firm before any signal arrives. So the investors give less weight to the feed. Hence, the price responds less to the feed. That implies that the response of next-quarter gains to dispersion should be the weakest among the most predictable firms. Neither of the model's return channels (the volatility channel of Proposition~\ref{prop:vol} and the correction of surviving overpricing of Proposition~\ref{prop:drift}) produces that effect.

\begin{table}[!htbp]\centering
\caption{Gains on dispersion holding cash-flow predictability fixed (R4B)}\label{tab:r4_innovvar}
\footnotesize
\begin{tabular}{@{}p{0.40\textwidth}cccc@{}}
\hline\hline
 & \multicolumn{3}{c}{Informed-trading-intensity tercile} & \\
Specification & 1 & 2 & 3 & $N$ \\
\hline
\multicolumn{5}{@{}l}{\textit{Panel A. Row R4 design with the firm's log cash-flow innovation variance}} \\
Row R4, full sample & 0.128 (3.15) & 0.003 (0.11) & 0.014 (0.34) & 89{,}982 \\
Row R4, firms with the measure & 0.134 (3.13) & $-0.001$ ($-0.03$) & 0.011 (0.29) & 77{,}988 \\
Adding the control & 0.152 (3.54) & 0.022 (0.65) & 0.026 (0.68) & 77{,}988 \\
Adding the control and its interaction with dispersion & 0.154 (3.57) & 0.021 (0.63) & 0.022 (0.48) & 77{,}988 \\
\addlinespace
\multicolumn{5}{@{}l}{\textit{Panel B. Row R4 design within each cash-flow-predictability tercile}} \\
Predictability tercile 1 (most predictable) & 0.136 (3.88) & 0.048 (1.26) & $-0.084$ ($-0.98$) & 28{,}086 \\
Predictability tercile 2 & 0.119 (2.31) & $-0.001$ ($-0.02$) & 0.109 (1.68) & 27{,}284 \\
Predictability tercile 3 (least predictable) & 0.075 (0.95) & 0.014 (0.26) & $-0.001$ ($-0.04$) & 22{,}618 \\
\addlinespace
 & \multicolumn{3}{c}{Cash-flow-predictability tercile} & \\
 & 1 & 2 & 3 & $N$ \\
\multicolumn{5}{@{}l}{\textit{Panel C. Dispersion slope by predictability tercile within each intensity tercile}} \\
Intensity tercile 1 & 0.109 (4.21) & 0.056 (1.46) & $-0.039$ ($-0.86$) & 26{,}873 \\
Intensity tercile 2 & 0.078 (1.95) & 0.006 (0.14) & $-0.009$ ($-0.18$) & 26{,}367 \\
Intensity tercile 3 & $-0.024$ ($-0.28$) & 0.111 (1.67) & 0.057 (1.99) & 24{,}748 \\
\hline\hline
\end{tabular}
\begin{minipage}{0.95\textwidth}\footnotesize\medskip
\textit{Notes:} Each cell is the slope of next-quarter total gains on dispersion, with the $t$ statistic in parentheses, from the design of row R4 (dispersion interacted with the informed-trading-intensity tercile, tercile fixed effects, the controls of row R2, quarter fixed effects, standard errors double clustered by firm and quarter). The firm's cash-flow innovation variance is the residual variance of a firm-level AR(1) regression of the quarter-demeaned ratio of operating income before depreciation over total assets on its one-quarter lag across the primary sample, for firms with at least twelve usable quarters (4,640 firms). In Panel A we add its log as a control and, in the last row, also its standardized log interacted with dispersion (interaction coefficient 0.006, $t = 0.29$). Cash-flow-predictability terciles are formed on the log innovation variance within geography, tercile 1 being the lowest innovation variance. In Panel B we run the row R4 design within each predictability tercile. In Panel C we interact dispersion with the predictability tercile within each intensity tercile. Of the stock-quarters in intensity tercile 1, 49 percent are in predictability tercile 1 and 13 percent in predictability tercile 3.
\end{minipage}
\end{table}

We repeat the main regressions, each time replacing one variable with an alternative (row R10). We reconstruct the outcomes (losses and gains) from daily returns after removing the three Fama-French factors out of sample. We construct the terciles from the high-low spread instead of the quoted spread. Among the controls used, we have residual idiosyncratic volatility, filtered standard deviation of daily returns instead of realized volatility, sentiment noise from quarterly sentiment volatility (QSV) instead of VAQSV. We also cap the quoted spread at its adverse-selection bounds, standardize the noise within each group, and drop the first quarter of sentiment data. The ranking into terciles is 98.1 percent unchanged under the cap. For all these alternatives, the domestic-foreign difference maintains its sign and the terciles keep their order. The exception is when we use QSV, the slope for foreign stocks becomes positive and the sign of the difference reverses, but is not significant ($+0.12$, $t=0.97$).

\begin{table}[!htbp]\centering
\caption{Conventions at the seams between data frequencies, with their diagnostics}\label{tab:seams}
\begin{tabularx}{\textwidth}{@{}>{\RaggedRight}p{0.24\textwidth}>{\RaggedRight}X>{\RaggedRight\arraybackslash}p{0.30\textwidth}@{}}
\hline\hline
Seam & Convention & Diagnostic \\
\hline
Price source for informativeness & CRSP capitalization over book assets & negative foreign informativeness holds under both price sources \\
Quarterly variance bridge & direct variance of the accrual sum & bridge ratios between 0.766 and 0.916 (Table~\ref{tab:psi}) \\
Day counts & sentiment days for precision, trading days for outcomes & distinct day sets by construction, each used only for its own object \\
Earnings seasonality & calendar demeaning with a fiscal-quarter variant & persistence moves by less than 0.001 \\
Turnover units & per-day turnover, the quarterly sum divided by the number of return days, both counted on the days with a valid return and valid turnover & 98.93 percent of the firm-quarters checked match a reconstruction from daily data. The remaining 1{,}048 are recomputed from daily data on that day set. \\
Announcement timing & announcement-aligned outcome quarter as the headline alignment & coverage 0.990, median lag 35 and 40 days \\
\hline\hline
\end{tabularx}
\begin{minipage}{0.95\textwidth}\footnotesize\medskip
\textit{Notes:} Each aggregation seam between daily market data, quarterly fundamentals, and the sentiment feed is a convention we adopt in advance, paired with a diagnostic of its effect. The fiscal and calendar alignment of book assets has zero mismatch by construction because fundamentals enter at the calendar quarter of their statement date. We take the timing convention from the reported announcement date.
\end{minipage}
\end{table}

\begin{table}[!htbp]\centering
\caption{Pre-specified estimation variants re-estimated from the headline estimate}\label{tab:variants}
\begin{tabular}{@{}l*{5}{>{\centering\arraybackslash}p{0.135\textwidth}}@{}}
\hline\hline
Parameter & Headline & Informativeness block excluded & Illiquid-foreign second moments dropped & Illiquid-foreign group dropped & Shrunk full-matrix weighting \\
\hline
$\pi$, Domestic T1 & 0.1744 & 0.1744 & 0.1744 & 0.1744 & 0.1690 \\
$\pi$, Domestic T2 & 0.1573 & 0.1573 & 0.1573 & 0.1573 & 0.1805 \\
$\pi$, Domestic T3 & 0.1447 & 0.1447 & 0.1447 & 0.1447 & 0.1718 \\
$\pi$, Foreign T1 & 0.1820 & 0.1820 & 0.1820 & 0.1866 & 0.2089 \\
$\pi$, Foreign T2 & 0.1637 & 0.1637 & 0.1637 & 0.1703 & 0.1927 \\
$\pi$, Foreign T3 & 0.1698 & 0.1699 & 0.1698 & 0.1698 & 0.1799 \\
$\sigma_b$, domestic & 0.0273 & 0.0273 & 0.0273 & 0.0272 & 0.0252 \\
$\sigma_b$, foreign & 0.0293 & 0.0293 & 0.0293 & 0.0303 & 0.0269 \\
$\kappa$ & 394 & 404 & 355 & 345 & 645 \\
\hline\hline
\end{tabular}
\begin{minipage}{0.95\textwidth}\footnotesize\medskip
\textit{Notes:} We re-estimate each variant from the headline estimate with the posting intensity set at 1.5. The comparison is therefore like with like. The mismatch parameters move by less than $10^{-5}$ under the moment-exclusion variants. The conversion constant $\kappa$ moves between 345 and 645 across the variants, inside the shallow part of the profile in Table~\ref{tab:kappa_profile}. The shrunk full-matrix weighting is the pre-specified robustness weighting. Under it, the repeat-author shares and the fad scales stay higher for foreign than for domestic.
\end{minipage}
\end{table}

\begin{table}[!htbp]\centering
\caption{Turning points of the loss-minus-gain response in measured noise units by informed-trading-intensity tercile}\label{tab:r3_turning}
\begin{tabular}{lcc}
\hline\hline
Cell & Turning point & SE \\
\hline
Tercile 1, calendar & 0.079 & 0.016 \\
Tercile 2, calendar & 0.070 & 0.012 \\
Tercile 3, calendar & 0.085 & 0.095 \\
Tercile 1, announcement & 0.076 & 0.016 \\
Tercile 2, announcement & 0.062 & 0.009 \\
Tercile 3, announcement & 0.071 & 0.070 \\
\hline\hline
\end{tabular}
\begin{minipage}{0.95\textwidth}\footnotesize\medskip
\textit{Notes:} We obtain the turning points by fitting, within each tercile, the excess of next-quarter losses over gains on measured noise and its square, with delta-method standard errors. The middle tercile turns left of the low tercile in both cases, which is consistent with the leftward shift the model predicts holding the other group primitives fixed. The high tercile is imprecise. These shape and ordering tests are the drift evidence under the pre-specified design.
\end{minipage}
\end{table}

\begin{table}[!htbp]\centering
\caption{Decomposition of the objective by moment family at the headline estimate}\label{tab:perblock_j}
\begin{tabular}{lccc}
\hline\hline
Moment family & Moments & Share of objective & Largest $|t|$ \\
\hline
Liquidity & 24 & 0.848 & 28.5 \\
Cash-flow persistence & 12 & 0.048 & 11.0 \\
Informativeness & 12 & 0.018 & 5.8 \\
Dispersion & 6 & 0.009 & 4.3 \\
Signal-noise levels & 6 & 0.027 & 7.2 \\
Sampling-noise slope & 1 & 0.001 & 1.8 \\
Quarterly variance & 6 & 0.018 & 7.1 \\
Gain and loss levels & 12 & 0.004 & 2.5 \\
Precision slopes & 12 & 0.028 & 6.5 \\
\hline\hline
\end{tabular}
\begin{minipage}{0.95\textwidth}\footnotesize\medskip
\textit{Notes:} The Hansen $J$ test rejects the full 91-moment specification (\ref{app:fit_detail}). Given the large effective sample, we therefore emphasize the decomposition of the objective and the moment-level fit. The liquidity family is 84.8 percent of the weighted distance and the principal source of model misfit. The gain and loss family is 0.4 percent of the weighted distance. The largest gain and loss residuals come from the Gaussian model of the quarterly gain and loss accrual. We report them without adjustment.
\end{minipage}
\end{table}

\begin{table}[!htbp]\centering
\caption{Implied price-informativeness data term by group and horizon against the median signal-quality index}\label{tab:r6_dataterm}
\begin{tabular}{lccccc}
\hline\hline
Group & Median signal-quality index & $s=1$ & $s=2$ & $s=3$ & $s=4$ \\
\hline
Domestic T1 & 159{,}003 & 3.20 & 4.29 & 3.99 & 6.41 \\
Domestic T2 & 147{,}407 & 0.26 & 0.20 & 0.05 & $-0.35$ \\
Domestic T3 & 130{,}836 & $-0.06$ & $-0.05$ & $-0.10$ & $-0.17$ \\
Foreign T1 & 131{,}910 & 0.12 & $-0.06$ & 0.34 & $-0.03$ \\
Foreign T2 & 123{,}989 & $-0.37$ & $-1.32$ & $-1.97$ & $-3.42$ \\
Foreign T3 & 122{,}056 & $-0.23$ & $-0.16$ & $-0.31$ & $-0.47$ \\
\hline\hline
\end{tabular}
\begin{minipage}{0.95\textwidth}\footnotesize\medskip
\textit{Notes:} This table is descriptive and has no test statistics. In each cell we divide the signed price-informativeness estimate at horizon $s$ by the growth and volatility components built from the first-stage estimates, leaving the implied information term that the model equates to one minus the posterior-to-prior variance ratio. For each group, the second column reports the median of the signal-quality index ($T_q/\mathrm{VAQSV}_{fq}^2$) across the firm-quarters in that group. We run the price-informativeness regressions over horizons one through four, with sector fixed effects, a lagged-earnings control, ratios winsorized at the first and ninety-ninth percentiles, and signed coefficients. We count the negative foreign cells as misfit. The formal test is the informativeness moment of the joint estimation.
\end{minipage}
\end{table}

\begin{table}[!htbp]\centering
\caption{Bridge ratios between the realized quarterly variance and the independence benchmark, winsorized and unwinsorized}\label{tab:psi}
\begin{tabular}{lccc}
\hline\hline
Group & $N$ & $\psi_j$ & $\psi_j$ unwinsorized \\
\hline
Domestic T1 & 27,024 & 0.881 & 0.860 \\
Domestic T2 & 27,035 & 0.906 & 0.860 \\
Domestic T3 & 27,048 & 0.800 & 0.748 \\
Foreign T1 & 5,031 & 0.916 & 0.890 \\
Foreign T2 & 5,040 & 0.870 & 0.835 \\
Foreign T3 & 5,051 & 0.766 & 0.732 \\
\hline\hline
\end{tabular}
\begin{minipage}{0.95\textwidth}\footnotesize\medskip
\textit{Notes:} The ratio is the cross-sectional variance of the quarterly accrual sum over the group mean of trading days times the squared daily standard deviation. Values below one mean the independence benchmark overstates the quarterly variance. The unwinsorized ratio is lower in each of the six groups, which rules out winsorization asymmetry as the cause. Measured first-order autocorrelation explains under six percentage points of the shortfall below one. Multi-lag within-quarter reversal is the remaining explanation and ranks correctly with illiquidity.
\end{minipage}
\end{table}

\subsection{The tercile pattern of the reliability gap}\label{app:tercile_pattern}

This subsection gives the full discussion of the tercile pattern summarized in Section~\ref{sec:rf_reliability}. The difference closes in the tercile where the contributor-pool composition of the two feeds differs most, since the domestic-foreign contrast in the repeat-author share of Table~\ref{tab:h1_contrasts} is widest in tercile 3. The model ties the size of the loss-on-noise slope to what investors know about a stock from sources other than the feed, the outside knowledge about the stock. The precision of investors' outside knowledge about the stocks of a group ($\tau_0$ in equation~\eqref{eq:peak}) is the sum of the prior precision ($\tau_d$), the price-signal precision ($\tau_p$), and the informed-trading intensity ($K_j$), and it takes one value for each geography at each tercile. Investors give less weight to the feed for a stock about which their outside knowledge is stronger. Hence, the price responds less to the feed ($G_j$ in equation~\eqref{eq:coefs}). The feed's noise then matters less for the stock's next-quarter losses (Section~\ref{sec:model_micro}). The loss-on-noise slope is closer to zero. The share of quarters with a negative slope is blind to the magnitude of a nonzero slope. A slope that the model places near zero, however, has its sign decided by sampling error in each quarterly cross-section. Its share is then near one half. Thus, the model implies a lower share of quarters with a negative slope for the stocks about which investors have stronger outside knowledge. Which tercile holds those stocks depends on which part of $\tau_0$ varies across the terciles. Holding the other group primitives fixed, the precision of outside knowledge rises with the informed-trading intensity (equation~\eqref{eq:peak}). The ranking statistic of Section~\ref{sec:model_micro} forms the terciles by that intensity inside each geography-quarter. Investors' outside knowledge is then strongest about the stocks in tercile 3. Hence, the share of quarters with a negative slope declines from tercile 1 to tercile 3. However, at the headline estimates, the precision of outside knowledge is almost entirely the prior precision, which is largest for the stocks in tercile 1 and smallest for the stocks in tercile 3 in both geographies. Investors' outside knowledge is then strongest about the stocks in tercile 1. Hence, the share of quarters with a negative slope rises from tercile 1 to tercile 3. Against the data, the two readings fare as follows. For foreign stocks the share rises at both steps, which agrees with the reading at the estimates at every step and with the intensity-only reading at none. For domestic stocks the share rises from tercile 1 to tercile 2, which agrees with the reading at the estimates, and is lowest in tercile 3, which agrees with the intensity-only reading. Neither reading accounts for the domestic pattern as a whole.

\section{Secondary verification of composition}\label{app:composition_check}

We calibrate the contributor composition against concentrations reported in independent Twitter samples. The exercise is descriptive, and the results in Sections~\ref{sec:structural} and~\ref{sec:reducedform} do not depend on it. The external ratios agree with the ordering we assume in the dispersion identification, with specialists the most concentrated posters and attention-driven posters the least.

\textit{Bot prevalence over the sample period.}\label{app:bot_prevalence} In the platform's own disclosures, false or spam accounts were below five percent of monetizable daily active users, its reported measure of its active user base. In an independent audit of September 2022, Similarweb found the same figure in its own account-level measurements, and attributed roughly twenty to thirty percent of U.S. content on the platform to bots (similarweb.com/corp/blog/insights/twitter-bot-research, accessed June 2026). These are industry measurements rather than peer-reviewed estimates, and detecting bots is subject to substantial classification error. In \citet{alarfaj2023twitter}, the best content-feature classifiers are about seventy-seven percent accurate, and weaker combinations of classifier and features are below fifty percent.

We compare the repeat-author share (Proposition~\ref{prop:dispersion}) with the posting concentration (the ratio of a group's account share to its tweet share, the population-level analogue of dispersion) reported for the three contributor populations (specialists, attention-driven posters, and bots) in independent data. The ratio is small when a group posts heavily and near one when it posts once. We take these ratios from three sources. For specialists and attention-driven posters we use the experience field of \citet{fallahgoul2020covid}. Self-reported professional users are $4.72$ percent of users and $8.73$ percent of messages (a ratio of $0.54$). Novice users are $4.92$ percent of users and $5.10$ percent of messages (a ratio of $0.96$). For bot accounts we use the ground-truth annotation of \citet{gilani2017bots}. Labeled bot accounts are $43.14$ percent of accounts and $51.85$ percent of statuses (a ratio of $0.83$). We see the same in the per-account counts, with $305$ tweets for bot accounts against $231$ for human accounts.

After we normalize the bot ratio to one, the relative account-to-tweet-share ratios for specialists, bot accounts, and attention-driven posters are $0.65$, $1.00$ and $1.16$, respectively. A feed can have higher noise even when a few informed users are posting. For validation, we use the account-to-tweet ratios from outside and use them to compare the direction of the difference in contributor composition behind foreign stocks' higher nonfundamental variance ($\Phi_j$) with those from our structural estimates. Our shares implied by dispersion agree with them. However, our main results do not depend on that agreement with the outside parameters. In the main model we assign specialists and bot accounts a common repeat intensity ($\lambda_I$, the per-account posting rate of a repeat group divided by the rate of an account that posts once), even though the three ratios differ. Specialists are $4.72$ percent of users but $8.73$ percent of messages (post $1.85$ times the average). Novice users are $4.92$ percent of users and $5.10$ percent of messages (post $1.04$ times the average). Automated accounts are $43.14$ percent of accounts but $51.85$ percent of posts (post $1.20$ times the average) and from their raw counts of $305$ against $231$, we get $1.32$. Thus, $\lambda_I$ lies between about $1.2$ and $1.85$. We set the value of $\lambda_I$ in that range and show the estimation results across that range (Section~\ref{sec:structural}). A common value of $\lambda_I$ is not a problem, since every $\pi_j$ is multiplied by the same number and the comparison across groups stays the same.

\section{Proofs of the Propositions}\label{app:model_proofs}

Throughout, we fix a stock $f$ in group $j$. We drop the stock and group subscripts where a single stock and group are understood. We restore the group subscript in the steps where the contrast across groups is the point. We write $\gamma = r/(r-\upsilon)$ for the present-value multiplier, $\tau_d = \Sigma_{jd}^{-1}$, $K$, $\tau_p$, and $\tau_s$ for the four precisions, $T = \tau_d + \tau_p + K + \tau_s$ for total precision, $\overline{\Sigma} = 1/T$ for the posterior variance at that precision, and $\tau_0 \equiv \tau_d + \tau_p + K$. For the contributor results we write $\pi_{\mathbb{S}}$ and $\pi_{\mathbb{B}}$ for the specialist and bot account shares, and $\pi = \pi_{\mathbb{S}} + \pi_{\mathbb{B}}$ for the repeat-author share. The common repeat intensity is $\lambda_I > 1$. The tweet shares are $\omega_{\mathbb{S}}$, $\omega_{\mathbb{A}}$, and $\omega_{\mathbb{B}}$ for the specialist, attention, and bot types. The fad variance carried by attention-driven contributors is $\sigma_{b,j}^2$, and the variance of the bot content process is $\sigma_{\beta,j}^2$. The nonfundamental variance is $\Phi_j = \omega_{\mathbb{A},j}^2\sigma_{b,j}^2 + \omega_{\mathbb{B},j}^2\sigma_{\beta,j}^2$. Throughout, $r$ denotes the gross riskless rate and starred symbols denote model value units. All random variables are jointly normal and mutually independent unless stated otherwise.

\subsection{Proof of Proposition \ref{prop:dispersion}}\label{app:proof_dispersion}

Given the feed structure, on a day with $A_t$ posting accounts, specialists number $\pi_{\mathbb{S}} A_t$ and bot accounts number $\pi_{\mathbb{B}} A_t$, each posting $\lambda_I$ tweets. Attention-driven accounts number $(1-\pi_{\mathbb{S}}-\pi_{\mathbb{B}})A_t$ and post one tweet each. Total tweets are
\[
N_t = A_t\left[(\pi_{\mathbb{S}}+\pi_{\mathbb{B}})\lambda_I + (1-\pi_{\mathbb{S}}-\pi_{\mathbb{B}})\right] = A_t\left[\pi\lambda_I + (1-\pi)\right],
\]
with $\pi = \pi_{\mathbb{S}} + \pi_{\mathbb{B}}$. Dispersion, the vendor's ratio of unique accounts to tweets, is
\begin{equation}
D \;=\; \frac{A_t}{N_t} \;=\; \frac{1}{\pi\lambda_I + (1-\pi)} \;=\; \frac{1}{1 + \pi(\lambda_I - 1)},
\end{equation}
which is deterministic given $(\pi, \lambda_I)$ and strictly decreasing in $\pi$ for $\lambda_I > 1$, with range $D \in (1/\lambda_I, 1]$ as $\pi$ ranges over $[0,1)$. Solving $1/D = 1 + \pi(\lambda_I - 1)$ gives the inversion $\pi = (1-D)/\left[D(\lambda_I - 1)\right]$, equation \eqref{eq:pi_inversion}, with $\pi = 0$ at $D = 1$. Values $D \leq 1/\lambda_I$ would imply $\pi \geq 1$ and correspond to the $\pi = 1$ boundary. With heterogeneous repeat intensities $\lambda_{\mathbb{S}}$ and $\lambda_{\mathbb{B}}$ the same count gives $D = \left[1 + \pi_{\mathbb{S}}(\lambda_{\mathbb{S}} - 1) + \pi_{\mathbb{B}}(\lambda_{\mathbb{B}} - 1)\right]^{-1}$. Therefore, dispersion identifies the repeat-author contribution to tweets per account, $\pi_{\mathbb{S}}(\lambda_{\mathbb{S}}-1) + \pi_{\mathbb{B}}(\lambda_{\mathbb{B}}-1)$, and the common-intensity case is the two-parameter reduction used in estimation. Hence, in either case the split of the repeat margin between $\pi_{\mathbb{S}}$ and $\pi_{\mathbb{B}}$ does not enter the count and is not identified from dispersion, which is the final claim. The tweet shares follow by the same counting, $\omega_{\mathbb{S}} = \pi_{\mathbb{S}}\lambda_I/\left[\pi\lambda_I + 1 - \pi\right]$, $\omega_{\mathbb{B}} = \pi_{\mathbb{B}}\lambda_I/\left[\pi\lambda_I + 1 - \pi\right]$, and $\omega_{\mathbb{A}} = (1-\pi)/\left[\pi\lambda_I + 1 - \pi\right]$, which sum to one, and the combined repeat tweet share is $\omega = \omega_{\mathbb{S}} + \omega_{\mathbb{B}} = \pi\lambda_I/\left[\pi\lambda_I + 1 - \pi\right]$. The derivation above treats the posting counts as fixed. If instead accounts post a random (Poisson) number of tweets, $D$ becomes a ratio of two random sums. The expectation of a ratio is not exactly the ratio of expectations. The gap is a correction of order $1/A_t$, which shrinks to zero as the number of posting accounts $A_t$ grows. Hence, the deterministic formula holds in the large-account limit.\footnote{For independent numerator and denominator, $E[A_t/N_t] = E[A_t]/E[N_t] + O(1/A_t)$ by a Taylor expansion of the ratio around the means, since both sums scale with $A_t$ and their coefficients of variation fall like $1/\sqrt{A_t}$.} The quarterly mean $\overline{D}$ used in estimation averages this ratio over days. $\square$

\subsection{Proof of Proposition \ref{prop:vaqsv}}\label{app:proof_vaqsv}

We first record an averaging fact used throughout. The mean of $n$ independent draws, each of variance $\sigma_e^2$, has variance $\sigma_e^2/n$, because the variance of the sum is $n\sigma_e^2$ and dividing by $n$ scales it by $1/n^2$.

Equation \eqref{eq:sbar_body} gives the day-$t$ sentiment, restated here,
\begin{equation}\label{eq:sbar_decomp}
\bar{s}_t \;=\; (1-\omega_{\mathbb{B},j})\,\epsilon_{f1} \;+\; \omega_{\mathbb{A},j}\, b_t \;+\; \omega_{\mathbb{B},j}\, \beta_t \;+\; \bar{u}_t,
\end{equation}
where $\bar{u}_t$ collects the sampling noise. Each tweet carries per-tweet noise of variance $\sigma_e^2$, and $\bar{s}_t$ averages all $N_t$ of them. Thus, $\bar{u}_t$ is the mean of $N_t$ such draws, and $\mathrm{Var}(\bar{u}_t) = \sigma_e^2/N_t$.

Take the variance of \eqref{eq:sbar_decomp} for a single day, conditional on the innovation $\epsilon_{f1}$. The first term is then constant and drops out. Under \ref{assump:dist} the fad $b_t$, the bot process $\beta_t$, and the noise $\bar{u}_t$ are mutually independent, so
\begin{equation}
\mathrm{Var}(\bar{s}_t \mid \epsilon_{f1}) \;=\; \omega_{\mathbb{A},j}^2\,\sigma_{b,j}^2 \;+\; \omega_{\mathbb{B},j}^2\,\sigma_{\beta,j}^2 \;+\; \frac{\sigma_e^2}{N_t} \;=\; \Phi_j \;+\; \frac{\sigma_e^2}{N_t},
\end{equation}
where $\Phi_j \equiv \omega_{\mathbb{A},j}^2\sigma_{b,j}^2 + \omega_{\mathbb{B},j}^2\sigma_{\beta,j}^2$ is the nonfundamental variance.

The vendor's series adds a weighting refinement to the sampling term. The series $S_t$ is an exponentially time-weighted sum of the day's tweet scores. The per-tweet average $\bar{s}_t = S_t/N_t$ is then a weighted mean of the $N_t$ scores. With weights $w_m$, the sampling variance of that weighted mean is $\sigma_e^2 \sum_m w_m^2/(\sum_m w_m)^2$, which equals $c_{1,t}\,\sigma_e^2/N_t$ with $c_{1,t} \equiv N_t \sum_m w_m^2/(\sum_m w_m)^2 \geq 1$.\footnote{With equal weights $\sum_m w_m^2 = N_t$ and $(\sum_m w_m)^2 = N_t^2$, so $c_{1,t} = 1$. Any departure from equal weights raises it, by the Cauchy-Schwarz inequality $(\sum_m w_m)^2 \leq N_t \sum_m w_m^2$, with equality only when the weights are equal.} The factor $c_{1,t}$ is one under equal weights and depends on the day's weight profile in general. Proposition~\ref{prop:vaqsv} treats it as a single constant $c_w$ common across days under its stable-profile condition. The exact statement replaces $c_w\overline{(1/N_t)}$ by the quarter average of $c_{1,t}/N_t$. The constant-$c_w$ form is the empirical specification. We do not separate $c_w$ from $\sigma_e^2$, since only the product $c_w\sigma_e^2$, the sampling-noise coefficient, enters anything the paper reports. Writing $v_t \equiv \Phi_j + c_w\sigma_e^2/N_t$ for the daily conditional variance and $\bar{v}_{fq} \equiv T_q^{-1}\sum_{t\in q} v_t = \Phi_j + c_w\sigma_e^2\,\overline{(1/N_t)}$ for its within-quarter average completes the daily object.

The measured $\mathrm{VAQSV}_{fq}^2$ is a realized sample variance, computed around the sample mean of the observed daily readings, not the average daily variance $\bar{v}_{fq}$ itself. The two coincide only in conditional mean and only under cross-day independence. Let $d_t \equiv \bar{s}_t - (1-\omega_{\mathbb{B},j})\epsilon_{f1}$ be the day-$t$ deviation from the constant, so that conditional on $\epsilon_{f1}$ the readings equal that constant plus $d_t$ and $\mathrm{Var}(d_t) = v_t$. For the sample variance around the estimated mean,
\begin{equation}\label{eq:vaqsv_cond}
\begin{split}
E\!\left[\mathrm{VAQSV}_{fq}^2 \mid \epsilon_{f1}, \{N_t\}\right] &=  \frac{1}{T_q}\sum_{t \in q} v_t \;-\; \frac{2}{T_q(T_q-1)}\sum_{t<t'}\mathrm{Cov}(d_t, d_{t'}) \\
&=  \bar{v}_{fq} \;-\; \frac{2}{T_q(T_q-1)}\sum_{t<t'}\mathrm{Cov}(d_t, d_{t'}).
\end{split}
\end{equation}
The within-quarter standard deviation is computed with the $T_q-1$ divisor. Hence, the sample variance is unbiased for the average daily variance under \ref{assump:dist}, and no degrees-of-freedom correction enters the mapping.\footnote{For deviations $d_t$ with $\sum_t (d_t - \bar{d})^2$ divided by $T_q-1$, the expectation of $\sum_t (d_t-\bar{d})^2$ is $(1-1/T_q)\sum_t \mathrm{Var}(d_t) - (2/T_q)\sum_{t<t'}\mathrm{Cov}(d_t,d_{t'})$, and dividing by $T_q-1$ gives \eqref{eq:vaqsv_cond}.} Under \ref{assump:dist} the daily deviations are independent across days, all covariances are zero, and the conditional mean reduces to $\bar{v}_{fq} = \Phi_j + c_w\sigma_e^2\,\overline{(1/N_t)}$, which is equation \eqref{eq:vaqsv}. Estimating \eqref{eq:vaqsv} across firm-quarters by ordinary least squares of $\mathrm{VAQSV}_{fq}^2$ on $\overline{(1/N_t)}$ recovers the intercept $\Phi_j$ and the slope $c_w\sigma_e^2$, because least squares fits this conditional mean.

A one-parameter version of $\Phi_j$ is available under a restriction. Impose that bot tweets carry no nonfundamental variance of their own, that is $\sigma_{\beta,j} = 0$. The second term of $\Phi_j$ then drops and $\Phi_j = \omega_{\mathbb{A},j}^2\sigma_{b,j}^2$. The three tweet shares sum to one, so with $\omega \equiv \omega_{\mathbb{S},j} + \omega_{\mathbb{B},j}$ the combined specialist and bot share, $\omega_{\mathbb{A},j} = 1-\omega$ and $\Phi_j = (1-\omega)^2\sigma_{b,j}^2$. The estimation baseline imposes this restriction and the attribution convention of Section~\ref{sec:model_sent}. The two-term composite is the general object it restricts. The falsification battery probes departures from the attribution (see Section~\ref{sec:reducedform}).

The precision in \eqref{eq:tau_s} comes from the quarterly average of the daily series. Averaging \eqref{eq:sbar_decomp} over the quarter's $T_q$ days gives $s = (1-\omega_{\mathbb{B},j})\epsilon_{f1} + u$, where $u = T_q^{-1}\sum_{t\in q} d_t$ is the average of the daily deviations and carries the signal's noise. Its variance is
\begin{equation}
\mathrm{Var}(u) \;=\; \frac{1}{T_q^2}\left[\,\sum_{t\in q}\mathrm{Var}(d_t) \;+\; \sum_{t\neq t'}\mathrm{Cov}(d_t,d_{t'})\,\right] \;=\; \frac{1}{T_q}\left[\,\bar{v}_{fq} \;+\; \frac{2}{T_q}\sum_{t<t'}\mathrm{Cov}(d_t,d_{t'})\,\right].
\end{equation}
Define the dependence multiplier $\kappa_{0,fq}$ against the average daily variance $\bar{v}_{fq}$,
\begin{equation}\label{eq:kappa0}
\mathrm{Var}(u) \;=\; \kappa_{0,fq}\,\frac{\bar{v}_{fq}}{T_q}, \qquad \kappa_{0,fq} \;\equiv\; 1 \;+\; \frac{2}{T_q\,\bar{v}_{fq}}\sum_{t<t'}\mathrm{Cov}(d_t,d_{t'}) \;\geq\; 1,
\end{equation}
which equals one under the independence of \ref{assump:dist} and exceeds one under positive serial correlation.\footnote{We restrict attention to non-negative serial correlation. The covariance sum, and with it the second term, is then non-negative. Negative serial correlation would require sentiment to reverse systematically from one day to the next, against how attention and coordinated posting behave, since a fad builds and decays gradually and a bot campaign runs over consecutive days. The multiplier is defined against the average daily variance $\bar{v}_{fq}$ rather than the realized $\mathrm{VAQSV}_{fq}^2$, which keeps the serial-dependence adjustment separate from the sampling of the variance itself.} The reader's signal about $\epsilon_{f1}$ rescales $s$ by $(1-\omega_{\mathbb{B},j})^{-1}$, so its precision in the model's units is
\begin{equation}
\tau_s \;=\; \frac{(1-\omega_{\mathbb{B},j})^2}{\mathrm{Var}(u)} \;=\; \frac{(1-\omega_{\mathbb{B},j})^2\,T_q}{\kappa_{0,fq}\,\bar{v}_{fq}}.
\end{equation}
The population variance $\bar{v}_{fq}$ is not observed. Combining \eqref{eq:vaqsv_cond} with \eqref{eq:kappa0} gives
\begin{equation}\label{eq:vaqsv_bias}
E\!\left[\mathrm{VAQSV}_{fq}^2 \mid \epsilon_{f1}, \{N_t\}\right] \;=\; \bar{v}_{fq}\,\frac{T_q - \kappa_{0,fq}}{T_q - 1},
\end{equation}
so the realized statistic is unbiased for $\bar{v}_{fq}$ under \ref{assump:dist}, where $\kappa_{0,fq} = 1$, and understates it when the serial dependence $\kappa_{0,fq}$ matters. Putting the realized statistic in place of the population variance is therefore exact only under independence. We do not use the decomposition above in the estimation. Instead we define a single conversion factor $\kappa$ through equation \eqref{eq:tau_s}, $\tau_s = T_q/(\kappa\,\mathrm{VAQSV}_{fq}^2)$, and use it to turn the observable signal-quality index into the signal's precision in the model's units. We hold $\kappa$ common across firm-quarters. Its components are the serial-dependence multiplier $\kappa_{0,fq}$ (\eqref{eq:vaqsv_bias}), the bot-dilution factor $(1-\omega_{\mathbb{B},j})^{-2}$, and the squared scale between the vendor's per-tweet score and the cash-flow innovation (from normalizing the specialist loading to one in \eqref{eq:sbar_decomp}). These are not separately identified, and \ref{app:kappa_cases} takes up which of them can differ across groups. $\square$
\subsection{The conversion constant and the bot share across groups}\label{app:kappa_cases}

We compare the feed's precision for domestic and foreign stocks and for the three terciles of informed-trading-intensity. The conversion factor ($\kappa$) has three components---the multiplier for serial dependence in the daily series ($\kappa_{0,fq}$), the bot-dilution factor ($(1-\omega_{\mathbb{B},j})^{-2}$), and the scale which converts the vendor's per-tweet sentiment score into the innovation's cash-flow units and which is normalized to one (equation~\eqref{eq:sbar_body}). $\kappa$ differs across these groups only if one of the three components differs across these groups. The scale does not differ across the groups as it has been normalized to one. Since the serial dependence in the daily series is small and almost the same for foreign and domestic stocks (\ref{app:serial_dependence}), the serial-dependence multiplier does not differ across the groups too. Hence, of the three components, only the bot-dilution factor can differ across the groups (through the bot tweet share ($\omega_{\mathbb{B},j}$)). Hence, $\kappa$ can differ across the groups only through the bot share. However, a difference in the bot share moves the domestic-foreign comparison only when it is systematic (when the bot share differs between domestic and foreign stocks, or across the terciles). The domestic-foreign comparison is not affected by a bot share that is common across the groups but changes over time, because the bot share is absorbed by the quarter fixed effects in our reduced-form regressions and we treat it as a common term in the structural estimation. We test for such a systematic difference in the bot share with a falsification battery and find no evidence of one. Its dispersion coefficient is statistically insignificant (row R13a, Section~\ref{sec:reducedform}).

We conduct sensitivity analysis to put a bound on the extent to which a systematic domestic-foreign difference in the bot share could reach while holding the other two components (the scale and the serial-dependence multiplier) common across the groups. At a common $\kappa$, the feed for domestic stocks is more precise than the feed for foreign stocks (by a median ratio of about $1.17$ in the pooled sample). This would reverse only if the feed for foreign stocks were less diluted by automated content than the feed for domestic stocks, which happens only if the share of the foreign feed which has the firm's signal ($1-\omega_{\mathbb{B},F}$) were about 8 percent higher than the share of the domestic feed. However, we find that that doesn't hold, and in fact, its opposite holds. Foreign stocks carry heavier automation, so their $\kappa_F$ is higher than the common $\kappa$. Hence, the feed for foreign stocks is less precise than it is at a common $\kappa$.

\subsection{Proof of Proposition \ref{prop:price}}\label{app:proof_price}

\textit{Posterior beliefs.} Conjecture the price function \eqref{eq:price}. Investor $i$ observes the prior, a private signal $\eta^i = \epsilon + e^i$ of precision $k^i$, with $e^i \sim N(0, 1/k^i)$, the public signal $s = \epsilon + u$ with $u \sim N(0, \tau_s^{-1})$, and the price. Given the conjecture, the pair $(P, s)$ is informationally equivalent to $(\zeta, s)$ where
\begin{equation}\label{eq:zeta}
\zeta \;\equiv\; \frac{rP - A - G\,s}{B - G} \;=\; \epsilon \;+\; \frac{C}{B - G}\,\tilde{x},
\end{equation}
a signal about $\epsilon$ with precision $\tau_p = \left[(B-G)/C\right]^2 \Sigma_{jx}^{-1}$.\footnote{From the displayed expression, the noise in $\zeta$ is $\frac{C}{B-G}\tilde{x}$ with $\tilde{x} \sim N(0,\Sigma_{jx})$, so its variance is $\left[C/(B-G)\right]^2 \Sigma_{jx}$ and its precision is the reciprocal, $\left[(B-G)/C\right]^2 \Sigma_{jx}^{-1}$.} The construction nets the publicly observed $s$ out of the price, which removes the common trading on $s$ from the price's incremental information. Bayesian updating with normal priors and signals gives posterior variance $\Sigma^i = (\tau_d + \tau_p + k^i + \tau_s)^{-1}$ and posterior mean $E^i[\epsilon] = \Sigma^i\left(\tau_p \zeta + k^i \eta^i + \tau_s s\right)$.

\textit{Demands and market clearing.} With $V = \gamma(\upsilon d_0 + \bar{\epsilon} + \epsilon)$ and mean-variance preferences, demand is $q^i = \left(E^i[V] - rP\right)/\left(\bar{\rho}\gamma^2 \Sigma^i\right)$.\footnote{An investor with risk aversion $\bar{\rho}$ choosing a share to hold maximizes $E^i[V]q - rPq - \frac{1}{2}\bar{\rho}\,\mathrm{Var}^i[V]\,q^2$. The first-order condition is $E^i[V] - rP - \bar{\rho}\,\mathrm{Var}^i[V]\,q = 0$, and $\mathrm{Var}^i[V] = \gamma^2 \Sigma^i$ because $V = \gamma(\upsilon d_0 + \bar{\epsilon} + \epsilon)$ and only $\epsilon$ is uncertain with posterior variance $\Sigma^i$. Solving gives the stated demand.} Imposing $\int q^i\,di = \bar{x} + \tilde{x}$, multiplying through by $\bar{\rho}\gamma^2$, and using $\int (\Sigma^i)^{-1} di = \overline{\Sigma}^{-1}$ together with $\int k^i \eta^i\,di = K\epsilon$ (idiosyncratic signal noise averages out across the continuum),
\begin{equation}
\gamma\left(\upsilon d_0 + \bar{\epsilon}\right)\overline{\Sigma}^{-1} \;+\; \gamma\left(\tau_p \zeta + K\epsilon + \tau_s s\right) \;-\; rP\,\overline{\Sigma}^{-1} \;=\; \bar{\rho}\gamma^2\left(\bar{x} + \tilde{x}\right),
\end{equation}
so that
\begin{equation}\label{eq:rP_solved}
rP \;=\; \gamma\left(\upsilon d_0 + \bar{\epsilon}\right) \;+\; \gamma\overline{\Sigma}\left(\tau_p \zeta + K\epsilon + \tau_s s\right) \;-\; \bar{\rho}\gamma^2\overline{\Sigma}\left(\bar{x} + \tilde{x}\right).
\end{equation}
Substituting $\zeta = \epsilon + \frac{C}{B-G}\tilde{x}$ and $s = \epsilon + u$ and matching coefficients with the conjecture,
\begin{align}
A &= \gamma\left(\upsilon d_0 + \bar{\epsilon}\right) - \bar{\rho}\gamma^2\overline{\Sigma}\,\bar{x}, \label{eq:Acoef}\\
B &= \gamma\overline{\Sigma}\left(\tau_p + K + \tau_s\right) \;=\; \gamma\left(1 - \overline{\Sigma}\,\tau_d\right) \;=\; \gamma\left(1 - \overline{\Sigma}/\Sigma_{jd}\right), \label{eq:Bcoef}\\
G &= \gamma\overline{\Sigma}\,\tau_s, \label{eq:Gcoef}\\
C &= \gamma\overline{\Sigma}\,\tau_p\,\frac{C}{B-G} \;-\; \bar{\rho}\gamma^2\overline{\Sigma}. \label{eq:Cfix}
\end{align}
From \eqref{eq:Bcoef} and \eqref{eq:Gcoef}, $B - G = \gamma\overline{\Sigma}(\tau_p + K) > 0$, so $\gamma\overline{\Sigma}\tau_p/(B-G) = \tau_p/(\tau_p + K)$ and \eqref{eq:Cfix} becomes $C\,K/(\tau_p + K) = -\bar{\rho}\gamma^2\overline{\Sigma}$, that is
\begin{equation}\label{eq:Csol}
C \;=\; -\,\bar{\rho}\,\gamma^2\,\overline{\Sigma}\,\frac{\tau_p + K}{K} \;<\; 0 .
\end{equation}
Then $(B - G)/C = -K/(\bar{\rho}\gamma)$, hence
\begin{equation}\label{eq:taup_sol}
\tau_p \;=\; \left(\frac{B-G}{C}\right)^{2}\Sigma_{jx}^{-1} \;=\; \frac{K^2}{\gamma^2\bar{\rho}^2\,\Sigma_{jx}},
\end{equation}
a closed form free of $\tau_s$ and of any fixed point. $\tau_p$ is determined by primitives, $\overline{\Sigma} = (\tau_d + \tau_p + K + \tau_s)^{-1}$ is then explicit, and \eqref{eq:Acoef} through \eqref{eq:Csol} deliver all coefficients uniquely within the linear class, establishing existence and the stated independence of $\tau_p$ from $\tau_s$.

\textit{Moments of the quarterly excess return.} With $R = V - rP$ and \eqref{eq:Acoef}, $R = (\gamma - B)\epsilon - G u - C\tilde{x} + \bar{\rho}\gamma^2\overline{\Sigma}\bar{x}$ and $\gamma - B = \gamma\overline{\Sigma}\tau_d$, so $E[R] = \bar{\rho}\gamma^2\overline{\Sigma}\bar{x}$ and
\begin{equation}
\mathrm{Var}[R] \;=\; \gamma^2\overline{\Sigma}^2\tau_d^2\,\Sigma_{jd} \;+\; \gamma^2\overline{\Sigma}^2\tau_s^2\,\tau_s^{-1} \;+\; C^2\Sigma_{jx} \;=\; \gamma^2\overline{\Sigma}^2\left(\tau_d + \tau_s\right) \;+\; C^2\Sigma_{jx}.
\end{equation}
Using \eqref{eq:Csol} and $\Sigma_{jx} = K^2/(\gamma^2\bar{\rho}^2\tau_p)$ from \eqref{eq:taup_sol}, $C^2\Sigma_{jx} = \gamma^2\overline{\Sigma}^2(\tau_p + K)^2/\tau_p$, which gives equation \eqref{eq:varR}.

\textit{Monotonicity in $\tau_s$.} Write $\mathrm{Var}[R] = \gamma^2 N(\tau_s)/T(\tau_s)^2$ with $N = \tau_d + \tau_s + (\tau_p + K)^2/\tau_p$ and $T = \tau_0 + \tau_s$, both free of $\tau_s$ elsewhere by \eqref{eq:taup_sol}. Then $\partial \mathrm{Var}[R]/\partial \tau_s = \gamma^2\left(T - 2N\right)/T^3$, and since $(\tau_p + K)^2/\tau_p \geq \tau_p + K$,\footnote{Dividing both sides by $\tau_p + K > 0$, the inequality is $(\tau_p + K)/\tau_p \geq 1$, which holds because $K \geq 0$.}
\begin{equation}
T - 2N \;=\; (\tau_p + K) - \tau_d - \tau_s - 2\,\frac{(\tau_p+K)^2}{\tau_p} \;\leq\; -\left(\tau_d + \tau_s + \tau_p + K\right) \;=\; -T \;<\; 0,
\end{equation}
so $\mathrm{Var}[R]$ is strictly decreasing in $\tau_s$. $\square$

\subsection{Proof of Proposition \ref{prop:pinf}}\label{app:proof_pinf}

The asset's earnings stream from the pricing date $q$ is $E_{f,q+s} = \upsilon^{s} d^{*}_{f1}$ for $s \geq 0$, the Gordon stream consistent with $V^{*}_{f1} = \gamma d^{*}_{f1}$, since $\sum_{s\geq 0}(\upsilon/r)^s = r/(r-\upsilon)$. Its firm-specific innovation is $\upsilon^{s}\epsilon$. The price varies only through its shock terms $(B\epsilon + Gu + C\tilde{x})/r$, conditional on $d_0$ and the common component. Therefore
\begin{equation}
\mathrm{Cov}\!\left(E_{f,q+s},\, P^{*}_{f1}\right) \;=\; \upsilon^{s}\,\frac{B}{r}\,\Sigma_{jd} \;=\; \upsilon^{s}\,\frac{1 - \overline{\Sigma}/\Sigma_{jd}}{\,r - \upsilon\,}\,\Sigma_{jd},
\end{equation}
using $B/r = \left(1 - \overline{\Sigma}/\Sigma_{jd}\right)/(r-\upsilon)$ from \eqref{eq:Bcoef}. Dividing by $\mathrm{SD}(P^{*}_{f1})$ gives equation \eqref{eq:pinf}. For the information component, $1 - \overline{\Sigma}/\Sigma_{jd} = (\tau_p + K + \tau_s)/T$\footnote{Since $\overline{\Sigma} = 1/T$ with $T = \tau_d + \tau_p + K + \tau_s$, and the prior precision is $\tau_d = 1/\Sigma_{jd}$, we have $1 - \overline{\Sigma}/\Sigma_{jd} = 1 - \tau_d/T = (T - \tau_d)/T = (\tau_p + K + \tau_s)/T$.} and
\begin{equation}
\frac{\partial}{\partial \tau_s}\,\frac{\tau_p + K + \tau_s}{T} \;=\; \frac{T - (\tau_p + K + \tau_s)}{T^2} \;=\; \frac{\tau_d}{T^2} \;>\; 0,
\end{equation}
so the information component is strictly increasing in $\tau_s$. $\square$

\subsection{Proof of Proposition \ref{prop:vol}}\label{app:proof_vol}

For $X \sim N(\mu, \sigma^2)$, substituting $z = (x-\mu)/\sigma$ and using $\int_a^{\infty} z\,\phi(z)\,dz = \phi(a)$,
\begin{equation}\label{eq:truncids}
E[\max(X,0)] = \mu\,\Phi\!\left(\tfrac{\mu}{\sigma}\right) + \sigma\,\phi\!\left(\tfrac{\mu}{\sigma}\right), \qquad
E[\min(X,0)] = \mu\,\Phi\!\left(-\tfrac{\mu}{\sigma}\right) - \sigma\,\phi\!\left(\tfrac{\mu}{\sigma}\right),
\end{equation}
whose sum is $\mu$. Under \ref{assump:daily}, the quarterly gain and loss are the sums of the positive and negative parts of the $T_{days}$ i.i.d.\ daily excess returns, so $E[\mathrm{Gain}] - E[|\mathrm{Loss}|] = T_{days}\,\mu_d$. Differentiating \eqref{eq:truncids} with respect to $\sigma$ and using $\phi'(z) = -z\phi(z)$,
\begin{equation}
\frac{\partial\,E[\mathrm{Gain}]}{\partial \sigma_d} \;=\; \frac{\partial\,E[|\mathrm{Loss}|]}{\partial \sigma_d} \;=\; T_{days}\,\phi\!\left(\tfrac{\mu_d}{\sigma_d}\right) \;>\; 0,
\end{equation}
so both tails rise with daily volatility at the same rate. Differentiating \eqref{eq:truncids} with respect to $\mu$ instead gives $\partial E[\mathrm{Gain}]/\partial \mu_d = T_{days}\,\Phi(\mu_d/\sigma_d)$ and $\partial E[|\mathrm{Loss}|]/\partial \mu_d = -T_{days}\,\Phi(-\mu_d/\sigma_d)$. The mean daily excess return $\mu_d$ moves the two tails in opposite directions, which is why conditioning on the observed daily volatility isolates it. Since $T_{days}\sigma_d^2 = \mathrm{Var}[R]$ under \ref{assump:daily} and $\mathrm{Var}[R]$ is strictly decreasing in $\tau_s$ (\ref{app:proof_price}), at fixed $\mu_d$ both $E[\mathrm{Gain}]$ and $E[|\mathrm{Loss}|]$ are strictly decreasing in $\tau_s$ through the volatility channel. $\square$

\subsection{Proof of Proposition \ref{prop:drift}}\label{app:proof_drift}
Under \ref{assump:ss}(ii), the part of the optimistic price impact that survives the quarter is $\delta\,G\,\max(u,0)$. Under \ref{assump:ss}(i), pessimistic realizations are corrected within the quarter and leave no persistent term. The terminal value $V^{*}_{f1}$ is realized at the dividend date. The dividend date is the model counterpart of the next earnings date in the data. Since no buyer pays more than a known payoff, no nonfundamental price component survives that revelation. Therefore, expected overpricing equals the expected correction over the following quarter, \begin{equation} \mathrm{OP} \;=\; \delta\,G\,E[\max(u,0)] \;=\; \delta\,G\,\frac{\sigma_u}{\sqrt{2\pi}} \;=\; \frac{\delta\,\gamma\,\overline{\Sigma}\,\tau_s\,\tau_s^{-1/2}}{\sqrt{2\pi}} \;=\; \frac{\delta\,\gamma}{\sqrt{2\pi}}\,\frac{\sqrt{\tau_s}}{\tau_0 + \tau_s}.
\end{equation}
The first equality takes the mean of the surviving component, with the constants $\delta$ and $G$ outside the expectation. The second applies \eqref{eq:truncids} at $\mu = 0$, where the term $\mu\Phi(\mu/\sigma)$ vanishes and the term $\sigma\phi(\mu/\sigma)$ becomes $\sigma_u\,\phi(0) = \sigma_u/\sqrt{2\pi}$, since $\phi(0) = 1/\sqrt{2\pi}$. The third substitutes $\sigma_u = \tau_s^{-1/2}$, the standard deviation of $u$, and $G = \gamma\overline{\Sigma}\tau_s$ from \eqref{eq:coefs}. The fourth uses $\overline{\Sigma} = 1/(\tau_0 + \tau_s)$ and $\tau_s\,\tau_s^{-1/2} = \sqrt{\tau_s}$. This is equation \eqref{eq:op}, and $\mu^{drift} = -\mathrm{OP}$ accrues across the next quarter's daily excess returns under \ref{assump:daily}. For the shape, write $\mathrm{OP} = (\delta\gamma/\sqrt{2\pi})\,f(\tau_s)$ with $f(\tau_s) = \sqrt{\tau_s}/(\tau_0 + \tau_s)$. Its derivative is
\begin{equation}
f'(\tau_s) \;=\; \frac{\tau_0 - \tau_s}{2\sqrt{\tau_s}\,\left(\tau_0 + \tau_s\right)^2}.
\end{equation}
The derivative is positive for $\tau_s < \tau_0$, zero at $\tau_s = \tau_0$, and negative for $\tau_s > \tau_0$. The function $f$ tends to zero as $\tau_s \to 0$ and as $\tau_s \to \infty$. Therefore, expected overpricing is a hump with an interior peak at $\tau_s^{*} = \tau_0$. This is equation \eqref{eq:peak}.

In measured-noise units, substitute $\tau_s = T_q/(\kappa w^2)$ from \eqref{eq:tau_s} with $w = \mathrm{VAQSV}_{fq}$ and let $a \equiv T_q/\kappa$, so that
\begin{equation}\label{eq:OPw}
\mathrm{OP}(w) \;=\; \frac{\delta\gamma\sqrt{a}}{\sqrt{2\pi}}\;\frac{w}{\tau_0\,w^2 + a}, \qquad
\frac{\partial\,\mathrm{OP}}{\partial w} \;=\; \frac{\delta\gamma\sqrt{a}}{\sqrt{2\pi}}\;\frac{a - \tau_0 w^2}{\left(\tau_0 w^2 + a\right)^2}.
\end{equation}
The peak in $w$ is at $w^{*} = \sqrt{a/\tau_0} = \sqrt{T_q/(\kappa\,\tau_0)}$. The peak is strictly decreasing in $\tau_0$. Holding $T_q$, $\kappa$, and the remaining group primitives fixed, the turning point is strictly decreasing in $K_j$, since $\tau_0 = \tau_d + K + K^2/(\gamma^2\bar{\rho}^2\Sigma_{jx})$ is strictly increasing in $K$ by \eqref{eq:taup_sol}.\footnote{Differentiating, $\partial \tau_0/\partial K = 1 + 2K/(\gamma^2\bar{\rho}^2\Sigma_{jx}) > 0$.} We therefore read the empirical ordering of turning points as consistent with this ceteris-paribus comparative static, not as mechanically implied by the tercile ranking alone. Across groups, $w^{*}$ carries the common factor $\sqrt{T_q/\kappa}$. For comparisons evaluated at a common $T_q$, the common conversion factor $\kappa$ does not affect the ordering. Finally, the correction realized in a given quarter is $\delta\,G\,u\,\mathbf{1}\{u > 0\}$. It is nonzero after optimistic realizations alone. The quarter's share of positive-sentiment days proxies these realizations. Therefore, returns following high-optimism quarters carry the overpricing correction. $\square$

\subsection{The symmetric-mechanism alternative and the two predictions}\label{app:symmetric}

Section~\ref{sec:model_gl} states the two predictions and the symmetric mechanism that is the alternative to the model. In each quarter the noise $u_{f1}$ is a single draw. The draw is positive in some quarters and negative in others. Symmetry imposes three equalities between the two signs. A positive draw and a negative draw are equally likely. A positive draw and a negative draw of the same size move the price by the same amount. Trading removes the same share of the price impact whether the draw is positive or negative. The share that trading does not remove is left in the price at quarter end. This remainder points upward after a positive draw and downward after a negative draw. The remainder is corrected in quarter $q+1$, when the value is revealed. The correction is a fall in the price after a positive draw and a rise after a negative draw. Under symmetry the two cases are equally likely. The expected correction in quarter $q+1$ is zero. Therefore, noise in quarter $q$ predicts no direction for quarter $q+1$ returns. The noise still moves the price while it is present. A larger amount of noise raises return volatility as well as both expected gains and expected losses (Proposition~\ref{prop:vol}). Therefore, under a symmetric mechanism, noise matters through its amount, not through its sign.

Section~\ref{sec:model_gl} states the first prediction, about shape, and why a symmetric mechanism produces no interior peak, and the second prediction, about sign. The optimistic case is a positive realization of the quarter $q$ signal noise $u_{f1}$. Its price impact $G_j u_{f1}$ is positive by \eqref{eq:price}. This impact is a part of the quarter $q$ price, and it is noise, not news about the dividend. While it is in the price, the price stays above the firm's value ($V^{*}_{f1}$) by that amount. As this part builds up over quarter $q$, the quarter's daily excess returns rise. A higher mean daily excess return raises the gains $\mathrm{Gain}_{fq}$ (the sum of the quarter's positive daily excess returns) and lowers the losses $|\mathrm{Loss}_{fq}|$ (the sum of the magnitudes of the negative ones), so the difference $|\mathrm{Loss}_{fq}| - \mathrm{Gain}_{fq}$ falls (\ref{app:proof_vol}). The higher gains are not the overpricing. They come only from the price rising during quarter $q$. The overpricing is the impact $G_j u_{f1}$ itself, the part of it still in the price at the end of quarter $q$. Trading over quarter $q$ corrects part of this impact, and under \ref{assump:ss}(ii) the fraction $\delta$ survives to the end of the quarter.

At the end of quarter ($q$), the price exceeds the firm's value ($V^{*}_{f1}$) by the surviving overpricing ($\delta\,G_j u_{f1}$). In the next quarter, when the firm's value becomes known, no buyer is willing to pay more than it. Hence, the price falls to $V^{*}_{f1}$. The fall occurs during quarter ($q+1$), on the days when the price adjusts to the announcement. Whether it is concentrated in one day or spread over several, it lowers the sum of quarter ($q+1$)'s daily excess returns by $\delta\,G_j u_{f1}$, the surviving overpricing. The quarter's mean daily excess return is this sum divided by the number of trading days ($T_{days}$). Thus, the mean falls. Under \ref{assump:daily}, the daily excess returns scatter around this lower mean. A lower mean shrinks $\mathrm{Gain}_{f,q+1}$ and raises $|\mathrm{Loss}_{f,q+1}|$ (\ref{app:proof_vol}). If the fall is instead concentrated in a few days, their negative returns enter the loss sum directly. In both cases the difference $|\mathrm{Loss}_{f,q+1}| - \mathrm{Gain}_{f,q+1}$ rises.

Quarter $q+1$ also carries its own new noise. Price movements from that noise that begin and reverse inside quarter $q+1$ raise $\mathrm{Gain}_{f,q+1}$ and $|\mathrm{Loss}_{f,q+1}|$ equally in expectation and cancel in the difference (Proposition~\ref{prop:vol}). The fall from quarter $q$'s overpricing has no offsetting rise inside quarter $q+1$, because the offsetting rise occurred in quarter $q$. Thus, one optimistic draw moves the two quarters in opposite directions, toward gains in quarter $q$ and toward losses in quarter $q+1$. Therefore, the response of the quarter $q+1$ outcome to $\mathrm{VAQSV}_{fq}$ should be stronger after quarters in which more of the sentiment days were positive.

In the tests, the share of positive-sentiment days among quarter $q$'s classified sentiment days proxies the sign of quarter $q$'s realization, measured on the vendor-classified days with thin coverage screened in the estimation (see Section~\ref{sec:reducedform}).

\subsection{The daily dealer market and the liquidity vector}\label{app:dealer_detail}

The informed-trading intensity $K_j$ is a precision parameter and is not observed in the data.\footnote{One common approach proxies informed-trading intensity with the share of a stock held by institutional investors. That proxy typically comes from subscription holdings databases and treats all institutional holders as informed.} This subsection recovers an ordinal ranking of stocks by $K_j$ from the daily dealer market, whose trading costs rise with it. In each trading day a dealer takes the other side of the day's buy and sell orders without knowing whether an order comes from an informed investor or an uninformed one. The dealer's two defenses are the bid-ask spread and the market-impact cost, the price change per share caused by the trader's own order.\footnote{The market-impact cost is a different object from the price impact of the signal noise in \eqref{eq:price}, which is a component of the price level. Throughout the paper, the price impact of a noise or supply realization refers to this price-level component. Price impact listed beside the spread as a liquidity measure refers to the dealer's market-impact object, the market-impact cost or its per-volume form, the price-impact ratio of Proposition~\ref{prop:micro}.} Both are the dealer's compensation for facing better-informed counterparties \citep{glosten1985bid, kyle1985continuous}. The dealer's exposure is the variance of the informed investor's estimate of the payoff-relevant value $v = \gamma_j\,\epsilon_{f1}$, conditional on the dealer's information set of the prior and the public signal (\ref{app:dealer_detail}),
\begin{equation}\label{eq:sigmaI}
\sigma_{I,j}^2 \;=\; \mathrm{Var}\!\left(E[v\,|\,\mathcal{I}_{inf}] \,\middle|\, \mathcal{I}_{dealer}\right)
\;=\; \gamma_j^2\, \frac{K_j}{(\tau_d + \tau_s)(\tau_d + \tau_s + K_j)}.
\end{equation}
The exposure rises with the private precision $K_j$ and falls with the public precision $\tau_s$. It falls because a more informative feed lets the dealer share in what the informed side knows. In the batch market of \citet{kyle1985continuous}, the informed investor's optimal position and the dealer's break-even condition tie the spread and the market-impact cost to this exposure and to the amount of noise trading, measured by its standard deviation $\sigma_{x,j}$ with $\sigma_{x,j}^2 = \Sigma_{jx}$ (\ref{app:dealer_detail}). Proposition~\ref{prop:micro} collects the liquidity measures that the dealer's problem implies.

\proposition{prop:micro}~\textit{(The liquidity vector.) With noise-trading standard deviation $\sigma_{x,j}$, the Kyle market-impact coefficient is $\lambda_j = \sigma_{I,j}/(2\sigma_{x,j})$. The price-impact ratio $\mathrm{impact}_j$ is the model counterpart of the Amihud (\citeyear{amihud2002illiquidity}) illiquidity measure. The spread is the model counterpart of the relative quoted spread. The expected adverse-selection cost per unit traded is proportional to $\sigma_{I,j}$ alone,}
\begin{equation}\label{eq:micro}
\mathrm{spread}_j = a_s\, \sigma_{I,j}, \qquad \mathrm{impact}_j = a_\lambda\, \frac{\sigma_{I,j}}{\sigma_{x,j}}, \qquad \mathrm{turnover}_j = a_T\, \sigma_{x,j},
\end{equation}
\textit{The positive constants $a_s$, $a_\lambda$, and $a_T$ are scale factors. The estimation sets $a_T = 1$ as a normalization, so $\sigma_{x,j}$ is measured in per-day turnover units. It estimates $a_s$ and $a_\lambda$ (see Section~\ref{sec:structural}). Hence the spread increases in $K_j$, decreases in $\tau_s$, and does not depend on $\Sigma_{jx}$, while the price-impact ratio also loads on the amount of noise trading. The spread and the price-impact ratio jointly invert to $(\sigma_{I,j}, \sigma_{x,j})$, and}
\begin{equation}\label{eq:Kinv}
K_j \;=\; \frac{\sigma_{I,j}^2\,(\tau_d + \tau_s)^2}{\gamma_j^2 - \sigma_{I,j}^2\,(\tau_d + \tau_s)} ,
\end{equation}
\textit{recovers informed-trading intensity from the liquidity vector, with turnover over-identifying $\sigma_{x,j}$ and the zero-return share serving as an auxiliary over-identifying check.} \textit{(Proof in \ref{app:proof_micro}.)}

We refer to the order flow from traders who trade for reasons unrelated to the firm's value, such as liquidity needs, as noise trading. The spread and the price-impact ratio identify the two unknowns, the informed investor's advantage ($\sigma_{I,j}$) and the amount of noise trading ($\sigma_{x,j}$). The spread reflects the adverse-selection cost incurred by the dealer per unit traded and does not depend on the amount of noise trading. This is because heavier noise trading spreads the informed investor's advantage over a larger order flow. Hence the market-impact cost falls while the trade volume rises by the same factor. The two effects cancel out. Thus, the spread identifies $\sigma_{I,j}$ on its own. The price-impact ratio depends on both unknowns. Given $\sigma_{I,j}$ from the spread, it identifies $\sigma_{x,j}$. $\sigma_{I,j}$ is then converted to the informed-trading intensity $K_j$ (Equation~\eqref{eq:Kinv}, \ref{sec:model_to_estimation}, \ref{app:liq_measures}). The identification carries one caveat. The spread's independence from the amount of noise trading is a property of the Kyle batch market. Thus, we treat the recovery as model-based and report over-identifying checks (\ref{app:liq_kyle}).

\ref{app:proof_micro} derives the dealer's exposure (equation~\eqref{eq:sigmaI}) with its comparative statics in $K_j$ and $\tau_s$, the Kyle equilibrium and the liquidity vector. The dealer conditions on the prior and the public signal alone. This information set is written $\mathcal{I}_{dealer}$. An informed investor observes in addition a private signal about the innovation $\epsilon_{f1}$. Thus, the informed investor knows everything in $\mathcal{I}_{dealer}$ and the private signal in addition. This larger information set is written $\mathcal{I}_{inf}$. The daily game endows both the dealer and the informed investor with the prior and the public signal but not with the quarterly price signal, whose precision is $\tau_p$. This is an abstraction. Because $\tau_p$ would enter both information sets, adding it lowers the level of $\sigma_{I,j}$ without changing the signs of its comparative statics in $K_j$ and $\tau_s$. The model measures the amount of noise trading in group $j$ by its standard deviation $\sigma_{x,j}$. We set $\sigma_{x,j}^2 = \Sigma_{jx}$, so the daily noise trading has the same variance as the quarterly noise trading in the same group.

The level of $\widehat{K}$ for foreign and domestic stocks is not comparable because $\widehat{K}$ is recovered from the spread in the daily dealer market where neither the dealer nor the informed investor conditions on the quarterly price signal, as a model simplification (\ref{app:dealer_detail}). The precision of that omitted signal ($\tau_p = K_j^2/(\gamma_j^2\bar{\rho}^2\Sigma_{jx})$ from equation~\eqref{eq:taup}) depends on $K_j$ and on the group parameters $\gamma_j$ and $\Sigma_{jx}$. Thus, the precision differs between domestic and foreign stocks. A group with more informed trading has a more informative price. Omitting that quarterly price signal loses more information for that group. Thus, this omission shifts the level of $\widehat{K}$ by an amount that differs across groups while leaving the comparative statics of the spread in $K_j$ unchanged. The order of stocks within a group is therefore preserved.

\subsection{Proof of Proposition \ref{prop:micro}}\label{app:proof_micro}

Throughout this proof the group index $j$ is suppressed, so $\gamma$, $K$, $\tau_s$, $\sigma_I$, and $\lambda$ denote the group-$j$ quantities $\gamma_j$, $K_j$, $\tau_s$, $\sigma_{I,j}$, and $\lambda_j$ of Proposition~\ref{prop:micro}. \textit{Informational disadvantage.} In the daily dealer market the dealer conditions on the prior and the public signal, so the dealer's conditional variance of the asset value is $\gamma^2/(\tau_d + \tau_s)$. An informed trader, who additionally holds the private precision $K$, faces $\gamma^2/(\tau_d + \tau_s + K)$.\footnote{The value is $v = \gamma(\upsilon d_0 + \bar{\epsilon} + \epsilon)$, so its one uncertain part is $\gamma\epsilon$. Conditioning on a set of independent normal signals with total precision $\tau$ leaves $\epsilon$ with posterior variance $1/\tau$, hence $v$ with variance $\gamma^2/\tau$. The dealer's total precision is $\tau_d + \tau_s$ and the informed trader's is $\tau_d + \tau_s + K$.} By the law of total variance, the variance of the informed trader's conditional expectation conditional on the dealer's information is
\begin{equation}
\sigma_I^2 \;=\; \mathrm{Var}\!\left(E[v \mid \mathcal{I}_{inf}] \,\middle|\, \mathcal{I}_{dealer}\right)
\;=\; \frac{\gamma^2}{\tau_d + \tau_s} \;-\; \frac{\gamma^2}{\tau_d + \tau_s + K}
\;=\; \frac{\gamma^2 K}{(\tau_d + \tau_s)(\tau_d + \tau_s + K)},
\end{equation}
equation \eqref{eq:sigmaI}, strictly increasing in $K$ (derivative $\gamma^2/(\tau_d+\tau_s+K)^2 > 0$) and strictly decreasing in $\tau_s$ (both factors of the denominator increase). If the dealer also conditions on past prices or any further public precision $\tau_{pub}$, all appearances of $\tau_d + \tau_s$ below become $\tau_d + \tau_s + \tau_{pub}$. The comparative statics in $K$ and $\tau_s$ and the inversion logic are unchanged. Level differences are absorbed in the estimated scales. The recovered levels of $\widehat{K}$, and with them $\widehat{\tau}_p$, do depend on this convention. This dependence is one reason the empirical stages condition on $\widehat{K}$ terciles rather than on levels and the joint estimation re-estimates $K$ together with the scale parameters. The tercile assignment ranks stocks on a scale-free statistic, the lagged spread multiplied by $\sqrt{\tau_d + \widehat{\tau}_s}$ at the same lag. The statistic is invariant to the estimated scales and orders with $\widehat{K}$ (see Section~\ref{sec:structural}).

\textit{Kyle equilibrium.} Let $m \equiv E[v \mid \mathcal{I}_{inf}]$ with $\mathrm{Var}(m \mid \mathcal{I}_{dealer}) = \sigma_I^2$, and the uninformed order quantity $x_n \sim N(0, \sigma_x^2)$. With linear pricing $p = \mu_D + \lambda q$, the informed trader maximizes $(m - \mu_D - \lambda q)q$, giving $q_{inf} = (m - \mu_D)/(2\lambda) \equiv \psi(m - \mu_D)$ with $\psi = 1/(2\lambda)$. Dealer efficiency requires $\lambda = \mathrm{Cov}(v, q)/\mathrm{Var}(q) = \psi\sigma_I^2/(\psi^2\sigma_I^2 + \sigma_x^2)$. Substituting $\psi = 1/(2\lambda)$ and solving yields $\lambda^2 = \sigma_I^2/(4\sigma_x^2)$, that is
\begin{equation}
\lambda \;=\; \frac{\sigma_I}{2\sigma_x}, \qquad \psi \;=\; \frac{\sigma_x}{\sigma_I},
\end{equation}
the standard solution of \citet{kyle1985continuous}. The total order quantity $q = \psi(m-\mu_D) + x_n$ has variance $\psi^2\sigma_I^2 + \sigma_x^2 = 2\sigma_x^2$, so $E|q| = \sqrt{2/\pi}\cdot\sqrt{2}\,\sigma_x = 2\sigma_x/\sqrt{\pi}$.\footnote{For a mean-zero normal variable with standard deviation $s$, the mean absolute value is $\sqrt{2/\pi}\,s$. Here $s = \sqrt{2}\,\sigma_x$, so $E|q| = \sqrt{2/\pi}\cdot\sqrt{2}\,\sigma_x$.}

\textit{Liquidity vector.} The expected effective half-spread, the adverse-selection cost per unit traded under the regret-free linear quotes, is
\begin{equation}
\lambda\,E|q| \;=\; \frac{\sigma_I}{2\sigma_x}\cdot\frac{2\sigma_x}{\sqrt{\pi}} \;=\; \frac{\sigma_I}{\sqrt{\pi}},
\end{equation}
proportional to $\sigma_I$ alone and exactly independent of $\sigma_x$. The data counterpart, the relative quoted spread, equals $a_s\sigma_I$ with $a_s$ absorbing $1/\sqrt{\pi}$, the division by the midpoint, the quoted-versus-effective wedge, and the daily-to-quarterly averaging. The Amihud ratio is the price move per unit of volume, $E|\Delta p|/E[\mathrm{vol}] \propto \lambda$, so $\mathrm{amihud} = a_\lambda\,\sigma_I/\sigma_x$. Expected volume is proportional to $E|q| \propto \sigma_x$, so $\mathrm{turnover} = a_T\,\sigma_x$, and the zero-return share, in the spirit of the measure of \citet{lesmond1999new}, provides an auxiliary trading-cost indicator used as an over-identifying check on the same pair. These are equations \eqref{eq:micro}. The signs in the proposition follow from the monotonicity of $\sigma_I^2$ established above together with $\partial\,\mathrm{spread}/\partial\Sigma_{jx} = 0$.

\textit{Inversion.} Given $(\widehat{\sigma}_I, \widehat{\sigma}_x)$ from spread and Amihud, rearranging $\sigma_I^2(\tau_d+\tau_s)(\tau_d+\tau_s+K) = \gamma^2 K$ gives $K\left[\gamma^2 - \sigma_I^2(\tau_d+\tau_s)\right] = \sigma_I^2(\tau_d+\tau_s)^2$, hence equation \eqref{eq:Kinv}. The denominator is positive whenever the data are generated by the model, since $\sigma_I^2 = \gamma^2 K/\left[(\tau_d+\tau_s)(\tau_d+\tau_s+K)\right] < \gamma^2/(\tau_d+\tau_s)$. The estimation reports any violation of this bound rather than truncating the estimate at it. Finally $\Sigma_{jx}$ follows from $\widehat{\sigma}_x$ up to the aggregation constants in $(a_\lambda, a_T)$, and $\tau_p = K^2/(\gamma^2\bar{\rho}^2\Sigma_{jx})$ completes the parameter map. $\square$

\subsection{Time-varying uncertainty}\label{sec:model_state}

This subsection deals with the state dependence of the drift channel (the condition under which next-quarter losses respond more to measured noise). That response depends on how much investors know from outside the feed. The prior is the part of that knowledge that moves with the state of the economy. Thus, this subsection lets the prior weaken in bad times.

\refstepcounter{assump}\label{assump:tvprior}\textbf{Assumption \theassump\ (time-varying prior precision).} $\tau_{d,q} = \tau_d \cdot h(\mathrm{FU}_q)$ with $h' < 0$, where $\mathrm{FU}_q$ is the aggregate financial-uncertainty index \citep{ludvigson2021uncertainty}.\footnote{The specifications use the financial-uncertainty index as the state variable. It is observed in all quarters of the panel. The quarter fixed effects absorb variation that is common to the stocks in a quarter, including the level of this index. Hence, the index enters the specifications only through its interaction with measured noise. The standard alternative state variable, the consumption-wealth ratio ($cay$) of \citet{lettau2001consumption}, is excluded from all the specifications. Its available vintage ends in 2019Q3. Conditioning on it would therefore discard the COVID period and after.}

The assumption allows the strength of prior beliefs to move with the state of the economy. The index $\mathrm{FU}_q$ measures financial uncertainty in the economy during quarter $q$. The prior precision $\tau_{d,q}$ is the confidence investors place in their prior view of the firm's dividend in that quarter. The prior precision in quarter $q$ is the baseline prior precision $\tau_d$ times the factor $h(\mathrm{FU}_q)$. The economic intuition here is that the prior view of the firm's dividend rests on the firm's cash-flow history (Section~\ref{sec:model_env}). When financial uncertainty is high, that history tells less about the next dividend, which makes investors then place less confidence in their prior view. Hence, the prior precision $\tau_{d,q}$ is lower in a high-uncertainty quarter.

\proposition{prop:state}~\textit{(Uncertainty amplification.) For measured noise below the turning point, the slope of expected overpricing, $\partial\,\mathrm{OP}/\partial w$, is decreasing in $\tau_0$. In high-uncertainty quarters $\tau_{d,q}$, and hence $\tau_0$, are low. The response of next-quarter losses to measured noise is steeper. The turning point $w^{*}$ shifts toward higher measured noise.} \textit{(Proof in \ref{app:proof_state}.)}

This proposition makes the drift channel (Proposition~\ref{prop:drift}) state dependent. Expected overpricing is the average amount by which optimism holds the quarter-end price above the firm's value $V^{*}_{f1}$ (equation~\ref{eq:op}). Measured noise ($w$) is the observed within-quarter volatility of the day's per-tweet sentiment average (equation~\ref{eq:tau_s}). The object of the proposition is the derivative of expected overpricing with respect to measured noise ($\partial\,\mathrm{OP}/\partial w$, the rate at which expected overpricing rises as measured noise rises) at values of measured noise below the turning point ($w^{*}$, the level of measured noise at which expected overpricing peaks, equation~\ref{eq:peak}). This derivative depends on the outside knowledge (what investors know from sources other than the feed), whose precision ($\tau_0$) is the sum of the precisions of the prior, of the price signal and of the private signals (equation~\ref{eq:peak}). The precision of the public signal ($\tau_s$) is therefore not part of $\tau_0$. Each of the three precisions in the sum takes one value for each stock group (Section~\ref{sec:model_env}). The three do not vary across investors. The private-signal precision enters as the aggregate $K_j$ over all investors. When outside knowledge is strong, the signal noise has a small impact on price (equation~\ref{eq:coefs}). When outside knowledge is weak, the signal noise then produces a large impact on price (equation~\ref{eq:coefs}). A larger price impact of an optimistic error (a positive realization of the signal noise $u_{f1}$) means a larger expected overpricing (equation~\ref{eq:op}). Thus, the derivative is larger when outside knowledge is weaker. Of the three precisions in $\tau_0$, only the prior precision depends on the state of the economy (\ref{app:proof_state}). Quarters with higher uncertainty weaken the prior by assumption (\ref{assump:tvprior}), which lowers $\tau_0$. Hence, the derivative is larger in high-uncertainty quarters. A larger expected overpricing implies a deeper correction the following quarter (Proposition~\ref{prop:drift}). Thus, the next-quarter losses respond more to measured noise in high-uncertainty quarters. The turning point also shifts. The peak of expected overpricing lies where the feed's precision equals the precision of outside knowledge ($\tau_0$, equation~\ref{eq:peak}). Weaker outside knowledge places the peak at a noisier feed, which in measured units is a higher turning point (\ref{app:proof_state}). The result rationalizes the empirical pattern in \citet{garcia2013sentiment} that the return-predictive content of sentiment concentrates in bad times. Proposition~\ref{prop:state} is tested by adding an interaction of measured noise with the financial-uncertainty index to the regression of next-quarter losses on current-quarter measured noise (row R7, Table~\ref{tab:battery}).

\subsection{Proof of Proposition \ref{prop:state}}\label{app:proof_state}

Under \ref{assump:tvprior}, a rise in financial uncertainty lowers $\tau_{d,q}$ and hence $\tau_0 = \tau_{d,q} + \tau_p + K$, since $\tau_p$ in \eqref{eq:taup_sol} does not involve $\tau_{d,q}$. Two claims follow. First, the peak $w^{*} = \sqrt{a/\tau_0}$ from \eqref{eq:OPw} is strictly decreasing in $\tau_0$, because $\tau_0$ enters under the square root in the denominator. Therefore, the peak shifts toward higher measured noise when uncertainty is high. Second, take the slope $\partial\,\mathrm{OP}/\partial w$ displayed in \eqref{eq:OPw} and differentiate it with respect to $\tau_0$ at fixed $w$. Writing $c \equiv \delta\gamma\sqrt{a}/\sqrt{2\pi}$ for the constant in \eqref{eq:OPw},
\begin{equation}
\frac{\partial}{\partial \tau_0}\!\left(\frac{\partial\,\mathrm{OP}}{\partial w}\right)
= c\,\frac{-w^2\left(\tau_0 w^2 + a\right)^2 - \left(a - \tau_0 w^2\right)\cdot 2\left(\tau_0 w^2 + a\right)w^2}{\left(\tau_0 w^2 + a\right)^4}
= -\,c\,\frac{w^2\left(3a - \tau_0 w^2\right)}{\left(\tau_0 w^2 + a\right)^3}.
\end{equation}
For $w < w^{*}$ we have $\tau_0 w^2 < a$, so $3a - \tau_0 w^2 > 2a > 0$ and the expression is strictly negative. Therefore, the slope of expected overpricing in measured noise is larger when $\tau_0$ is lower. Low $\tau_0$ is the high-uncertainty state (\ref{assump:tvprior}). $\square$

\subsection{Partial observation of the public signal}\label{app:theta_lemma}

In this subsection, we derive the claim in the footnote to \ref{hyp:mismatch}. We take common observation as the baseline, since the feed's aggregate sentiment is a freely observable public series while producing firm-specific knowledge is the costly activity \citep{van2009information}. Fix a group and suppress its index. A mass $\theta \in (0,1]$ of investors observe the feed's public signal ($s$). We call them the observers, and the remaining investors, who do not observe $s$, the non-observers. We add one behavioral restriction. Every investor conditions on the netted price signal ($\zeta$) of \ref{app:proof_price} and does not invert the price further. This restriction is behavioral instead of informational as $\zeta$ and the observed price would reveal $s$. Without such a restriction, a rational non-observer could recover a mixed signal from the price, with precision that depends on $\tau_s$. We impose the restriction to rule this out, which is what makes the reduction below exact. We keep the other primitives as in \ref{app:proof_price}. An observer has posterior variance $\Sigma^i = (\tau_d + \tau_p + k^i + \tau_s)^{-1}$ and precision-weighted posterior mean $\tau_p \zeta + k^i \eta^i + \tau_s s$ (\ref{app:proof_price}). A non-observer has posterior variance $(\tau_d + \tau_p + k^i)^{-1}$, with the signal term $\tau_s s$ gone from the mean. The market-clearing condition of \ref{app:proof_price} depends on investor information only through two sums across investors, the sum of the precision-weighted posterior means and the sum of the posterior precisions. Summing the precision-weighted posterior means gives
\[
\tau_p\,\zeta \;+\; K\,\epsilon \;+\; \theta\,\tau_s\, s .
\]
The private terms aggregate to $K\epsilon$ as in \ref{app:proof_price}. Only the mass $\theta$ of observers has the signal term, leaving $\theta\tau_s s$. The posterior precisions sum across investors to
\[
\bar{T} \;=\; \tau_d + \tau_p + K + \theta\,\tau_s .
\]
Every appearance of $\tau_s$ in the market-clearing condition of \ref{app:proof_price} lies in these two sums, and the price-signal precision $\tau_p = K^2/(\gamma^2\bar{\rho}^2\Sigma_{jx})$ does not involve $\tau_s$ (equation~\ref{eq:taup}). Thus, the market-clearing condition coincides with that of a full-observation economy with signal precision, $\theta\tau_s$. The price coefficients of \eqref{eq:coefs} then follow with $\tau_s$ replaced by $\theta\tau_s$, so the equilibrium price function is exactly that of the $\theta\tau_s$ economy. $\square$

This equivalence is only about the price function. The realized return has the sentiment shock, so the return variance and the expected overpricing under partial observation are not those of the $\theta\tau_s$ economy. We use no return-level quantity under partial observation in the estimation. This scope is stated in the footnote to \ref{hyp:mismatch}.

We now make the footnote's bound precise. Write the price-relevant precision of group $j$ as $\theta_j\tau_{s,j}$. When $\theta_F \le \theta_D$, we have $\theta_D\tau_{s,D}/(\theta_F\tau_{s,F}) \ge \tau_{s,D}/\tau_{s,F}$, so the measured precision ratio $\tau_{s,D}/\tau_{s,F}$ is a lower bound on the combined price-relevant ratio. For the level difference instead, with a common $\theta < 1$, the combined difference is $\theta(\tau_{s,D}-\tau_{s,F})$, which stays below the measured difference $\tau_{s,D}-\tau_{s,F}$ when $\tau_{s,D} > \tau_{s,F}$, the direction \ref{hyp:mismatch} maintains.

\section{Supporting Derivations}\label{app:basic_derivations}
\subsection{Value of the dividend claim}\label{app:proof_value}
We derive two environment statements. The first is the equivalence between the dividend law and its one-step recursion. The second is the value of the claim.

For the first, fix $s \geq 2$ and factor one power out of the law, $d^{*}_{fs} = \upsilon_j^{\,s-1} d^{*}_{f1} = \upsilon_j\,(\upsilon_j^{\,s-2} d^{*}_{f1}) = \upsilon_j\, d^{*}_{f,s-1}$. Conversely, the recursion $d^{*}_{f,s+1} = \upsilon_j\, d^{*}_{fs}$ with the period-1 dividend as initial condition gives $d^{*}_{fs} = \upsilon_j^{\,s-1} d^{*}_{f1}$ by induction on $s$. The law carries no error term. Hence, the stream after period 1 is deterministic given $d^{*}_{f1}$.

For the second, the position pays the period-1 value of the stream, $V^{*}_{f1} = \sum_{s \geq 1} d^{*}_{fs}/r^{\,s-1}$. A dividend paid in period $s$ arrives $s-1$ periods after period 1. One unit invested at the gross rate grows to $r^{\,s-1}$ over those periods, which is the discounting. Substituting the law and writing $q \equiv \upsilon_j/r$ gives $V^{*}_{f1} = d^{*}_{f1} \sum_{k \geq 0} q^k$. The partial sum $S_N = 1 + q + \cdots + q^N$ satisfies $S_N - q S_N = 1 - q^{N+1}$, so $S_N = (1 - q^{N+1})/(1-q)$. With $\upsilon_j < r$ the ratio $q$ lies strictly below one, the tail $q^{N+1}$ vanishes, and the sum equals $1/(1-q) = r/(r-\upsilon_j) = \gamma_j$. Hence $V^{*}_{f1} = \gamma_j\, d^{*}_{f1}$. All steps are identities. Thus, the formula is exact. The gross convention is what makes the per-period ratio exactly $\upsilon_j/r$, with no expansion of one plus a net rate anywhere.

\section{Additional Related Literature}\label{app:lit_extra}

\citet{hong2007disagreement} survey the literature on heterogeneous beliefs, where divergence of opinions raises prices under short-sale constraints because optimists set the price \citep{miller1977risk}. \citet{diether2002differences} show how the analyst-forecast dispersion predicts lower subsequent returns. \citet{cookson2020agree} show that disagreement driven by different information is more tightly related to trading volume than disagreement driven by different investment approaches. \citet{cookson2020partisan} show that partisan identity shapes how investors read the same signals. The last three papers measure the variance of stated beliefs. Our dispersion measure instead helps capture the composition of the contributor pool. Dispersion is the ratio of the number of unique posting accounts to the number of tweets. A low ratio indicates that a few accounts are posting repeatedly about a stock. A high ratio indicates many accounts each posting rarely about a stock. Accounts that post repeatedly about a firm are specialists \citep{van2009information} or automated accounts, while accounts which post rarely are attention-driven contributors \citep{barber2008all}. In the model, the tweets of accounts that post rarely carry the firm-specific news together with the day's fad (a transient component that carries none of that news), while the tweets of specialists carry the firm-specific news without the fad. Thus, at a given fad variance, a larger share of tweets from accounts that post rarely implies a larger nonfundamental component of the feed (Section~\ref{sec:model_sent}). The presence of a fad in such tweets is further supported by field evidence that people's beliefs and decisions reflect overconfidence, limited attention, persuasion and emotion \citep{dellavigna2009psychology}.

Our paper also relates to the literature on cross-listing. \citet{karolyi1998companies} surveys the evidence and reports favorable share-price reactions in the first month after listing, an overall improvement in liquidity, and a reduction in domestic market risk that can lower the cost of equity capital. \citet{foerster1999effects} find that firms cross-listing as American Depositary Receipts earn abnormal returns before and at listing and lose part of them in the year after. The pattern is related to an expansion of these firms' shareholder base and to the capital raised at listing. It supports the market segmentation hypothesis and the investor recognition hypothesis of \citet{merton1987simple}. \citet{eun2003cross} find that for Canadian stocks listed in both Toronto and the United States, prices in the two markets are cointegrated and mutually adjusting, with the home market contributing most of the price discovery on average. \citet{gagnon2010multi} find that deviations from price parity between U.S. cross-listed shares and their home-market shares are small on average but volatile and are related to proxies for holding costs that can impede arbitrage. Our producer-knowledge mismatch hypothesis rests on the premise that the home-market investors of a foreign firm tend to possess an informational advantage about it (\ref{hyp:mismatch}). Since ADR prices are linked to home-market prices through the depositary structure, the finding that the domestic-foreign difference in the slope of next-quarter losses on sentiment noise is strongest for ADRs admits a mechanical explanation (Section~\ref{sec:reducedform}).

\end{document}